\documentclass[12pt,amsmath,amssymb,showkeys,prb]{revtex4}
\usepackage{xcolor}
\usepackage{setspace}
\usepackage{graphicx}
\usepackage{mathrsfs}

\DeclareSymbolFont{matha}{OML}{txmi}{m}{it}
\DeclareMathSymbol{\varw}{\mathord}{matha}{119}   
\newcommand{\citen}[1]{%
  \begingroup
    \romannumeral-`\x 
    \setcitestyle{numbers}%
    \cite{#1}%
  \endgroup
}
\begin{document}

$\null$
\hfill {October 6, 2020}
\vskip 0.3in

\begin{center}

{\Large\bf Comparative Roles of Charge, $\pi$ and Hydrophobic}\\

\vskip 0.3cm

{\Large\bf Interactions in Sequence-Dependent Phase}\\

\vskip 0.3cm

{\Large\bf Separation of Intrinsically Disordered Proteins}\\

\vskip .5in
{\bf Suman D{\footnotesize{\bf{AS}}}},$^{1}$
{\bf Yi-Hsuan L{\footnotesize{\bf{IN}}}},$^{1,2}$
{\bf Robert M. V{\footnotesize{\bf{ERNON}}}},$^{2}$
{\bf Julie D. F{\footnotesize{\bf{ORMAN}}}-K{\footnotesize{\bf{AY}}}}$^{2,1}$
 and
{\bf Hue Sun C{\footnotesize{\bf{HAN}}}}$^{1,*}$

$\null$

$^1$Department of Biochemistry,
University of Toronto, Toronto, Ontario M5S 1A8, Canada;\\
$^2$Molecular Medicine, Hospital for Sick Children, Toronto, 
Ontario M5G 0A4, Canada\\

\vskip 1.3cm

%

\end{center}

\vskip 1.3cm

\noindent
$*$Corresponding author\\
{\phantom{$^\dagger$}}
E-mail: chan@arrhenius.med.utoronto.ca;
Tel: (416)978-2697; Fax: (416)978-8548\\
{\phantom{$^\dagger$}}
Mailing address:\\
{\phantom{$^\dagger$}}
Department of Biochemistry, University of Toronto,
Medical Sciences Building -- 5th Fl.,\\
{\phantom{$^\dagger$}}
1 King's College Circle, Toronto, Ontario M5S 1A8, Canada.\\

\vfill\eject

\noindent
{\large\bf Abstract}\\

\noindent
Endeavoring toward a transferable, predictive coarse-grained explicit-chain
model for biomolecular condensates underlain by liquid-liquid phase separation
(LLPS) of proteins and nucleic acids, we conducted multiple-chain
simulations of the N-terminal intrinsically disordered region (IDR) of 
DEAD-box helicase Ddx4, as a test case, to assess the roles of electrostatic,
hydrophobic, cation-$\pi$, and aromatic interactions in amino acid
sequence-dependent LLPS. We evaluated three different residue-residue 
interaction schemes
with a shared electrostatic potential. Neither a common
hydrophobicity scheme nor one augmented with arginine/lysine-aromatic 
cation-$\pi$
interactions consistently accounted for the available experimental LLPS data
on the wildtype, a charge-scrambled, a phenylalanine-to-alanine (FtoA), and an
arginine-to-lysine (RtoK) mutant of Ddx4 IDR. In contrast, interactions
based on contact statistics among folded globular protein structures reproduce
the overall experimental trend, including that the RtoK mutant has a much
diminished LLPS propensity. 
Consistency between simulation and LLPS experiment
was also found for RtoK mutants of P-granule protein LAF-1, 
underscoring that, to a degree, the important LLPS-driving
$\pi$-related interactions are embodied in classical statistical potentials.
Further elucidation will be necessary, however, especially of
phenylalanine's role in condensate assembly because experiments on FtoA and
tyrosine-to-phenylalanine mutants suggest that LLPS-driving phenylalanine
interactions are significantly weaker than those posited by common
statistical potentials. Protein-protein electrostatic interactions are
modulated by relative permittivity, which in general depends on aqueous protein
concentration. Analytical theory suggests that this dependence entails enhanced
inter-protein interactions in the condensed phase but more favorable
protein-solvent interactions in the dilute phase. The opposing trends lead to
only a modest overall impact on LLPS.

\vfill\eject

\noindent
{\large\bf Significance Statement}\\

\noindent
Mesoscopic condensates of proteins and nucleic acids, including various
``membraneless organelles'', serve myriad 
biological functions, whereas dysregulation
of condensates can cause disease. 
Deciphering how condensates are
governed by genetically encoded biomolecular sequences is of central
importance. Critical assessment of how various interactions reproduce 
experimental
phase behaviors indicates that $\pi$-interactions' notable facilitation of
condensate formation is well reflected in established contact statistics among
folded protein structures, although the condensate-driving capability of the 
large hydrophobic/aromatic residue phenylalanine is weaker than expected. By 
elucidating physical aspects of the residue-residue contacts in condensates,
highlighting these contacts' largely solvent-exposed character with 
ramifications such as reduced hydrophobic strengths relative to buried
nonpolar contacts, fundamental conceptual and quantitative progress is made 
toward predictive models for biomolecular condensates.

\vfill\eject

\noindent
{\large\bf INTRODUCTION}\\

A preponderance of recent advances demonstrate
that liquid-liquid phase separation
(LLPS) of intrinsically disordered proteins (IDPs), proteins containing 
intrinsically disordered regions (IDRs), folded proteins, and nucleic acids 
is a general biophysical mechanism to achieve
functional spatial and temporal organization of biomolecules
in both intra- and extra-cellular organismal 
space.\cite{brangwynne2009,Rosen12,McKnight12,Nott15,tanja2015,parker2015,mingjie2018,keeley2018,nedelsky2019}
LLPS underpins formation of a variety of biomolecular 
condensates,\cite{rosen2017} including intracellular bodies, such as 
nucleoli and stress granules, that are often referred to as membraneless 
organelles,\cite{Nott15,shorter2019} 
and precursor of extracellular materials as in 
the case of sandcastle worm adhesive\cite{jones2017} and
elastin in vertebrate tissues\cite{keeley2018}.
These dynamic, phase-separated condensates perform
versatile functions, as underscored by their recently elucidated roles in
synapse formation and plasticity,\cite{mingjie2018,mingjie2020}
organization of chromatin,\cite{rosen19}
regulation of translation,\cite{BrianTsang2019,julie2019}
B cell response,\cite{griesinger2020}
and autophagosome formation.\cite{noda2020}
The pace of discovery in this very active area of research has 
been accelerating.\cite{cliff2017,biochemrev,Monika2018Rev,julieRev,AlexRev,perryRev,jeetainRev,tanja2019,Roland2019,choiRev2020}

While experimental progress has been tremendous, theory for the
physico-chemical basis of biomolecular condensates is still in 
its infancy. Biomolecular condensates in vivo are complex, involving
many species of proteins and nucleic acids maintained often by 
non-equilibrium processes,\cite{cliff2017,rosen2017,chong2016,babu2018,lee2018} 
rendering atomistic modeling impractical. Facing this challenge, 
promising initial theoretical steps using coarse-grained approaches 
were made to tackle
simpler in-vitro LLPS systems, as their elucidation is a prerequisite for 
physical insights into more complex in vivo condensates. These recent efforts 
encompass analytical theories at various levels of 
approximation,\cite{CellBiol,NatPhys,linPRL,linJML,lin2017,njp2017,singperry2017,moleculargrammar,schmitJACS2020,kings2020,Alan2020}
field theory simulations,\cite{joanElife,joanJPCL,joanPNAS,joanJCP}
and lattice\cite{suman1,stefan2019,lassi,anders2020} or 
continuum\cite{dignon18,suman2,jeetainACS,Panag2020} 
coarse-grained explicit-chain simulations that account for either
individual amino acid residues\cite{dignon18,suman1,jeetainPNAS,suman2} 
or, at lower structural resolution,
groups of residues\cite{Ruff15,harmonNJP}---including 
using patchy particle representations.\cite{hxzhou2018,hxzhouPNAS2019}
The different theoretical/computational approaches are complementary, and
were applied to address how amino acid composition (number/fraction of 
hydrophobic,\cite{Panag2020} 
aromatic,\cite{moleculargrammar,TanjaScience2020} or charged\cite{Nott15} 
residues) and the sequence pattern of 
charge,\cite{linPRL,suman2,stefan2019,koby2020} 
hydrophobic,\cite{jeetainACS,anders2020,Panag2020} or 
aromatic\cite{TanjaScience2020} 
residues affect LLPS propensity of heteropolymers
as well as pertinent impact of 
temperature,\cite{biochemrev,jeetainACS,joanElife,Panag2020} 
hydrostatic pressure,\cite{roland18,rolandJACS19,roland20} 
salt,\cite{joanJCP,kings2020} and osmolyte,\cite{Roland2019,rolandJACS19} 
offering physical insights into the 
LLPS behaviors of, for example, the DEAD-box RNA helicase 
Ddx4,\cite{linPRL,jacob2017} 
RNA-binding protein fused in sarcoma (FUS),\cite{dignon18}
prion-like domains,\cite{TanjaScience2020} and 
postsynaptic densities.\cite{roland20}

Developing LLPS models with transferable interaction 
potentials applicable to a wide range of amino acid sequences is essential 
for advancing fundamental physical understanding of natural biomolecular 
condensates and engineering of bio-inspired materials.\cite{chilkoti2020}
In this endeavor, the rapidly expanding repertoire of experimental data offers
critical assessment of theoretical and computational approaches.
Building on aforementioned progress,\cite{linPRL,kings2020,dignon18,suman2} 
the present study evaluates a variety of 
interaction schemes for coarse-grained residue-based chain simulations 
of LLPS of IDPs or IDRs,
including but not limited to schemes proposed in the literature.\cite{dignon18} 
We do so by first comparing their sequence-specific predictions against 
experiment on the RNA helicase Ddx4 
for which extensive LLPS data on the
wildtype (WT) and three mutant sequences are avaialable to probe the
contribution of hydrophobic, electrostatic,\cite{Nott15} 
cation-$\pi$, and possibly other $\pi$-related\cite{Nott15,jacob2017,robert} 
interactions.
We use these data to benchmark the relative strengths of different 
types of interaction in our model.
Of particular interest are the aromatic\cite{diederich} and other
$\pi$-related\cite{robert} interactions, which have significant impact on folded
protein structure, conformational distribution of IDPs and 
LLPS properties,\cite{dougherty1999,dougherty2000,crowley2005,kaw2013,cosb15,Nott15,linPRL,moleculargrammar,TanjaScience2020} 
but are often not adequately accounted for in model potentials.\cite{robert}
Interestingly, a simple statistical potential based upon
folded protein structures\cite{MJ85,MJ96} accounts for the general trend
of LLPS properties of the four Ddx4 IDR sequences, including LLPS is
more favored by arginine than lysine despite their essential identical
electric charges, but a model potential that rely solely on 
hydrophobicity\cite{rossky2014} does not. This finding indicates that, at 
the coarse-grained level of residue-residue interactions, IDP/IDR LLPS is 
governed largely by similar forces---including the $\pi$-related ones---that 
drive protein folding. 
Analogous argreement between statistical potential 
model prediction and experiment with respect to the arginine/lysine 
contrast is also found for the N-terminal RGG domain of 
P-granule RNA Ddx3 helicase LAF-1.\cite{Schuster2020} 
However, experimental data on tyrosine-to-phenylalanine mutants
of LAF-1 and FUS\cite{moleculargrammar}
indicate that the contribution of the large aromatic
residue phenylalanine to LLPS is overestimated by statistical potentials, 
most likely because the interactions involving phenylalanine in the 
sequestered hydrophobic core of globular proteins are not sufficiently 
representative of more solvent-accessible LLPS-driving interactions,
pointing to a crucial aspect of LLPS energetics that future investigations
should be directed. To gain further insights into the electrostatic 
driving forces for LLPS, we have conducted explicit-water 
simulation and develop new analytical 
theory which suggest, at variance with previous analyses,\cite{linJML,njp2017} 
that the physically expected dependence of effective permittivity on IDR 
concentration may have a modest instead of drastic impact on LLPS propensity 
because of a tradeoff between solvent-mediated electrostatic interchain 
interactions and self-interactions. These findings and their 
ramifications are discussed below.

$\null$

\noindent
{\large\bf RESULTS AND DISCUSSION}\\

As described in Materials and Methods and SI Appendix, SI Text,
our coarse-grained protein chain model for IDP LLPS basically
follows the approaches in Refs.~ \citen{dignon18,suman2}, which
in turn are based on a recently proposed efficient simulation
protocol.\cite{panag2017}
In the first step of our analysis, we critically assess the models 
against the experimental data on the Ddx4 IDR (SI Appendix, Fig.~S1), 
which indicate that 
all three Ddx4 IDR mutants---the charge scrambled (CS), 
phenylalanine-to-alanine (FtoA), and arginine-to-lysine (RtoK) 
variants---have significantly reduced
LLPS propensities relative to the WT.\cite{Nott15,jacob2017,robert} 
The CS, FtoA, and RtoK variants are useful probes for LLPS energetics.
They were constructed specifically to study the experimental
effects of sequence charge pattern (the arrangement of charges 
along CS sequence is less blocky than that in WT while the 
amino acid composition is unchanged), the relative importance of 
aromatic/$\pi$-related vs hydrophobic/nonpolar interactions 
(all 14 Phe residues in WT Ddx4 IDR are mutated to Ala in FtoA), 
and the role of Arg vs Lys (all 24 Arg residues in WT IDR are mutated 
to Lys in RtoK) on the LLPS of Ddx4.

{\bf Assessing Biophysical Perspectives Embodied by Different
Coarse-Grained Interaction Schemes For Modeling Biomolecular Condensates.}
We consider the potential functions in the hydrophobicity-scale (HPS)
and the Kim-Hummer (KH) models in Dignon et al.\cite{dignon18} 
as well as the HPS potential with augmented cation-$\pi$ terms,\cite{kaw2013}
all of which share the same bond energy term, $U_{\rm bond}$, for chain 
connectivity and screened electrostatic term, $U_{\rm el}$, for pairs of
charged residues, as described in SI Appendix, SI Text. We focus
first on the pairwise contact interactions between amino acid residues, 
which correspond to the $U_{\rm aa}$ energies of either the HPS or KH model
(excluding $U_{\rm bond}$ and $U_{\rm el}$).

The HPS model assumes that the dominant driving force for IDP LLPS is
hydrophobicity as characterized by a scale for the 20 amino acid residues. 
Pairwise contact energy is taken to be the sum of the hydrophobicities of 
the two individual residues of the pair. The HPS model adopts the
scale of Kapcha and Rossky, in which the hydrophobicity of a residue 
is a composite quantity based on a binary hydrophobicity scale of the
atoms in the residue.\cite{rossky2014}

In contrast, the KH model\cite{KH} relies on knowledge-based potentials
derived from contact statistics of folded protein structures in the 
Protein Data Bank (PDB).
As such, it assumes that the driving forces for IDP LLPS are essentially 
identical to those for protein folding at a coarse-grained residue-by-residue 
level, as obtained by Miyazawa and Jernigan,\cite{MJ96} without singling 
out a priori a particular interaction type as being dominant.

The HPS model has been applied successfully to study the
FUS low-complexity-domain,\cite{fawzi-mittal2019}
the RNA-binding protein TDP-43,\cite{fawzi-mittal2020}
and the LAF-1 RGG domain as well as its charge shuffled 
variants.\cite{Schuster2020} A temperature-dependent version of HPS
(HSP-T)\cite{jeetainACS} was also able to rationalize the experimental
LLPS properties of artificial designed sequences.\cite{chilkoti2015}
When both the HPS and KH models were applied to FUS and LAF-1, the
predicted phase diagrams were qualitatively similar for a given sequence
though they exhibited significantly different critical 
temperatures,\cite{dignon18} which should be attributable to the 
difference in effective energy/temperature scale of the two models.
Here we conduct a systematic assessment of the two models' 
underlying biophysical assumptions by evaluating their ability to provide 
a consistent rationalization of the LLPS properties of a set of IDR sequences.

The scatter plot in Fig.~1a of HPS and KH energies indicates that, despite 
an overall correlation, there are significant outliers.
The most conspicuous outliers are interactions involving Arg (red),
which are much less favorable in HPS than in KH. 
By comparison, most of the interactions involving Pro, as depicted by the 
16 outlying blue circles as well as the single yellow and single green 
circles to the left of the main trend, are considerably more favorable in HPS 
than in KH. Interactions involving Phe (yellow) and Lys (green) 
essentially follow the main trend. Those involving Phe are favorable
to various degrees in both models. However, some interactions involving Lys
are attractive in HPS but repulsive in KH. For example, Lys-Lys 
interaction is attractive at $\approx -0.1$ kcal mol$^{-1}$ for HPS but 
is repulsive at $\approx +0.2$ kcal mol$^{-1}$ for KH.
Figure~1b underscores the difference in interaction pattern of the
two models for the WT Ddx4 IDR. The KH pattern is
clearly more heterogeneous with both attractions and repulsions,
whereas the HPS pattern is 
more uniform with no repulsive interactions. These differences should
lead to significantly different predictions in
sequence-dependent LLPS properties, as will be explored below.

Because of the importance of cation-$\pi$ interactions in protein 
folded structure\cite{dougherty1999} as well as conformational distribution 
of IDP and LLPS,\cite{kaw2013,cosb15,Nott15,moleculargrammar,Schuster2020} 
we study another set of model interaction schemes---in addition to HPS 
and KH, referred to as HPS+cation-$\pi$---that augment the HPS potentials
with terms specific for cation-$\pi$ interactions between Arg or Lys and the 
aromatic Tyr, Phe, or Trp (Fig.~2). As explained in the SI Appendix, SI Text, 
we consider two alternate scenarios: (i) the cation-$\pi$ interaction 
strength is essentially uniform, irrespective of the cation-aromatic 
pair (Fig.~2a), as suggested by an earlier analysis;\cite{dougherty1999}
and (ii) the cation-$\pi$ interaction strength is significantly stronger
for Arg than for Lys (Fig.~2b). The latter scheme is motivated by
recent experiments showing that Arg to Lys substitutions reduce LLPS 
propensity, as in the cases of the RtoK mutant of Ddx4 IDR\cite{robert} 
and variants of FUS\cite{moleculargrammar} 
and LAF-1,\cite{Schuster2020} as well as a recent 
theoretical investigation pointing to different roles of Arg and Lys 
in cation-$\pi$ interactions.\cite{kumar2018} Contact statistics of PDB 
structures, including those of Miyazawa and Jernigan\cite{MJ85,MJ96}
on which the KH potential is based, may also suggest that Arg-$\pi$ attractions 
are stronger than Lys-$\pi$'s. Indeed, among a set of 6,943 high-resolution 
X-ray protein structures,\cite{robert} we find that an Arg-aromatic pair
is at least 75\% more likely than a Lys-aromatic pair to be within a
C$_\alpha$--C$_\alpha$ distance of $\le 6.5$ \AA~(SI Appendix, Fig.~S2),
a separation that is often taken as a criterion for 
residue-residue contact.\cite{MJ85} On top of that, given an Arg-aromatic 
and a Lys-aromatic pair are separated by the same C$_\alpha$--C$_\alpha$ 
distance (Fig.~2c), the Arg-aromatic pair (solid curves) are more likely than 
the Lys-aromatic pair (dashed curves) to adopt configurations consistent 
with a cation-$\pi$ interaction.
We emphasize, however, that although a significantly stronger 
Arg- than Lys-associated cation-$\pi$ interaction 
is explored here as an alternate scenario, 
it is probable, as argued by Gallivan and Dougherty using a comparison 
between Lys-like ammonium-benzene and Arg-like guanidinium-benzene
interactions, that the strengths of the ``pure'' cation-$\pi$ parts
of Arg- and Lys-aromatic interactions are similar despite the relative 
abundance of Arg-aromatic contacts due to other factors\cite{dougherty1999} 
such as $\pi$-$\pi$ effects.\cite{robert}

{\bf Hydrophobicity, Electrostatics And Cation-$\pi$ Interactions
Are Insufficient By Themselves To Rationalize Ddx4 LLPS Data In Their Entirety.}
We begin our assessment of models by applying the HPS and HPS+cation-$\pi$
potentials to simulate the phase diagrams of the four Ddx4 IDRs (SI Appendix,
Fig.~S1),
the sequence patterns of which are illustrated in Fig.~3a 
using a style
employed previously, e.g., in Refs.~\citen{Nott15,linPRL,moleculargrammar}. 
The phase diagrams presented here are coexistence curves. When the overall 
average IDR density lies in between the left and right arms of 
the coexistence curve at a given temperature, the system phase separates into
a dilute phase and a condensed phase with IDR densities given, respectively, 
by the low- and high-density values of the coexistence curve at the given 
temperature. When the average IDR density is not in the region underneath
the coexistence curve, the system is not phase separated
(see, e.g., Refs.~\citen{linJML,biochemrev,Rohit2008} 
for introductory discussions).
Consistent with experiments,\cite{Nott15,jacob2017} the simulated phase 
diagrams (Fig.~3b) exhibit upper critical solution temperatures, which
is a maximum temperature above which phase separation does not occur
(corresponding to the local maxima of the coexistence curves at IDR
density $\approx 200$ mg/ml in Fig.~3). We emphasize, however, that although
the simulated critical temperatures are assuringly in the same range as those 
deduced experimentally,\cite{jacob2017} the model temperature (in K) of our 
simulated phase diagrams in Figs.~3b and 4
should not be compared directly with experimental temperature. This is 
because not only of uncertainties about the overall model energy
scale but also because the models do not account for the temperature dependence 
of effective residue-residue interactions.\cite{biochemrev,jeetainACS,joanElife}
For simplicity, our models include only temperature-independent energies 
as they are purposed mainly for comparing the LLPS propensities of different 
amino acid sequences on the same footing rather than for 
highly accurate prediction of LLPS behaviors of any individual sequence.

The leftmost panel of Fig.~3b provides the HPS phase diagrams 
at relative permittivity $\epsilon_{\rm r}=80$ (approximately 
equal to that of bulk water, as in Ref.~\citen{dignon18}).
The predicted behaviors of the CS and FtoA variants
are consistent with experiments in that their LLPS propensities are reduced
relative to WT;\cite{Nott15,jacob2017} but the predicted enhanced LLPS 
propensity of RtoK is opposite to the experimental finding of Vernon et al. 
that the LLPS propensity of RtoK is lower than that of WT.\cite{robert}
This shortcoming of the HPS model is a consequence of its assignment of
much less favorable interactions for Arg than for Lys, as noted in Fig.~1a.

The other panels of Fig.~3b provide the HPS+cation-$\pi$ phase diagrams.
They are computed for $\epsilon_{\rm r}=80$, $40$, and $20$ to gauge the 
effect of electrostatic interactions relative to other types of interactions.
The $\epsilon_{\rm r}$-dependent results are also preparatory for 
our subsequent investigation of the effect
of IDR-concentration-dependent permittivity\cite{linJML,njp2017}
on predicted LLPS properties. 
All of the HPS+cation-$\pi$ phase diagrams
here still disagree with experiment\cite{jacob2017,robert} as they all 
predict a higher LLPS propensity for the RtoK variant than for WT. Apparently, 
the bias of the HPS potential against Arg interactions is so strong that 
it cannot be overcome by
additional Arg-aromatic interactions that are reasonably more 
favorable than Lys-aromatic interactions (Fig.~2b). 
The $\epsilon_{\rm r}=80$ results for both uniform and variable cation-$\pi$
strength exhibit an additional mismatch: Contrary to 
experiments,\cite{Nott15,jacob2017} they predict similar LLPS propensities
for the CS variant and WT, suggesting that under this dielectric condition,
electrostatic interactions are unphysically overwhelmed by the presumed 
cation-$\pi$ interactions. The $\epsilon_{\rm r}=20$ results for variable 
cation-$\pi$ also indicate an additional mismatch, in this case they fail
to reproduce the experimental trend of a significantly lower LLPS
propensity of the FtoA variant relative to that of WT,\cite{Nott15}
probably because the relatively weak cation-$\pi$ contribution is overwhelmed 
by strong electrostatic interactions in this low-$\epsilon_{\rm r}$ situation.
Taken together, although a perspective involving only 
electrostatic and cation-$\pi$
interactions was adequate to account for sequence-specific LLPS
trend of WT and CS (and possibly also FtoA) before the RtoK experiment
was performed,\cite{linPRL} such a perspective is incomplete when 
RtoK enters the picture. Fig.~3b shows that the HPS+cation-$\pi$
model, which takes into account hydrophobic, charged, and cation-$\pi$ 
interactions, cannot account for the general trend of available Ddx4 LLPS data.
It follows that these interactions---at least when hydrophobicity
is accorded by the particular scale\cite{rossky2014} adopted by HPS---are 
insufficient by themselves to account for LLPS of IDRs in general.

{\bf Structure-Based Statistical Potentials Provide An Approximate 
Account Of $\pi$-Related Driving Forces for Ddx4 LLPS.}
In contrast to the HPS and HPS+cation-$\pi$ models, direct application
of the KH model---without augmented cation-$\pi$ terms---leads to an overall 
trend largely in agreement with experiments\cite{Nott15,jacob2017,robert} 
for the $\epsilon_{\rm r}$ values tested, i.e., all three Ddx4 IDR variants 
are predicted by the KH potential to have lower LLPS propensities than WT 
(Fig.~4). 
For WT at $\epsilon_{\rm r}=80$ and $T=300$ K, the simulated condensed
phase density of $\sim 500$ mg ml$^{-1}$ is comparable to the 
experimental value of $\sim 400$ mg ml$^{-1}$ with [NaCl] = 100 mM at
the same temperature (Ref.~\citen{jacob2017}).
Similar KH- and HPS-simulated condensed phase densities were obtained 
for FUS and LAF-1 IDRs.\cite{dignon18}
Illustrations of our simulated chain
configurations are provided in Fig.~5.
Time-dependent mean-square deviation of molecular 
coordinates have been used to verify liquid-like chain dynamics in the 
condensed phase of HPS and KH models.\cite{dignon18} Examples
of similar calculation are provided in SI Appendix, Fig.~S3 and S4 for 
the Ddx4 IDRs examined in this work.

This success of the KH model suggests that empirical, knowledge-based
statistical potentials derived from the PDB may capture key effects governing
both protein folding and IDR LLPS without prejudging the dominance of,
or lack thereof, particular types of energetics such as hydrophobicity
in the HPS model. In this respect, it would not be surprising that cation-$\pi$ 
and other $\pi$-related interactions are reflected in these knowledge-based
potentials as well. After all, the importance of 
cation-$\pi$ interactions in folded protein structure\cite{dougherty1999}
and $\pi$-$\pi$ interactions in IDR LLPS\cite{robert} is recognized
largely by bioinformatics analyses of the PDB.

As discussed above, a major cause of the shortcoming of HPS in accounting
for the LLPS of Ddx4 IDRs (Fig.~3b) is the high degree of 
unfavorability it ascribes to Arg interactions. Its hydrophobicity scale, 
based on the atomic partial charges in the OPLS forcefield, posits that 
Arg has the least hydrophobicity value of $+14.5$, the 
next-least hydrophobic is Asp with $+7.5$, whereas Lys has $+5.0$, and the
most hydrophobic is Phe with $-4.0$ (Ref.~\citen{rossky2014}).
This assignment results in highly unfavorable Arg-associated 
interactions relative to Lys-associated interactions. In the HPS model, 
when one of the residues, $i$, in the pairwise energy $E_{ij}(r_0)$ (Fig.~1a) 
is Arg, the average of $E_{ij}(r_0)$ over $j$ for all amino acids 
except the charged residues Arg, Lys, Asp, and Glu 
is equal to $-0.0762$ in units of kcal mol$^{-1}$, whereas the 
corresponding average for Lys is much more favorable at $-0.1276$.
When the charged residues are included, the trend is the same
with the average being $-0.0677$ for Arg and $-0.119$ for Lys. 
In contrast, for the KH model, the trend is opposite with Arg-associated
interactions much more favorable: the corresponding 
average is $-0.123$ for Arg and $-0.041$ for Lys when charged residues
are excluded in the averaging and $-0.0990$ for Arg and $-0.0161$ 
for Lys when charge residues are included. This trend echoes an earlier
eigenvalue analysis of the Miyazawa-Jernigan energies\cite{MJ96} (which
underlie the KH potential) indicating that Arg has a significantly 
larger projection than Lys along the dominant eigenvector.\cite{alphabets}

Whereas correlation among hydrophobicity scales inferred from different
methods is limited\cite{devido,paKarplus,ChanDill97rev,ELS}
with significant variations especially for the nonhydrophobic 
polar and charged residues,\cite{paKarplus} the extremely low 
hydrophobicity assigned by HPS\cite{dignon18,rossky2014} to Arg relative to Lys 
is unusual.  For instance, Lys is substantially
less hydrophobic than Arg in two of the three scales tabulated and 
compared in Ref.~\citen{ELS}.
In a commonly-utilized scale based on the free energies of transfer of
amino acid derivatives from water to octanol measured by
Fauch\`ere and Pli{\v s}ka\cite{octanol}
(the second scale tabulated in Ref.~\citen{ELS}), Arg is only slightly less
hydrophobic ($+5.72$ kJ mol$^{-1}$) than Lys ($+5.61$ kJ mol$^{-1}$)
and thus, essentially, Arg and Lys are deemed to possess equally low 
hydrophobicities. Accordingly, this scale affords a better correlation with 
the Miyazawa-Jernigan energies\cite{MJ96} 
(Fig.~3b of Ref.~\citen{ELS}) than that exhibited in Fig.~1a.

It is reasonable to expect the $210$ (or more) residue-residue
contact energy parameters in PDB-structures-based potentials to
contain more comprehensive energetic information than merely the 
hydrophobicities of the 
20 types of amino acid residues. In this regard, it is notable
that a higher propensity for Arg than Lys to engage favorably with
another residue appears to be a robust feature of PDB statistics.
It holds for the cation-aromatic pairs we analyze in SI Appendix, Fig.~S2,
for the KH potential, and also for the original Miyazawa-Jernigan
energies put forth in 1985.\cite{MJ85} According to Table V of
Ref.~\citen{MJ85}, the contact energies $e_{ij}$ between Arg
and aromatic or negatively charged residues are
$-3.54$, $-3.56$, $-2.75$, $-2.07$, and $-1.98 k_{\rm B}T$, respectively,
for Arg-Phe, Arg-Trp, Arg-Tyr, Arg-Glu, and Arg-Asp
($k_{\rm B}$ is Boltzmann constant, $T$ is absolute temperature), whereas
the corresponding contact energies are weaker for Lys at 
$-2.83$, $-2.49$, $-2.01$, $-1.60$, and $-1.32 k_{\rm B}T$, respectively,
for Lys-Phe, Lys-Trp, Lys-Tyr, Lys-Glu, and Lys-Asp.
All twenty Arg interactions are more favorable than the corresponding 
Lys interactions. The average $e_{ij}$ over all Arg-associated pairs
is $-2.22 k_{\rm B}T$, which is substantially more favorable than
the corresponding average of $-1.4795 k_{\rm B}T$
for the Lys-associated pairs. 
It is apparent from the present application 
of KH to the Ddx4 IDRs that this feature is crucial, at least at a 
coarse-grained level, for an adequate accounting of the $\pi$-related 
energetics of biomolecular LLPS. 
Physically, the general preference for Arg-$\pi$ over Lys-$\pi$ contacts
is underpinned in part by the Arg geometry which allows for favorable 
guanidinium-aromatic coplanar packing,\cite{crowley2005} as 
discussed recently by Wang et al. 
(Ref.~\citen{moleculargrammar} and references therein)
and that Arg-aromatic is less attenuated by the aqueous dielectric medium
because it is less dominated than Lys-aromatic 
interactions by electrostatics.\cite{kumar2018}
Consistent with this perspective, RtoK mutants of FUS\cite{moleculargrammar} 
and LAF-1 RGG\cite{Schuster2020} also exhibit substantially reduced LLPS 
propensity relative to WT.


{\bf Phenylalanine Interactions in Liquid Condensates, Expected to be 
More Solvent-Exposed, are Weaker than Statistical Estimates Based on 
Mostly Core Phenylalanine Contacts in Folded Proteins.}
To assess further the generality of the KH model, we apply it to simulate 
the phase behavior of WT LAF-1 and its RtoK and tryosine-to-phenylalanine 
(YtoF) mutants (Fig.~6, and SI Appendix, Fig.~S5). 
Simulation studies of LLPS of full-length and the 
RGG IDR\cite{dignon18} of LAF-1, including the latter's charge shuffled 
variants,\cite{Schuster2020} have been conducted extensively using 
the HPS\cite{dignon18,Schuster2020} and KH\cite{dignon18} models to gain 
valuable insights; but
phase behaviors of the RtoK and YtoF LAF-1 RGG mutants have not been 
simulated using these models.
Recent experiments indicate that the RtoK mutant does not phase separate
under the conditions tested, whereas the LLPS propensity of the YtoF mutant
is reduced relative to that of WT.\cite{Schuster2020}
As for the Ddx4 IDR, the KH-predicted coexistence curve
for RtoK in Fig.~6 (purple curve) is consistent with the 
experiemtal trend, as it has a far lower critical temperature than
that of the WT (red curve). However, KH posits a significantly
higher LLPS propensity for YtoF (green dashed curve
with a much higher critical temperature),
which is opposite to experimental observation. On closer inspection,
the basic feature that causes this shortcoming is that KH deems
Phe interactions more favorable than Tyr interactions.
According to KH, $E_{ij}(r_0)$ for $i={\rm Phe}$ averaged 
over $j$ for all amino acids is $-0.453$ kcal mol$^{-1}$ for Phe, which
is significantly more negative than the corresponding 
average $E_{ij}(r_0)$ of $-0.281$ kcal mol$^{-1}$ for Tyr.
For this reason, KH is unlikely to reproduce the experimentally
observed reduced LLPS propensities for any other YtoF mutant either,
including the YtoF mutants of FUS.\cite{moleculargrammar}

This overestimation of the favorability of Phe-related interactions in
the LLPS context also causes the KH-predicted LLPS propensity of the 
FtoA mutant of Ddx4 IDR (Fig.~4, green curves) to be substantially lower 
than that of the RtoK mutant (purple curves). Experimentally, however, 
FtoA mutant LLPS is observed at $\sim 350$ mg ml$^{-1}$ 
protein concentration, but the RtoK mutant does not phase separate up
to 400 mg ml$^{-1}$ (in comparison, LLPS of WT is observed at
25 mg ml$^{-1}$).\cite{robert} In other words, the KH-predicted rank 
ordering of LLPS propensities of Ddx4 FtoA and RtoK is opposite to
experiment. This is because the KH energy $E_{ij}(r_0)$ for Ala averaged 
over all 20 residue types, at $-0.160$ kcal mol$^{-1}$, is much less 
favorable than the corresponding $-0.453$ kcal mol$^{-1}$ 
for Phe, and that the change in average KH $E_{ij}(r_0)$ from Phe to Ala
($-0.160+0.453=+0.293$ kcal mol$^{-1}$) is far larger than 
the corresponding change of $-0.0161+0.0990=+0.0829$ kcal mol$^{-1}$ 
from Arg to Lys.

Interestingly, while these Ala-, Tyr-, and Phe-related KH energies do not 
reproduce experimental observations for YtoF and FtoA mutants, the KH 
energies are largely in line with rank ordering in common hydrophobicity 
scales. For instance, in the water-to-octanol scale of Fauch\`ere and 
Pli{\v s}ka,\cite{octanol}
the transfer free energies for Ala, Tyr, and Phe are, respectively,
$-1.76$, $-5.44$, $-10.1$ kJ mol$^{-1}$.
By comparison, the average interaction energies in the HPS model
for Ala, Tyr, and Phe differ less---being $-0.141$, $-0.154$, and 
$-0.168$ kcal mol$^{-1}$, respectively; but Phe-related interactions
are still generally more favorable than Tyr-related interactions. Thus,
HPS is not only insufficient to account for RtoK experiments (Fig.~3), 
HPS is not expected to capture the observed YtoF trend either.

The above consideration underscores once again that the energetic contributions
of amino acid residues to LLPS are not necessarily dominated by their 
hydrophobicities. Phe is generally considered to be more hydrophobic than 
Tyr. However, even for the globular proteins ribonuclease Sa and Sa3, 
14 out of 16 YtoF single-site substitutions destabilize the folded 
state. One of the physical reasons for this behavior is that the Tyr hydroxyl 
group---which is absent in Phe---makes a hydrogen bond with
the N$^\epsilon$ atom of an Arg residue.\cite{NickPace2001} It would be 
enlightening to investigate whether similar hydrogen bonding effects 
account to any degree for the preference for Tyr--Arg over Phe--Arg 
interactions in the LLPS context.

For a relative large residue that possesses both 
hydrophobic and aromatic properties such as phenylalanine, the character 
of its LLPS-driving contributions can also differ 
significantly from those stabilizing the well-packed core of a folded 
protein, because 
most residue-residue contacts in the LLPS context are at least partially 
exposed to solvent even in the condensed phase.
This observation offers a perspective to understand KH's inability to
capture LLPS behaviors of YtoF and FtoA mutants because KH is derived 
from statistical potentials which in turn are based upon contact 
frequencies in folded proteins.
Since Phe is one of the most buried residues in folded proteins with
on average 0.88 of its surface area inaccessible to solvent upon folding
(the corresponding fractions for Tyr, Ala, Lys, and Arg are, respectively, 
0.76, 0.74, 0.52, and 0.64),\cite{GeorgeSci1985} the difference in 
character between Phe's LLPS-driving and folded-state-stabilizing
interactions is expected to be more prominent.

Indeed, it has long been recognized that ``pair'' interactions in 
solvent-exposed environments have properties---including thermodynamic
signatures\cite{Maria2005,Tieleman2007}---that are distinctly different 
from those of ``bulk'' interactions quantified by common hydrophobicity 
measurements.\cite{Thompson90} In this regard, computational evidence 
suggests that Phe-related interaction in a highly solvent accessible 
environment---which is more pertinent to LLPS---is not substantially
stronger than Ala-related interactions. For instance,
using ethane and ethylbenzene as models, respectively, for Ala and Phe, 
explicit-TIP3P water simulations of pairwise potentials of mean force (PMFs) 
at 298 K by Makowski et al. indicate that the lowest 
free energies at the ethane-ethane, ethane-ethylbenzene,\cite{harold2007} and 
ethylbenzene-ethylbenzene\cite{harold2008} contact minima are
all $\approx -1$ kcal mol$^{-1}$ (Tyr was not 
considered in these studies\cite{harold2007,harold2008}). Thus, in contrast to
the vastly different Ala-Ala, Ala-Phe, and Phe-Phe interaction strengths 
in KH, with $E_{ij}(r_0)=-0.142$, $-0.425$, and 
$-0.756$ kcal mol$^{-1}$, respectively,\cite{dignon18} these explicit-water 
PMFs suggest a more modest difference in interaction strength between Ala
and Phe in the LLPS context, which would be in line with the
FtoA experiment. With this knowledge, future investigations 
should tackle hitherto overlooked features of solvent-accessible contacts 
among amino acid residues to improve coarse-grained interaction 
potentials for LLPS.


{\bf IDR Concentration Can Significantly Affect The Dielectric Environment Of 
Condensed Droplets But Its Impact On LLPS Propensity Can Be Modest.}
In recent\cite{dignon18,suman2,jeetainPNAS} and the above
coarse-grained, implicit-solvent simulations of LLPS of IDRs,
electrostatic interactions are assumed, for simplicity, to operate in
a uniform dielectric medium with a position-independent $\epsilon_{\rm r}$. 
However, the dielectric environment is certainly nonuniform 
upon LLPS: The electrostatic interaction between two charges are affected to 
a larger extent by the intervening IDR materials in the condensed
phase---where there is a higher IDR concentration---than in the 
dilute phase. Protein materials have lower $\epsilon_{\rm r}$s
than bulk water.\cite{bertrand,vanGunsteren2001,warshel2001} Analytical 
treatments with effective medium 
theories suggest that a decrease in effective $\epsilon_{\rm r}$ with
increasing IDR concentration enhances polyampholytes LLPS in a cooperative
manner because the formation of condensed phase lowers $\epsilon_{\rm r}$ 
and that in turn induces stronger electrostatic attractions that favor
the condensed phase.\cite{linJML,njp2017}

In principle, LLPS of IDR chains in polarizable aqueous media
can be directly simulated using explicit-water atomic models wherein
partial charges are assigned to appropriate sites of the water and protein
molecules; but such simulations are computationally extremely costly
because a large number of IDR chains are needed to model LLPS.
Here we use explicit-water atomic simulation involving only a few IDR
molecules, not to model LLPS but to estimate how the effective 
$\epsilon_{\rm r}$ depends on IDR concentration. We will then combine this
information with analytical formulations to provide a more complete 
account of the electrostatic driving forces for LLPS. The dielectric 
properties of folded proteins,\cite{bertrand,vanGunsteren2001} their
solutions,\cite{rudas_etal2006} 
and related biomolecular\cite{HXZhou2008} and cellular\cite{tros_etal2017}
settings have long been of interest.\cite{wyman1931}
For the current focus on biomolecular condensates, their interior 
dielectric environments are expected to be of functional significance, 
e.g., as drivers for various ions and charged molecules to preferentially 
partition into a condensate.\cite{NMartin}

As outlined in Materials and Methods,
GROMACS \cite{gromacs2018} and the amber99sb-ildn
forcefield \cite{DEShaw2010} are used for our explicit-water
simulations using either the TIP3P \cite{jorgensen1983} or
the SPC/E \cite{berendsen1987} model of water in simulation boxes 
containing WT Ddx4 IDR chains.
Relative permittivities are estimated by 
fluctuations of the system dipole moment.\cite{vanGunsteren2001,rudas_etal2006}
Simulations are also performed on artificially constructed Ddx4 (aDdx4)
in which the sidechain charges of Arg, Lys, Asp, and Glu are neutralized
for possible applications when sidechain charges are treated separately from 
that of the background dielectric medium. Methodological details are 
provided in SI Appendix, SI Text.

Some of the simulated $\epsilon_{\rm r}$ values are plotted in Fig.~7a
to illustrate their dependence on IDR volume fraction $\phi$ (the
$\phi\propto$ concentration relation and an extended set of
simulated $\epsilon_{\rm r}$s are provided by SI Appendix, SI Text, 
Fig.~S6, and Table~S1).
The difference in simulated $\epsilon_{\rm r}(\phi)$ for Ddx4 and aDdx4 is
negligible except at very low IDR concentration (Fig.~7a and SI Appendix,
Fig.~S6), 
likely because the main contribution to the dielectric effect of IDR in 
the atomic model is from the partial charges on the protein backbone.
Consistent with expectation,\cite{linJML,njp2017} simulated 
$\epsilon_{\rm r}(\phi)$ in Fig.~7a decreases with increasing $\phi$
for all solvent conditions considered. Permittivity is known to decrease 
with salt.\cite{hasted1948,orland2012} Here this expected effect is 
observed for TIP3P solution of IDR at low but not at high IDR concentration. 
Interestingly, the $\epsilon_{\rm r}(\phi)$ simulated with SPC/E water 
and 100 mM NaCl exhibits nonlinear decrease with increasing $\phi$, 
which is akin to that predicted by the Bragg-Pippard\cite{bragg} and 
Clausius-Mossotti models; but the TIP3P-simulated $\epsilon_{\rm r}(\phi)$ 
appears to be linear in $\phi$, which is more in line with 
the Maxwell Garnett and Bruggeman models.\cite{njp2017}

We utilize the salient features of the coarse-grained KH chain model 
for Ddx4 (Fig.~4) and the IDR-concentration-dependent 
permittivities from explicit-water simulations (Fig.~7a) to inform
an analytical theory for IDR LLPS, referred to as RPA+FH, that combines a 
random-phase-approximation (RPA) of charge-sequence-specific electrostatics 
and Flory-Huggins (FH) mean-field treatment for the other 
interactions.\cite{linPRL,linJML} An in-depth analysis of our previous
RPA formulation for IDR-concentration-dependent $\epsilon_{\rm r}$
(Ref.~\citen{linJML}) indicates that only an $\epsilon_{\rm r}(\phi)$
linear in $\phi$ can be consistently treated by RPA (SI Appendix, SI Text).
In view of this recognition, and
considering the uncertainties of simulated $\epsilon_{\rm r}(\phi)$ for
different water models (Fig.~7a), three alternative linear forms of 
$\epsilon_{\rm r}(\phi)$ (straight lines in Fig.~7a) are used for the
present RPA formulation to cover reasonable variations in 
$\epsilon_{\rm r}(\phi)$.

The mean-field FH interaction parameters in our RPA+FH models for the 
four Ddx4 IDRs are obtained from the four sequences' average pairwise 
non-electrostatic KH contact energies.  For each of the 236-residue 
sequences, we calculate the average of the $E_{ij}(r_0)$ [KH] quantity 
(Fig.~1a), for a given pair of residue types, over all pairs
of residues on the sequence, including a residue with itself
($236\times 237/2 =$ 27,966 pairs total),
except those pairs involving two charged residues 
(Arg-Arg, Arg-Lys, Arg-Asp, Arg-Glu, Lys-Lys, Lys-Asp, Lys-Glu, Asp-Asp,
Asp-Glu, and Glu-Glu) 
because interactions of charged pairs are accounted for by RPA separately.
The resulting average energies in units of kcal mol$^{-1}$,
$-0.1047$ for WT and CS, $-0.0689$ for FtoA, and $-0.0924$ for RtoK,
are then input with an overall multiplicative scaling factor into
RPA+FH theories with $\phi$-independent $\epsilon_{\rm r}$
for three different fixed $\epsilon_{\rm r}=$ 80, 40, and 20. The computed
RPA+FH phase diagrams are then fitted to the corresponding phase diagrams 
simulated by coarse-grained KH chain models in Fig.~4 to determine a single
energy scaling factor from the best possible fit (SI Appendix, Fig.~S7).
The product of this factor and the sequence-dependent averages of
$E_{ij}(r_0)$ [KH] defined above is now used as the enthalpic FH $\chi$ 
parameters in the final RPA+FH theories with IDR-concentration-dependent
$\epsilon_{\rm r}(\phi)$. Details of unit conversion between our
explicit-chain simulation and our analytical RPA+FH formulation 
are in SI Appendix, SI Text.

In this connection, it is instructive to note that the corresponding averages 
of $E_{ij}(r_0)$ [HPS] for the HPS model are $-0.1214$ for WT
and CS, $-0.1179$ for FtoA, and $-0.1294$ for RtoK. In this case, the more
favorable (more negative) average energy of RtoK than WT underlies
the mismatch of HPS prediction with experiment seen in Fig.~3b.

Figure~7b and c show the phase diagrams of the four Ddx4 IDRs predicted 
using RPA+FH theories with three alternative IDR-concentration-dependent 
$\epsilon_{\rm r}(\phi)$ functions and KH-derived mean-field FH parameters 
as prescribed above. In all cases considered, the WT sequence (red curves) 
exhibit a higher propensity to LLPS than the three variants, indicating
that this general agreement with the experimental trend seen in Fig.~4
holds up not only under the simplifying assumption of a constant 
$\epsilon_{\rm r}$ but also when the dielectric effect of the IDRs is 
taken into account.
As discussed in SI Appendix, SI Text, we have previously
subtracted the self-energy term in the RPA formulation for numerical
efficiency because the term has no impact on the predicted phase
diagram when $\epsilon_{\rm r}$ is a constant independent of $\phi$
because the self-energy contribution is identical for the dilute and
condensed phases. However, with an IDR-concentration-dependent 
$\epsilon_{\rm r}(\phi)$, as for the cases considered here, 
the self energy---with the short-distance cutoff of Coulomb
interaction in the RPA formulation corresponding roughly to a finite
Born radius\cite{Wang2010}---is physically relevant as it decreases
with increasing $\epsilon_{\rm r}$, and therefore it affects the
predicted LLPS properties as manifested by the difference between
Fig.~7b and c. 
It follows that the self-energy term quantifies a tendency for
an individual polyampholyte chain to prefer the dilute phase
with a higher $\epsilon_{\rm r}$---because of its more favorable electrostatic
interactions with the more polarizable environment---over the condensed 
phase with
a lower $\epsilon_{\rm r}$. This tendency disfavors LLPS. At the same time,
the lower $\epsilon_{\rm r}$ in the condensed phase entails a stronger
inter-chain attractive electrostatic force that drives the association 
of polyampholyte
chains. Therefore, taken together, relative to the assumption of a constant 
$\epsilon_{\rm r}$, the overall impact of an IDR-concentration-dependent
$\epsilon_{\rm r}(\phi)$ is expected to be modest because it likely
entails a partial tradeoff between these two opposing effects. This 
consideration is borne out in Fig.~7b and c. When self energy is neglected 
in Fig.~7c, LLPS propensities predicted using 
$\epsilon_{\rm r}(\phi)$s are relatively high (as characterized by the
critical temperatures), comparable to those for a fixed $\epsilon_{\rm r}=40$
in Fig.~4b. When the physical effect of self energy is accounted for in Fig.~7b,
LLPS propensities predicted using
$\epsilon_{\rm r}(\phi)$s are significantly lower: overall they are comparable
but slightly lower than those for a fixed $\epsilon_{\rm r}=80$ in Fig.~4a.  
Consistent with this physical picture, whereas the $\epsilon_{\rm r}(\phi)$
with a sharper decrease with increasing $\phi$ leads to a higher LLPS
propensity when self energy is neglected (dashed curves have higher critical 
temperatures than dashed-dotted curves of the same color in Fig.~7c), for 
the physically appropriate formulation with self energy, a 
sharper decrease in $\epsilon_{\rm r}(\phi)$ with increasing $\phi$ 
leads to a lower LLPS propensity (dashed curves have lower 
critical temperatures than dashed-dotted curves of the same color in Fig.~7b).
\\

\noindent
{\large\bf CONCLUSION}\\

In summary, we have gained new insights into the physical forces that 
drive the formation of biomolecular condensates by systematically evaluating 
coarse-grained, residue-based protein chain models embodying different 
outlooks as to the types of interactions 
that are important for LLPS of IDRs by comparing model predictions
against experimental data on WT Ddx4 IDR and its three variants
as well as WT LAF-1 RGG and its two variants, acquiring essential knowledge
from agreement as well as disagreement between simulation and experiment.
Aiming
to account for all observed relative LLPS propensities of the 
sequences, we find that hydrophobicity, electrostatic, and cation-$\pi$ 
interactions are insufficient by themselves. Rationalization of 
experiment on arginine-to-lysine variants entails significantly
more favorable arginine-associated over lysine-associated contacts,
an effect that is most likely underpinned by $\pi$-$\pi$ interactions.
This perspective is in line with bioinformatics analysis of LLPS
propensities\cite{robert} and recent experiments on other 
IDRs.\cite{moleculargrammar,TanjaScience2020}
In this regard, it is reassuring that the balance of forces for LLPS of 
IDRs appears to be partly 
captured by common PDB-derived statistical 
potentials developed to study protein folding and binding. 
However, we found that the condensate-stabilizing contributions of 
phenylalanine interactions are significantly weaker than that estimated 
from statistical potentials, most likely because such interactions exist in 
a highly solvent accessible environments rather than in the sequestered cores 
or binding interfaces of globular proteins.
We have also highlighted the 
reduced electric permittivity inside condensed IDR phases. Although
this effect's overall influence on LLPS propensity may be modest because
of a tradeoff between its consequences on IDR self energies and on inter-IDR 
interactions, the effect of IDR-concentration-dependent permittivity by 
itself should be of functional importance in biology because of its potential 
impact on biochemical reactions and preferential partition of certain molecules 
into a given biomolecular condensate. All told, the present study
serves not only to clarify the aforementioned issues of general principles, 
it also represents a useful step toward a transferable coarse-grained model 
for sequence-specific biomolecular LLPS. Many questions remain to be further
investigated nonetheless.
These include---and are not limited to---an adequate
description of solvent-exposed interactions involving large 
hydrophobic/aromatic residues such as phenylalanine,
a proper balance between attractive and repulsive interactions,\cite{suman2}
devising temperature-dependent effective interactions,\cite{jeetainACS}
an accurate account of small ion 
effects,\cite{panagio2003,schmitKAD2010,kings2020}
and incorporation of nucleic acids into LLPS simulations.\cite{joanElife}
Progress in these directions will deepen our understanding of fundamental
molecular biological processes and will advance the design of novel IDR-like 
materials as well.

$\null$\\
{\large\bf Materials and Methods}\\ 

Our implicit-water coarse-grained explicit-chain modeling setup
used in the first part of the present investigation
for multiple IDP molecules
follows largely the Langevin dynamics formulations in
Refs.~\citen{dignon18,suman2} for IDP LLPS. The simulation protocol
features an initial slab-like condensed configuration that allows
for efficient equilibration to produce simulated LLPS data\cite{panag2017}.
As discussed, model energy functions embodying different physical
perspectives are considered. The subsequent explicit-water simulations
for estimating IDR-concentration-dependent
relative permittivity are conducted on five WT Ddx4 IDRs using
GROMACS\cite{gromacs2018} in conjunction with the amber99sb-ildn
forcefield\cite{DEShaw2010} and with TIP3P\cite{jorgensen1983} or
SPC/E\cite{berendsen1987} waters.
Details of our methodology and the development of analytical RPA
formulation are provided in SI Appendix, SI Text.

$\null$\\
{\large\bf Acknowledgments}\\ 
We thank Robert Best, Alex Holehouse, Jeetain Mittal, Rohit Pappu, and 
Wenwei Zheng for helpful discussions.
This work was supported by Canadian Institutes of Health Research grants
MOP-84281 and NJT-155930 to H.S.C., Natural Sciences and Engineering Research 
Council of Canada Discovery grants RGPIN-2016-06718 to J.D.F.-K. and
RGPIN-2018-04351 to H.S.C., and computational resources provided 
generously by Compute/Calcul Canada.

\vfill\eject
\begin{center}
   \includegraphics[width=0.9\columnwidth]{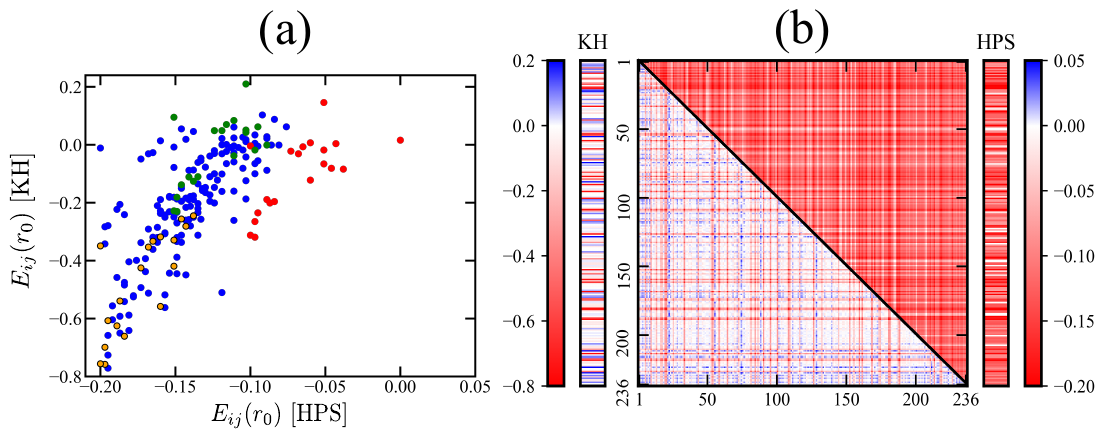}
\end{center}
\vskip -0.4cm
{\footnotesize{
\noindent
{\bf Fig. 1.\ } 
   {Comparing two amino acid residue-based coarse-grained potentials.
(a) Scatter plot of 210 pairwise contact energies (in units of kcal mol$^{-1}$)
in the HPS (horizontal variable) versus those in 
the KH (vertical variable) model.\cite{dignon18} $E_{ij}(r_0)$s are
the pairwise potential energies $U_{\rm aa|HPS}(r)$ or
$U_{\rm aa|KH}(r)$ (SI Appendix, SI Text), 
between two residues of types $i,j$ 
separated by $r_{ij}=r_0$ where the Lennard-Jones component of the potential is 
minimum ($i,j$ here stand for labels for the 20 amino acid types). 
Energies of contacts 
involving Arg (red circles), Lys (green circles) and 
Phe (yellow-filled black circles) are colored differently 
from others (blue circles).  (b) Contact energies between residue 
pairs at positions $i,j$ of the $n=236$ sequence of WT Ddx4 IDR 
(Ddx4$^{\rm N1}$, Ref.~\citen{Nott15}) in the two
potentials are color coded by the scales. The vertical 
and horizontal axes represent residue positions $i,j\le n$. 
The $i\ne j$ contact energies in the HPS and KH models are provided
in the two-dimensional plot, whereas the $i=j$ contact energies
are shown alongside the model potentials' respective color scales.
}
}}
\vfill\eject
\begin{center}
   \includegraphics[width=\columnwidth]{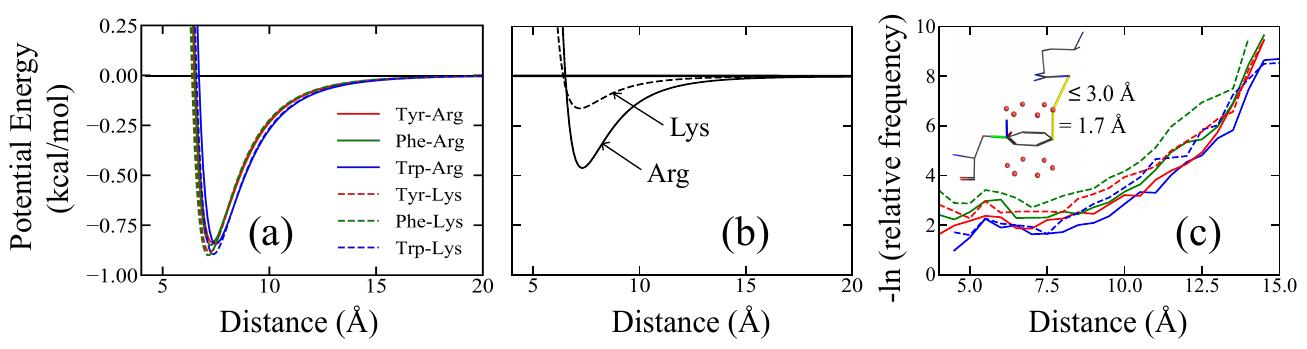}
\end{center}
\vskip -0.4cm
{\footnotesize{
\noindent{\bf Fig. 2.\ } 
   {Possible cation-$\pi$ interaction potentials.
(a) Sum of the coarse-grained HPS potential and a model cation-$\pi$ 
interaction with a uniform $(\epsilon_{{\rm c}\pi})_{ij}=3.0$ kcal mol$^{-1}$
as a function of residue-residue distance for the residue pairs
Arg-Tyr, Arg-Phe, Arg-Trp, Lys-Tyr, Lys-Phe and Lys-Trp, wherein
Tyr/Phe/Trp are labeled as red/green/blue and Arg/Lys are represented
by solid/dashed curves.
(b) An alternate cation-$\pi$ potential in which Arg-Tyr/Phe/Trp
is significantly more favorable (solid curve, 
$(\epsilon_{{\rm c}\pi})_{ij}=1.85$ kcal mol$^{-1}$) than Lys-Tyr/Phe/Trp 
(dashed curve, $(\epsilon_{{\rm c}\pi})_{ij}=0.65$ kcal mol$^{-1}$).
Note that the plotted curves here---unlike those in (a)---do not contain 
the HPS potential.
(c) Normalized C$_\alpha$--C$_\alpha$ distance-dependent contact frequencies 
for the aforementioned six cation-$\pi$ pairs (color coded as in (a)) 
computed using a set of 6,943 high-resolution X-ray protein crystal 
structures (resolution $\le 1.8$ \AA) from a published 
non-redundant set.\cite{robert} 
Contact pair statistics are collected from residues on different 
chains in a given structure and residues separated by $\ge 50$ 
amino acids along the same chain. C$_\alpha$--C$_\alpha$ distances are 
divided into 0.2 \AA~bins. For each bin, the relative frequency is the number
of instances of a cation-$\pi$-like contact (defined below) divided 
by the total number of residue pairs with C$_\alpha$--C$_\alpha$ distances 
within the narrow range of the bin. Thus, the shown curves quantify the 
tendency for a given pair of residues to engage in cation-$\pi$ interaction 
provided that the pair is spatially separated by a given 
C$_\alpha$--C$_\alpha$ distance. Here a cation-$\pi$-like contact is 
recognized if either a Lys NZ or an Arg NH1 nitrogen atom is within
3.0~\AA~of any one of the points 1.7~\AA~above or below a $sp^2$ carbon atom 
along the normal of the aromatic ring in a Tyr, Phe, or Trp residue.
This criterion is exemplified by the molecular drawing (inset) of a contact
between an Arg (top) and a Phe (bottom). Colors of the chemical bonds
indicate types of atom involved, with carbon in black, oxygen in red, and 
nitrogen in blue.  The red dots here are points on 
the exterior surfaces of the electronic orbitals farthest from the $sp^2$ 
carbons in the aromatic ring. The blue, green, and red lines emanating from 
a corner of the aromatic ring constitute a local coordinate frame, with 
the blue line being the normal vector of the plane of the aromatic ring 
determined from the positions of its first three atoms. The yellow lines 
mark spatial separations used to define the cation-$\pi$-like contact.
}}}
\vfill\eject
\begin{center}
   \includegraphics[width=0.86\columnwidth]{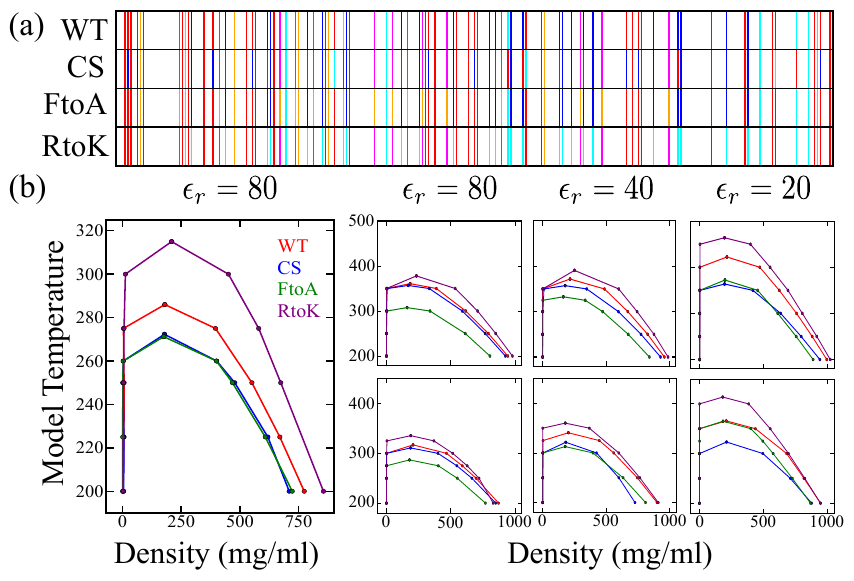}
\end{center}
\vskip -0.4cm
{\footnotesize{
\noindent
{\bf Fig. 3.\ }
   {Simulated phase behaviors of Ddx4 IDR variants in a 
hydrophobicity-dominant potential augmented by cation-$\pi$ interactions.
(a) Sequence patterns of the wildtype (WT) and its charge-scrambled (CS), 
Phe to Ala (FtoA) and Arg to Lys (RtoK) variants. 
Select residue types are highlighted: Ala (orange), Asp and Glu (red), 
Phe (magenta), Lys (cyan), and Arg (dark blue); other residue types are 
not distinguished.
(b) Simulated phase diagrams of WT, CS, FtoA and RtoK Ddx4 IDR at various
relative permittivities ($\epsilon_{\rm r}$) as indicated, using the
HPS model only (leftmost panel) and the HPS model augmented with
cation-$\pi$ interactions (other panels on the right) with either a uniform 
$(\epsilon_{{\rm c}\pi})_{ij}$ as described in Fig.~2a (top)
or with different $(\epsilon_{{\rm c}\pi})_{ij}$ values for Arg and Lys
as given in Fig.~2b (bottom).
}}} 
\vfill\eject
\begin{center}
   \includegraphics[width=0.80\columnwidth]{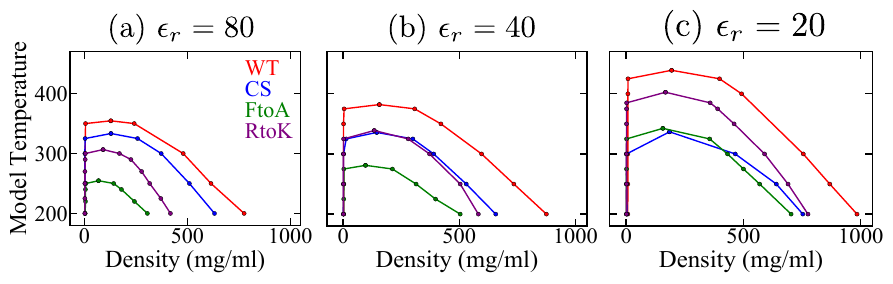}
\end{center}
\vskip -0.45cm
{\footnotesize{
\noindent
{\bf Fig. 4.\ }
   {Simulated phase behaviors of Ddx4 IDR variants using an interaction
scheme based largely on PDB-derived statistical potentials.
Phase diagrams were computed using the KH model at three different
relative permittivities ($\epsilon_r$).
}} }
\vfill\eject
\begin{center}
   \includegraphics[width=0.49\columnwidth]{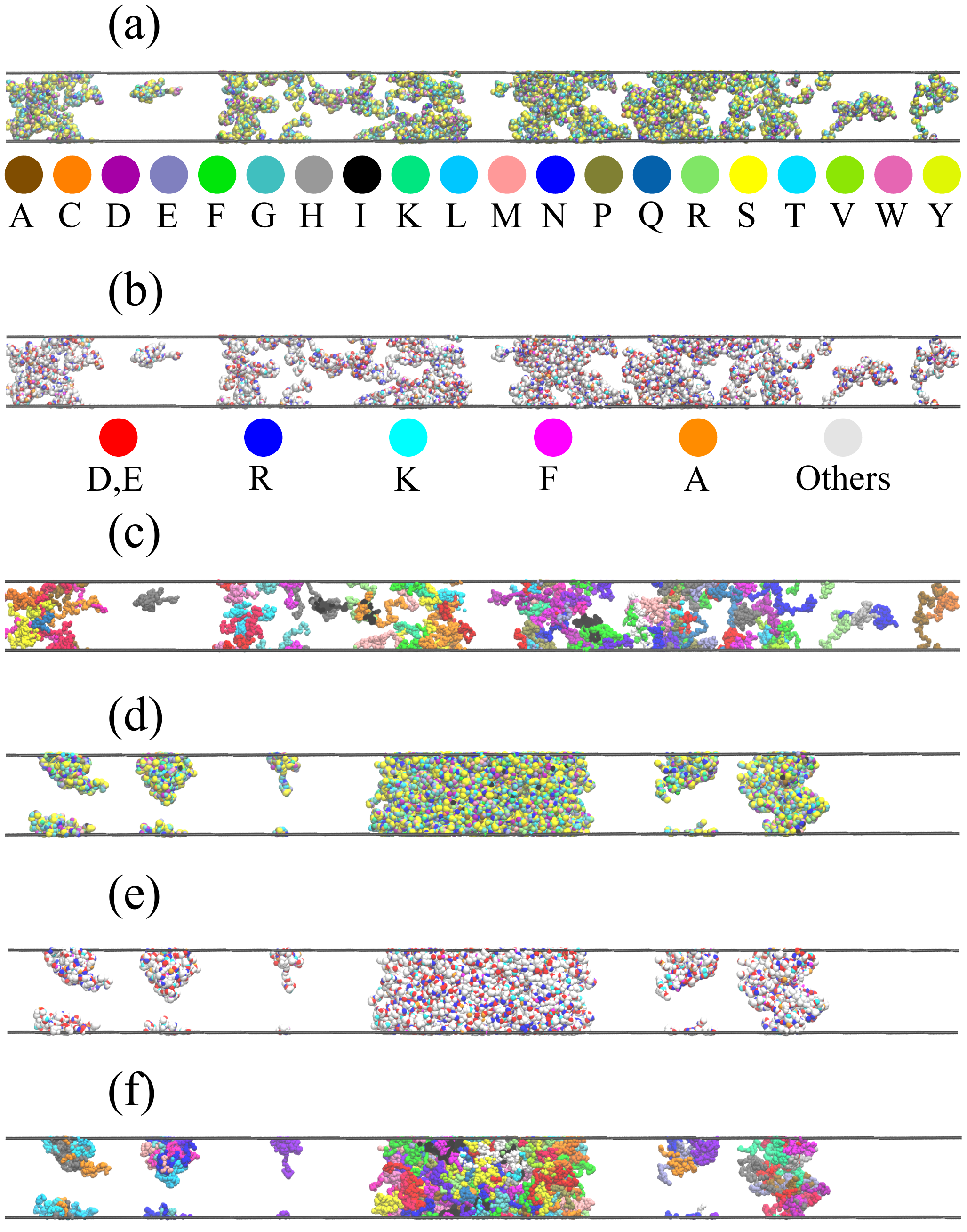}
\end{center}
\vskip -0.1cm
{\footnotesize{
\noindent
{\bf Fig. 5.\ }
   {Illustrative snapshots of Ddx4$^{\rm N1}$CS phase behaviors 
simulated using the KH potential for $\epsilon_{\rm r} = 40$.
(a) A non-phase-separated snapshot at model temperature 375 K, wherein the 
amino acid residues are colored using the default VMD scheme\cite{VMD1,VMD2}
as provided by the key below the snapshot.  
(b) Same as (a) except the color scheme (as shown) is essentially 
identical to that in Fig.~3a.
(c) Same as (a) and (b) except all residues along the same chain share 
the same color. Neighboring chains are colored differently to highlight the 
diversity of conformations in the system. (d--f) A phase-separated snapshot at
model temperature 325 K. The color schemes are the same, respectively,
as those in (a--c).
}}}
\vfill\eject
\begin{center}
   \includegraphics[width=0.96\columnwidth]{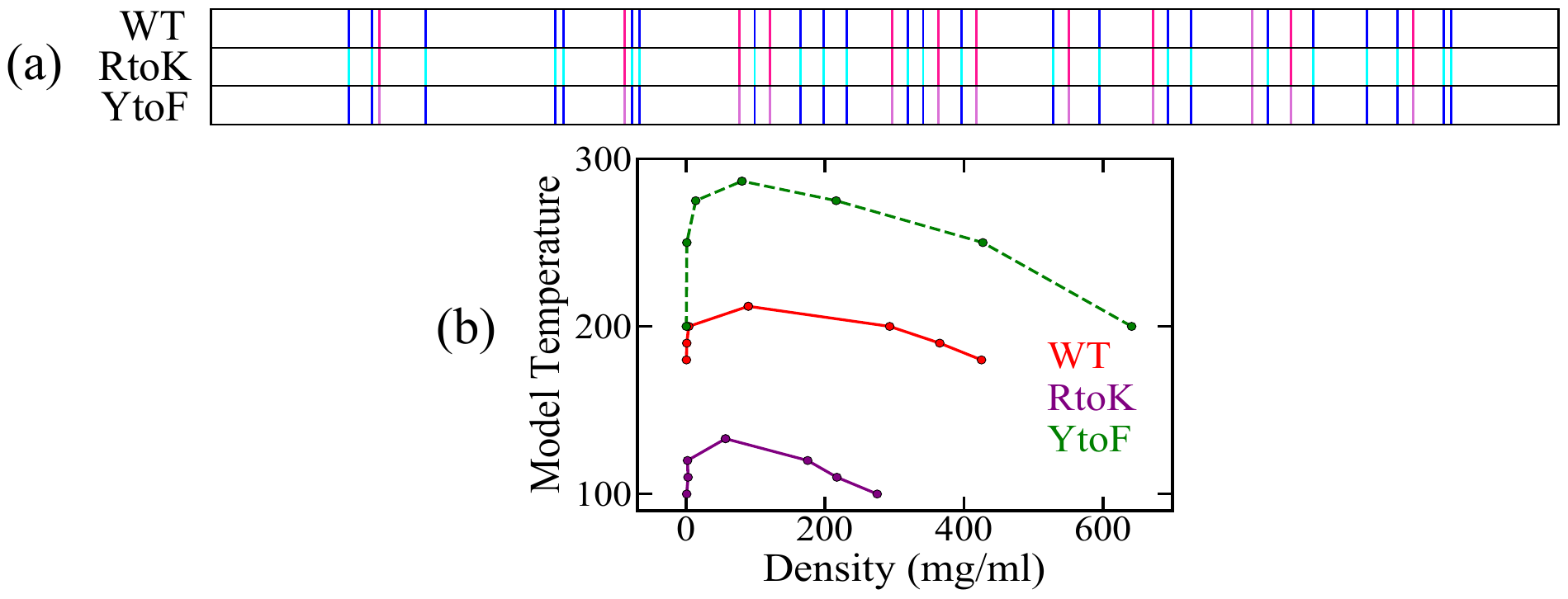}
\end{center}
\vskip -0.2cm
{\footnotesize{
\noindent
{\bf Fig. 6.\ } 
   {Simulated phase diagrams of LAF-1 IDRs computed
by the KH model at $\epsilon_{\rm r}=40$ for the three sequences 
in SI Appendix, Fig.~S5.
(a) Sequence patterns of the WT and two variants. 
Highlighted residues are Phe (magenta), Lys (cyan), Arg (dark blue), as in
Fig.~3a, and Tyr (pink); other residue types, including Ala, Asp, and Glu
which were highlighted in Fig.~3a, are not distinguished here.
(b) Phase diagrams for WT and mutant
sequences are plotted in different colors as indicated.
}
}}
\vfill\eject
\begin{center}
   \includegraphics[width=\columnwidth]{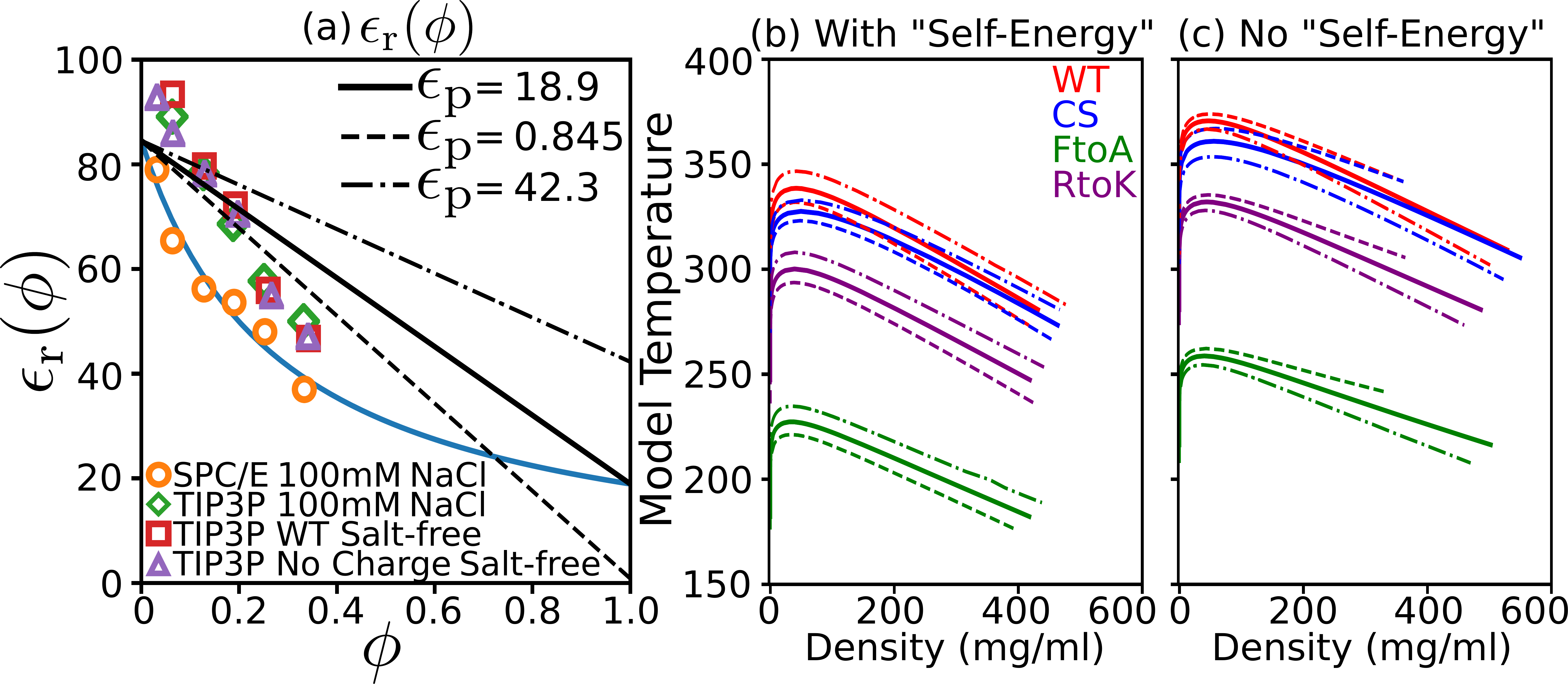}
\end{center}
\vskip -0.2cm
{\footnotesize{
\noindent
{\bf Fig. 7.\ }
{Effects of IDR-concentration-dependent relative permittivity
on phase behaviors.
(a) Relative permittivity $\epsilon_{\rm r}(\phi)$ values obtained by atomic 
simulations (symbols) using various explicit-water models (as indicated, 
bottom) are shown as functions of Ddx4 volume fraction $\phi$ ($\phi=1$
corresponds to pure Ddx4).
The blue curve is a theoretical fit of the SPC/E, [NaCl] = 100 mM 
explicit-water simulated data based on the Slab (Bragg and Pippard\cite{bragg}) 
model [Eq.~(34) of Ref.~\citen{njp2017}], viz., $1/\epsilon_{\rm r}(\phi) = 
\phi/\epsilon_{\rm p}+(1-\phi)/\epsilon_{\rm w}$ with the fitted
$\epsilon_{\rm p}=18.9$ and $\epsilon_{\rm w}=84.5$ 
where $\epsilon_{\rm p}$ and $\epsilon_{\rm w}$ are, respectively,
the relative permittivity of pure protein and pure water. 
The black solid, dashed, and dashed-dotted lines are approximate
linear models of $\epsilon_{\rm r}(\phi)=\epsilon_{\rm p} \phi +
\epsilon_{\rm w}(1-\phi)$ with the same $\epsilon_{\rm w}$ but
different $\epsilon_{\rm p}$ values as indicated (top-right), resulting
in $d\epsilon_{\rm r}(\phi)/d\phi$ 
slopes, respectively, of $-65.6$, $-83.9$, and $-42.2$.
(b, c) Theoretical phase diagrams of the four Ddx4 IDR variants were
obtained by a RPA theory that incorporates an $\epsilon_{\rm r}(\phi)$
linear in $\phi$. Solid, dashed, and dashed-dotted curves correspond, as
in (a), to the three different $\epsilon_{\rm p}$ values used in the theory.
The electrostatic contribution to the phase behaviors is calculated 
here using either (b) the expression for $f_{\rm el}$ given by Eq.~(S51) in
SI Appendix SI Text [i.e., Eq.~(68) of Ref.~\citen{linJML} with
its self-interaction term ${\cal G}_2({\tilde k})$ excluded] or (c) the 
full expression for $f_{\rm el}$ [Eq.~(68) in the same reference, or 
equivalently Eq.~(S2) in SI Appendix, SI Text]. Further details are 
provided in SI Appendix, SI Text.
}}}
\vfill\eject

\vfill\eject
\def\thebibliography#1{\section*{\ \markboth
 {References}{References}}\list
 {$^{{\arabic{enumi}}}$}
 {\settowidth\labelwidth{{#1}}\leftmargin\labelwidth
 \advance\leftmargin\labelsep
 \usecounter{enumi}}
 \def\newblock{\hskip .11em plus .33em minus -.07em}
 \sloppy
 \sfcode`\.=1000\relax}
\let\endthebibliography=\endlist
\renewcommand{\theequation}{{\rm S}\arabic{equation}}
\newcommand{\kr}{\tilde{k}}
\newcommand{\GM}{\hat{G}_\mathrm{M}}
%
%

%
\newcommand{\kB}{k_{\rm B}} 
\newcommand{\lb}{l_{\rm B}} 
\newcommand{\lbr}{u} 

\newcommand{\bra}[1]{\langle #1 |} 
\newcommand{\ket}[1]{| #1 \rangle} 
\newcommand{\braket}[2]{ \langle #1 | #2 \rangle} 

\newcommand{\np}{{n_p}}  
\newcommand{\nc}{{n_c}}   
\newcommand{\ns}{{n_s}}   
\newcommand{\nsp}{{n_+}}   
\newcommand{\nsm}{{n_-}}   

\newcommand{\zs}{z_s} 
\newcommand{\zc}{z_c} 

\newcommand{\nw}{n_w}   
\newcommand{\pc}{q_c}   

\newcommand{\R}{{\bf R}} 
\newcommand{\rr}{{\bf r}}   
\newcommand{\xx}{{\bf x}} 
\newcommand{\kk}{{\bf k}} 
\newcommand{\kzero}{{\bf 0}} 


\newcommand{\T}{\mathscr{T}}    
\newcommand{\U}{\mathscr{U}}   
\newcommand{\UU}{{\cal U}}        
\newcommand{\VV}{{\cal V}}         
\newcommand{\Eng}{\mathscr{H}} 
\newcommand{\Hp}{{\cal H}_p}      

\newcommand{\Z}{{\cal Z}} 
\newcommand{\Zel}{{\Z_{\rm el}}} 
\newcommand{\Zp}{{\cal Z}_p} 
\newcommand{\Zion}{{\cal Z}_{\rm ion}} 
\newcommand{\Zzero}{{\cal Z}_0} 
\newcommand{\Qp}{{\cal Q}_p} 

\newcommand{\fion}{f_{\rm ion}} 
\newcommand{\fzero}{f_{\rm 0}} 
\newcommand{\fpoly}{f_{\rm p}}   

\newcommand{\w}{\varw} 
\newcommand{\ffc}{\varphi}   
\newcommand{\ffv}{\Psi} 


\newcommand{\avg}[1]{\left\langle #1 \right\rangle} 
\newcommand{\DD}{\mathscr{D}} 

\newcommand{\epsr}{\epsilon_{\rm r}}
\newcommand{\epsp}{\epsilon_{\rm p}}
\newcommand{\epsw}{\epsilon_{\rm w}}
\newcommand{\deps}{\epsilon'}
%
%

\begin{center}


{\Huge\bf Supplementary Information}\\

\vskip 0.3cm

{\Large\bf (SI Appendix)}

\vskip 0.3cm

{\Large\it for}

\vskip 0.3cm

{\Large\bf Comparative Roles of Charge, $\pi$ and Hydrophobic}\\

\vskip 0.2cm

{\Large\bf Interactions in Sequence-Dependent Phase}\\

\vskip 0.2cm

{\Large\bf Separation of Intrinsically Disordered Proteins}\\

\vskip .2in

{\bf Suman D{\footnotesize{\bf{AS}}}},$^{\rm a}$
{\bf Yi-Hsuan L{\footnotesize{\bf{IN}}}},$^{\rm a,b}$
{\bf Robert M. V{\footnotesize{\bf{ERNON}}}},$^{\rm b}$
{\bf Julie D. F{\footnotesize{\bf{ORMAN}}}-K{\footnotesize{\bf{AY}}}}$^{\rm b,a}$
and
{\bf Hue Sun C{\footnotesize{\bf{HAN}}}}$^{{\rm a},1}$

\vskip .2in

$^{\rm a}$Department of Biochemistry,
University of Toronto, Toronto, Ontario M5S 1A8, Canada;\\
$^{\rm b}$Molecular Medicine, Hospital for Sick Children, Toronto,
Ontario M5G 0A4, Canada\\
\vskip 0.6cm

%

\end{center}


\noindent
$^1$Corresponding author\\
{\phantom{$^1$}}
E-mail: chan@arrhenius.med.utoronto.ca;
{\phantom{$^1$}}
Tel: (416)978-2697; Fax: (416)978-8548\\
{\phantom{$^1$}}
Department of Biochemistry, University of Toronto,
Medical Sciences Building -- 5th Fl.,\\
{\phantom{$^\dagger$}}
1 King's College Circle, Toronto, Ontario M5S 1A8, Canada.\\

\noindent
-----------------------------------------------------------------------------------------------------------------------

\noindent
{\Large\bf Contents}\\

\noindent
{\bf SI Text} \dotfill 25\\
$\null\quad$ Models and Methods \dotfill 25\\
$\null\quad$ RPA Theory \dotfill 31\\
{\bf SI Figures} \dotfill 47\\
$\null\quad$ Figure~S1 \dotfill 47\\
$\null\quad$ Figure~S2 \dotfill 48\\
$\null\quad$ Figure~S3 \dotfill 49\\
$\null\quad$ Figure~S4 \dotfill 50\\
$\null\quad$ Figure~S5 \dotfill 51\\
$\null\quad$ Figure~S6 \dotfill 52\\
$\null\quad$ Figure~S7 \dotfill 53\\
{\bf SI Table} \dotfill 54\\
$\null\quad$ Table~S1 \dotfill 54\\
$\null$\\
{\bf Combined Main-Text and SI References} \dotfill 55\\

\vfill\eject
\noindent
{\Huge\bf SI Text}\\

\noindent{\large\bf Models and Methods}
\\


\noindent{\bf C{\footnotesize{OARSE-}}G{\footnotesize{RAINED}} 
C{\footnotesize{HAIN}} M{\footnotesize{ODELS}}}

The coarse-grained protein chain models in the present study basically follow
those in Refs.~\citen{dignon18,suman2}, but with modified and additional 
features. In accordance with our previous notation for explicit-chain 
simulation studies,\cite{suman1,suman2} 
let $\mu,\nu = 1,2,\dots,n$ be the labels for the $n$ IDP chains in the 
system, and $i,j=1,2,\dots,N$ be the labels of the $N$ residues in each 
IDP chain. The total potential energy $U_{\rm T}$ is a function of
the residue positions, denoted here as $\{ {\bf R}_{\mu i} \}$. Writing
\begin{equation}
U_{\rm T} = U_{\rm bond} + U_{\rm el} + U_{\rm aa} \; ,
\label{eq:U-T}
\end{equation}
where $U_{\rm bond}$ is the bond-length term for chain connectivity:
\begin{equation}
U_{\rm bond} = \frac {K_{\rm bond}}{2} \sum_{\mu=1}^n \sum_{i=1}^{N-1}
(r_{\mu i,\mu i+1}-l)^2
\; 
\label{eq:U-bond}
\end{equation}
with $r_{\mu i,\nu j}\equiv |{\bf R}_{\mu i}-{\bf R}_{\nu j}|$,
$l=3.8$~\AA~ is the C$_\alpha$-C$_\alpha$ virtual bond length
[$l$ is equivalent to $a$ in Eq.~(3) of Ref.~\citen{suman2}],
$K_{\rm bond}=10$ kJ mol$^{-1}$\AA$^{-2}$ 
[this value would be identical to that used in Ref.~\citen{dignon18} 
if the 10 kJ/\AA$^2$ value quoted above Eq.~(1) in
this reference is a typographical error, i.e., it misses a ``/mol'';
by comparison, the much stiffer $K_{\rm bond}$ value used in Eq.~(3) in 
Ref.~\citen{suman2}, which follows Ref.~\citen{panag2017} with the aim
of comparing with fixed-bond-length Monte Carlo simulations,
is equivalent to $23.7$ MJ mol$^{-1}$\AA$^{-2}$], 
and $U_{\rm el}$ is the electrostatic interaction:
\begin{equation}
U_{\rm el} = 
\sum_{\mu,\nu=1}^n  
\sum_{\substack{i,j=1\\ \null\hskip -0.8cm(\mu,i)\ne(\nu,j)}}^N 
\frac {\sigma_{\mu i}\sigma_{\nu j}e^2}{4\pi\epsilon_0\epsilon_{\rm r}
r_{\mu i,\nu j}} \exp \Bigr(-\kappa r_{\mu i,\nu j} \Bigr)
\; ,
\label{eq:U-el}
\end{equation}
wherein $\sigma_{\mu i}$ is the charge of the $i$th residue in units of
elementary electronic charge $e$, ($\sigma_{\mu i}$ is independent of $\mu$),
$\epsilon_0$ is vacuum permittivity, $\epsilon_{\rm r}$ is relative
permittivity (dielectric constant), and $\kappa$ is the reciprocal
of the Debye screening length, which is taken to be $10.0$ \AA~ in
this study ($\kappa=0.1$ \AA$^{-1}$).
Following Table S1 of Ref.~\citen{dignon18}, $\sigma$ values for Arg and
Lys are assigned to be $+1$, those of Asp and Glu are $-1$, and that
of His is $+0.5$. All other residues are taken to be neutral, i.e.,
with $\sigma = 0$.

The $U_{\rm aa}$ in Eq.~(\ref{eq:U-T}) is the sum of pairwise 
interaction energies among the residues, viz.,
\begin{equation}
U_{\rm aa} = 
\sum_{\mu,\nu=1}^n  
\sum_{\substack{i,j=1\\ \null\hskip -0.8cm(\mu,i)\ne(\nu,j)}}^N 
(U_{\rm aa})_{\mu i,\nu j}
\; ,
\label{eq:U-aa}
\end{equation}
where $(U_{\rm aa})_{\mu i,\nu j}$ is the interaction between the
$i$th residue of the $\mu$th chain with the $j$th residue of the $\nu$th
chain. We investigate several physically plausible $U_{\rm aa}$ 
functions, as follows:
\\

\noindent
{\bf The HPS model}

The hydrophobicity scale (HPS) model is identical to the one introduced
by Dignon et al.\cite{dignon18} based on an atomic-level hydrophobicity
scale devised by Kapcha and Rossky.\cite{rossky2014} The interaction
between amino-acid pairs in this model is given by
\begin{equation}
(U_{\rm aa})_{\mu i,\nu j} =
(U_{\rm aa|HPS})_{\mu i,\nu j} \equiv
\begin{cases}
(U_{\rm LJ})_{\mu i,\nu j} + (1-\lambda_{ij}^{\rm HPS})\epsilon \; , 
& {\rm if\ } r\le 2^{1/6}a_{ij}\\
\lambda_{ij}^{\rm HPS} (U_{\rm LJ})_{\mu i,\nu j} & {\rm otherwise}
\end{cases}
\; 
\label{eq:U-HPS}
\end{equation}
where $\lambda_{ij}^{\rm HPS}\equiv (\lambda_i+\lambda_j)/2$,
$a_{ij}\equiv (a_i+a_j)/2$, with $\lambda_i$ and $a_i$ being
the hydrophobicity and diameter, respectively, 
of the model amino acid residue at sequence position $i$,
as given, respectively, by the $\lambda$ and $\sigma$ values
in Table~S1 of Ref.~\citen{dignon18};
$(U_{\rm LJ})_{\mu i,\nu j}$ is the Lennard-Jones (LJ) potential, 
\begin{equation}
(U_{\rm LJ})_{\mu i,\nu j} = 
4\epsilon_{ij} \left[\left(\frac {a_{ij}}{r_{\mu i,\nu j}}
\right)^{12} -
\left(\frac {a_{ij}}{r_{\mu i,\nu j}}\right)^6 \right] \; ,
\label{eq:LJ}
\end{equation}
where the LJ well depth $\epsilon_{ij}$ (not to be confused with the
permittivities) is set to be $\epsilon_{ij}=0.2$ kcal mol$^{-1}$ 
irrespective of $i,j$ for the HPS model, as in Ref.~\citen{dignon18}.
\\

\noindent
{\bf The HPS+cation-$\pi$ models}

In view of the importance of cation-$\pi$ interactions in protein
structure (see discussion in the main text), we consider also
a class of model potentials, $U_{{\rm aa|HPS+c}\pi}$s, that augment the 
HPS potential with cation-$\pi$ terms for Arg-Phe, Arg-Trp, Arg-Tyr, 
Lys-Phe, Lys-Trp, and Lys-Tyr residue pairs. In these interaction
schemes,
\begin{equation}
(U_{\rm aa})_{\mu i,\nu j} =
(U_{{\rm aa|HPS+c}\pi})_{\mu i,\nu j} \equiv
(U_{\rm aa|HPS})_{\mu i,\nu j} + (U_{{\rm aa|c}\pi})_{\mu i,\nu j}
\; ,
\label{eq:U-HPS_cp}
\end{equation}
where
\begin{equation}
(U_{\rm aa|c\pi})_{\mu i,\nu j} = 
(\epsilon_{{\rm c}\pi})_{ij} \left[\left(\frac {a_{ij}}{r_{\mu i,\nu j}}
\right)^{12} -
\left(\frac {a_{ij}}{r_{\mu i,\nu j}}\right)^6 \right] \; ,
\end{equation}
and $(\epsilon_{{\rm c}\pi})_{ij}$ is the cation-$\pi$ interaction strength,
$(\epsilon_{{\rm c}\pi})_{ij}>0$ only if residue pair $\mu i,\nu j$ is one 
of the aforementioned six cation-$\pi$ pairs, otherwise 
$(\epsilon_{{\rm c}\pi})_{ij}=0$.
This simple form is adopted from the cation-$\pi$ term in 
Eq.~(S1) of Ref.~\citen{kaw2013}.
\\

Two sets of $(\epsilon_{{\rm c}\pi})_{ij}$ values are analyzed in the
present study:\\

(i) $(\epsilon_{{\rm c}\pi})_{ij}=3.0$ kcal mol$^{-1}$ for all six
cation-$\pi$ pairs. 
The rationale for using a single $(\epsilon_{{\rm c}\pi})_{ij}$ value 
is the suggestion by statistical and other inferences that the
variations of interaction strengths among the six cation-$\pi$ amino
acid residue pairs could be relatively small,\cite{dougherty1999,kaw2013}
though subsequently we will also explore scenarios in which significant
variations in cation-$\pi$ interaction strengths exist among the pairs.
When combined with the $(U_{\rm aa|HPS})_{\mu i,\nu j}$ contribution in
Eq.~(\ref{eq:U-HPS_cp}), $(\epsilon_{{\rm c}\pi})_{ij}=3.0$ kcal mol$^{-1}$ 
leads to well depths for 
$(U_{\rm aa})_{\mu i,\nu j} = (U_{{\rm aa|HPS+c}\pi})_{\mu i,\nu j}$
of $\approx 0.85$ kcal mol$^{-1}$ for Arg-Phe, Arg-Trp, Arg-Tyr, 
and corresponding well depths of $\approx 0.90$ kcal mol$^{-1}$
for Lys-Phe, Lys-Trp, and Lys-Tyr (see Fig.~2a of the main text).
It should be noted here that
we have chosen an $(\epsilon_{{\rm c}\pi})_{ij}$ value significantly
smaller than those used in Ref.~\citen{kaw2013} in order for the model
cation-$\pi$ interactions to be
more compatiable with the shallow well depths of the 
$(U_{\rm aa|HPS})_{\mu i,\nu j}$ potentials in the HPS model, which has
a maximum well depth of $0.2$ kcal mol$^{-1}$.
Nonetheless, the $(\epsilon_{{\rm c}\pi})_{ij}=3.0$ kcal mol$^{-1}$
value still entails a cation-$\pi$ interaction strength which is about
double that of electrostatic interaction 
when $\epsilon_{\rm r}$ in Eq.~(\ref{eq:U-el})
corresponds to that of bulk water ($\epsilon_{\rm r}\approx 80$).
This ratio between the strengths of cation-$\pi$ and electrostatic 
interactions in an aqueous environment conforms to a similar ratio 
deduced computationally.\cite{dougherty2000}
\\

(ii) Different $(\epsilon_{{\rm c}\pi})_{ij}$ values for cation-$\pi$ pairs
involving Arg and pairs involving Lys, with
$(\epsilon_{{\rm c}\pi})_{ij}=1.85$ kcal mol$^{-1}$
for Arg-Phe, Arg-Trp, Arg-Tyr and
$(\epsilon_{{\rm c}\pi})_{ij}=0.65$ kcal mol$^{-1}$
for Lys-Phe, Lys-Trp, and Lys-Tyr.
This alternate model cation-$\pi$ interaction scheme is motivated
by observed trends of statistical potentials derived from PDB protein 
structures such as the Miyazawa-Jernigan energies\cite{MJ85,MJ96} 
used in the KH/MJ model\cite{dignon18} (described below) and
the new analysis presented in the main text as well as
recent experimental evidence,\cite{moleculargrammar,robert} 
all of which suggest that cation-$\pi$ 
interactions involving Arg is more favorable than those involving Lys.
The $(\epsilon_{{\rm c}\pi})_{ij}$ values in this scheme are chosen
such that the combined well depth of $(U_{{\rm aa|HPS+c}\pi})_{\mu i,\nu j}$
for cation-aromatic pairs are comparable to the deepest well depth
of $\approx 0.5$ kcal mol$^{-1}$ in the KH/MJ model. In particular,
$(\epsilon_{{\rm c}\pi})_{ij}=1.85$ kcal mol$^{-1}$
leads to a combined well depth of $\approx 0.55$ kcal mol$^{-1}$
for terms in $(U_{{\rm aa|HPS+c}\pi})_{\mu i,\nu j}$ involving Arg-aromatic
pairs, whereas $(\epsilon_{{\rm c}\pi})_{ij}=0.65$ kcal mol$^{-1}$
leads to a corresponding combined well depth of $\approx 0.3$ kcal mol$^{-1}$
for Lys-aromatic pairs (Fig.~2b of the main text). 
\\

\noindent
{\bf The KH (KH/MJ) model}

The Kim-Hummer/Miyazawa-Jernigan (KH/MJ) model corresponds to the 
KH-D model used by Dignon et al.,\cite{dignon18} and is based on the 
statistical potentials of Miyazawa and Jernigan (MJ).\cite{MJ96} Following
Ref.~\citen{dignon18}, we refer to this model as KH in the main text
and hereafter.
The basic functional form of the  KH potential, $U_{\rm aa|KH}$,
is similar to that for the HPS potential in Eq.~(\ref{eq:U-HPS}).
For the KH model,
\begin{equation}
(U_{\rm aa})_{\mu i,\nu j} =
(U_{\rm aa|KH})_{\mu i,\nu j} \equiv
\begin{cases}
(U_{\rm LJ})_{\mu i,\nu j} + (1-\lambda_{ij}^{\rm KH})\epsilon \; , 
& {\rm if\ } r\le 2^{1/6}a_{ij}\\
\lambda_{ij}^{\rm KH} (U_{\rm LJ})_{\mu i,\nu j} & {\rm otherwise}
\end{cases}
\; 
\label{eq:U-KH}
\end{equation}
where $(U_{\rm LJ})_{\mu i,\nu j}$ is given by Eq.~(\ref{eq:LJ}),
but now $\epsilon_{ij}$ depends on $i,j$. Specifically, 
for the KH model 
\begin{equation}
\epsilon_{ij}= |\alpha (e_{{\rm MJ},ij} - e_0)|
\; ,
\end{equation}
where $e_{{\rm MJ},ij}$ is the MJ statistical potential between
the residue type at position $i$ and the residue type at position $j$,
$e_0$ is a constant shift of the energies, and
\begin{equation}
\lambda_{ij}^{\rm KH} =
\begin{cases}
1 & {\rm if\ } e_{{\rm MJ},ij} \le e_0\\
-1 & {\rm otherwise} 
\end{cases}
\; .
\end{equation}
We use $\alpha=0.228$ and $e_0=-1.0$ kcal mol$^{-1}$ in the present
study. The resulting pairwise energies $e_{\rm MJ}$ correspond to the KH-D
parameter set for IDRs in Table~S3 of Ref.~\citen{dignon18}.
\\

\noindent
{\bf Simulation method}

Molecular (Langevin) dynamics simulations are carried out using the
protocol outlined in the ``Simulation framework'' section of 
Ref.~\citen{dignon18}, with parameters modified for the present
applications. For each simulation, we consider 100 copies of one
of the four Ddx4 IDR sequences (Fig.~S1) or the three LAF-1 
IDR sequences (Fig.~S5), governed by 
one of the above coarse-grained model potential functions. 
At the initial step, all the IDR chains are randomly placed in 
a relatively large, $300\times 300\times 300$ \AA$^3$ simulation box.
Energy minimization is then applied to minimize unfavorable
steric clashes among the amino acid residues. Equilibrating {\it NPT} 
simulation is then performed for 50 ns at a temperature of 100 K and pressure
of 1 bar, maintained by Martyna-Tobias-Klein (MTK) thermostat and 
barostat\cite{klein1994,tuckerman_etal2006} with a coupling constant of 1 ps. 
It should be noted that the simulation pressure
does not correspond to physical pressure because solvent (water) pressure
is not accounted for in the present coarse-grained, implicit-solvent model
setup. In this regard, pressure is used entirely as an efficient 
computational device
for achieving condensed configurations as starting point of subsequent
simulations. Throughout the dynamics simulation, equations
of motion are integrated with a timestep of 10 fs and periodic boundary
conditions are applied to all three spatial dimensions. After the initial 
{\it NPT} step, the simulation box is compressed again for 50 ns along
all three spatial dimensions at 100 K as successive {\it NVT}
ensembles ($P$ changes during the process) using Langevin thermostat with
friction coefficient 1 ps$^{-1}$. The extent of compression varies 
for different systems. Then the dimension along one of the three Cartesian axes 
of the simulation box is expanded 20 times relative to its initial value 
for a period of 50 ns while maintaing the temperature at 100 K.
Equilibration {\it NVT} simulation is then performed at the chosen 
temperature for 2 $\mu$s. Finally, production {\it NVT} runs are carried
out for 4 $\mu$s and the chain configurations are saved every 0.5
ns for subsequent analysis. During the production run, the friction 
coefficient of the Langevin thermostat is decreased to 0.01 ps$^{-1}$ for
sampling efficiency. All simulations are performed by the HOOMD-blue 
software package.\cite{anderson_etal2008,glaser_etal2008}
After the snapshots of simulated chain configurations are collected,
the procedure for constructing phase diagrams from the 
configurations follows that described in the ``Simulation framework'' 
section of Ref.~\citen{dignon18} and the ``Results and discussion'' 
section of Ref.~\citen{suman2}.
\\



\noindent{\bf E{\footnotesize{XPLICIT-}}W{\footnotesize{ATER}} 
S{\footnotesize{IMULATION OF}}
IDR-C{\footnotesize{ONCENTRATION}}-D{\footnotesize{EPENDENT}}\\
P{\footnotesize{ERMITTIVITY}}}\\

\noindent
{\bf Computational procedure}

We estimate the IDR-concentration-dependent relative 
permittivity\cite{linJML,njp2017} by atomistic explicit-water molecular dynamic 
simulations performed at six Ddx4 IDR (wildtype, WT) concentrations using
GROMACS, version 2016.5.\cite{gromacs2018} 
The simulation proceeds as follows. 
Initially, a fully extended configuration of a Ddx4 IDR
is prepared by PYMOL,\cite{pymol2002} to be used as input for 
Packmol\cite{packmol2009}
to place five Ddx4 IDRs at random locations in a cubic simulation box.
The size of the box is varied to achieve different Ddx4 IDR concentrations.
The Ddx4 IDRs are solvated by explicit water models in the simulation box.
Each of the systems so constructed is then charge neutralized by 
adding appropriate number of Na$^+$ ions.  This is followed
by energy minimization by steepest descent to minimize steric clashes.
Hydrogen bonds are constrained with the LINCS algorithm.\cite{lincs1997} 
Equation of motion is integrated using a time step of 2 fs with the 
leap-frog integrator\cite{hockney_etal1974} 
and cubic periodic boundary conditions. Long spatial-range
electrostatic interaction is treated with particle mesh Ewald (PME) 
method\cite{essmann_etal1995}
with a grid spacing of 0.16 nm and an interpolation order of 4. 
A cut-off of 1 nm is used for short-range van der Waals and electrostatic
interactions.
Initial equilibration is carried out for 2 ns under $NVT$ conditions at 300 K. 
Temperature is maintained by Velocity-rescale 
thermostat\cite{bussi2007} with
a time constant of 0.1 ps for all simulations.
This is followed by equilibration for 2 ns at 300 K under $NPT$ conditions
under 1 atm pressure, which is maintained by a Berendsen 
barostat\cite{berendsen1984} 
with a coupling constant of 2 ps. Since the Berendsen
barostat does not always yield an $NPT$ ensemble with high accuracy, 
the resulting system is equilibrated again for 1 ns as an $NPT$
ensemble using the Parrinello-Rahman 
barostat\cite{parrinello1981,nose1983} with the
same coupling constant, after which the production {\it NPT} run is carried out 
for 20 ns using the same Parrinello-Rahman barostat. 
Configurations are saved every 1.0 ps during the production run for
subsequent analysis. 
In addition to simulations of Ddx4 IDR in essentially pure water (except
a few Na$^+$ ions), we also conduct simulations with 
Na$^+$ and Cl$^-$ ions at [NaCl] = 100 mM.
\\

In order to enable a potentially more direct comparison with 
analytical theory that does not include the charges of amino acid
residues in the estimation of effective permittivity of the aqueoue
medium,\cite{linJML,njp2017}
we carry out another set of simulations with Ddx4 IDR concentrations
similar to the ones for which the above protocol is applied but with
the charges of the sidechains of the charged amino acids Arg, Lys, Asp, and
Glu artifically turned off.
This set of simulation data is referred to as
artificial Ddx4 or aDdx4. 
The same aforementioned procedure for equilibration and production
is applied for this set of simulations.
The amber99sb-ildn force field\cite{DEShaw2010} and the TIP3P water 
model\cite{jorgensen1983} are used for both sets of simulations.
To assess the robustness of the computed $\epsilon_{\rm r}$ values,
all simulations are also repeated using the SPC/E water 
model.\cite{berendsen1987}
\\

\noindent
{\bf Relative permittivity analysis}

Static relative permittivity $\epsilon_{\rm r}$ (dielectric constant) 
is determined by the fluctuation of 
the total dipole moment vector, ${\bf M}_{\rm T}$, of the system via the 
relation\cite{vanGunsteren2001}
\begin{equation}
\epsilon_{\rm r} = \frac {\langle M^2_{\rm T}\rangle - 
\langle M_{\rm T}\rangle^2}{3V\epsilon_0 k_{\rm B}T} + 1 
\; ,
\label{eq:MT}
\end{equation}
where $M_{\rm T}\equiv ({\bf M}_{\rm T}\cdot {\bf M}_{\rm T})^{1/2}$ 
is the magnitude of the system dipole moment,
$\langle\dots\rangle$ denotes averaging over system configurations
under equilibrium conditions, 
and $V$ is the volume of the simulation box.
This relation, Eq.~(\ref{eq:MT}), has been used to compute the static
dielectric constant of several biological 
systems.\cite{vanGunsteren2001,rudas_etal2006,papa2015} Following
the formulation in Ref.~\citen{rudas_etal2006}, ${\bf M}_{\rm T}$
is obtained as sum of dipole moments
of individual water molecules and individual Ddx4 IDR chain molecules.
Irrespective of the net charge of the molecule (water has net charge 0
whereas Ddx4 IDR has net charge $\approx -4e$),
the dipole moment, ${\bf m}$, of a molecule comprising of
$N_m$ atoms with masses $m_s$ ($s=1,2,\dots,N_m$) at positions ${\bf r}_s$ 
with point charges $q_s$ is given by
${\bf m}=\sum_s^{N_m} q_s ({\bf r}_s - {\bf r}_{\rm cm})$, where
${\bf r}_{\rm cm}\equiv \sum_s^{N_m} m_s{\bf r}_s/\sum_s^{N_m} m_s$ 
is the center-of-mass
position of the molecule. Accordingly, atomic ions, Na$^+$s and
Cl$^-$s in our case, 
have zero dipole moment in this formulation. Once the dipole moments
of the water and Ddx4 molecules are determined in this manner, they are
combined to yield ${\bf M}_{\rm T}$ which in turn provides 
the system relative permittivity through Eq.~(\ref{eq:MT}).
Our computed $\epsilon_{\rm r}$ for various concentrations of Ddx4 IDR
at different salt concentrations using both the TIP3P and SPC/E water
models are given in Table~S1.
\\

\vfill\eject

\noindent{\large\bf 
Random-Phase-Approximation (RPA) Theory of Phase
{\hbox{Separation}} with 
IDR-Concentration-Dependent Permittivity}\\

\noindent {\bf B{\footnotesize{ACKGROUND}}}

Our group has previously considered, within  
our RPA theory of liquid-liquid phase separation (LLPS),
the effects of 
relative permittivity $\epsilon_{\rm r}$ being dependent 
upon local protein concentration;\cite{linJML,njp2017} i.e., 
$\epsilon_{\rm r}=\epsilon_{\rm r}(\phi_m)$ where $\phi_m$ is polymer (IDR)
volume fraction. An $\epsilon_{\rm r}(\phi_m)$ necessitates
changes to our earlier RPA expressions for electrostatic interaction 
for a constant, position-independent 
$\epsilon_{\rm r}$, viz. [Eq.~(33) of Ref.~\citen{linJML}],
\begin{equation}
f_{\rm el} = \frac{1}{2}\int \frac{d^3 (ka)^3}{(2\pi)^3}
\left\{\ln[\det(1 + \hat{G}_k\hat{U}_k)] - \mathrm{Tr}
(\hat{\rho}\;\hat{U}_k) \right\} \; .
        \label{eq:fel_ori}
\end{equation}
Here, as in Ref.~\citen{linJML}, $a^3$ is unit volume, $\hat{G}_k$ is 
the position correlation matrix, 
$\hat{\rho}$ is the density operator that provides the densities of 
various molecular species in the system (accounting for matter, not
electric charge), and $\hat{U}_k$ accounts for 
sequence-dependent Coulumb interactions [the expression for $\hat{U}_k$ 
is provided by Eq.~(35) of Ref.~\citen{linJML}]. 
For the simple illustrative case here, which is a system
of only IDR polymers without salt or counterions,
$\hat{G}_k$ reduces to the monomer-monomer correlation 
$(\hat{G}_k)_{ij}=(\rho_m/N)(\hat{G}_{\rm M}(kl))_{ij}=$
$\rho_m\exp[-(kl)^2|i-j|/6]/N$, where 
$\rho_m$ is monomer density,
$l$ is the length of a polymer link (virtual bond length, denoted
as $b$ in Ref.~\citen{linJML}),
$i,j=1,2,\dots N$ are monomer labels along the 
polymer chain with $N$ being the length of a chain, and
$\hat{\rho}\;\hat{U}_k$ $=\rho_m\hat{U}_k/N$ [Eq.~(4) of Ref.~\citen{linPRL}].

When $\epsilon_{\rm r}=\epsilon_{\rm r}(\phi_m)$, we applied the following
modified version of $f_{\rm el}$
[Eq.~(68) of Ref.~\citen{linJML}]: 
\begin{equation}
f_{\rm el} = \int \frac{d\kr\kr^2}{4\pi^2} \left\{\frac{1}{\eta}\ln\left[
1 + \eta {\cal G}_1(\kr)\right] - {\cal G}_2(\kr) \right\} \; ,
        \label{eq:fel_dimless_dic}
\end{equation}
where $\kr = kb$, $\eta = (b/a)^3$ and, in the absence of salt and
counterions, Eqs.~(69a) and (69b) of Ref.~\citen{linJML} become
\begin{subequations}
\begin{align}
\!\!{\cal G}_1(\kr) & = \frac{4\pi}{\kr^2[1+\kr^2] T_0^* \epsilon_r(\phi_m)}
         \left( \frac{\phi_m}{N}
                \langle \sigma | \GM(\kr) | \sigma \rangle \right), \\
\!\!{\cal G}_2(\kr) & = \frac{4\pi}{\kr^2[1+\kr^2] T_0^* \epsilon_r(\phi_m)}
        \left(\frac{\phi_m}{N} \sum_{i=1}^N|\sigma_i|
\right) \; .
\end{align}
\label{eq:G_factors}
\end{subequations}
$\null\hskip -0.25cm$
As in Refs.~\citen{linJML} and \citen{njp2017},
column vector $|\sigma\rangle$ is the charge sequence---its $i$th element, 
$\sigma_i$, being the charge of the 
$i$th monomer (residue) of the IDR in units of the electronic charge $e$,
and $\langle\sigma\vert\equiv|\sigma\rangle^{\rm T}$;
$T_0^* \equiv {4\pi\epsilon_0 k_{\rm B} Tb}/{e^2}$
is a reduced temperature. As noted above, $\epsilon_0$
is vacuum permittivity, $k_{\rm B}$ is Boltzmann constant, and $T$ is
absolute temperature. Previously,\cite{linJML,njp2017} expressions
such as above Eqs.~(\ref{eq:fel_dimless_dic}) and 
(\ref{eq:G_factors}) for $\epsilon_{\rm r}(\phi_m)$ were obtained 
heuristically by replacing every instance of $\epsilon_{\rm r}$ in the 
corresponding constant-$\epsilon_{\rm r}$ expressions by 
$\epsilon_{\rm r}(\phi_m)$.
\\


$\null$

\noindent 
{\bf C{\footnotesize{ONCENTRATION}}-D{\footnotesize{EPENDENT}}}
{\bf P{\footnotesize{ERMITTIVITY}}}
{\bf {\footnotesize{IN THE}}}
{\bf RPA} {\bf C{\footnotesize{ONTEXT}}}

We now examine whether---and if so what---restrictive 
conditions have to be satisfied for the heuristic prescription 
Eqs.~(\ref{eq:fel_dimless_dic}) and (\ref{eq:G_factors}) to be valid.

When $\epsr$ is position-independent, the electrostatic interaction 
energy (potential), in units of $k_{\rm B}T$, between two unit point 
charges $e$ at positions $\rr$ and
$\rr'$ is given 
by $\UU(\rr,\rr')=\UU(\rr-\rr')=$
$e^2/(4\pi\epsilon_0\epsr k_{\rm B}T|\rr-\rr'|)$.
However, when $\epsr$ is position-dependent, i.e., $\epsr=\epsr(\rr)$, 
in general
the electrostatic potential $\UU$ is not expressible in a simple closed 
form because it is the solution to the generalized Poisson equation 
\begin{equation}
-\nabla_\rr \cdot \left[ \epsr(\rr) \nabla_\rr \; \UU(\rr-\rr') \right] 
= 4\pi\lb \delta(\rr-\rr') \; ,
\label{eq:poisson}
\end{equation}
as noted by Wang,~\cite{Wang2010}
where $\lb=e^2/(4\pi\epsilon_0 k_{\rm B}T)$ is vacuum Bjerrum length (unlike
Ref.~\citen{linJML}, here $\lb$ does not include $\epsr$).
Thus, position dependence of $\epsr$ can entail 
rather complex modifications of the charge-charge interactions.
It cannot be analytically treated, in general, by simply replacing the 
constant $\epsr$ in 
$\UU(\rr,\rr')=e^2/(4\pi\epsilon_0\epsr k_{\rm B}T|\rr-\rr'|)$
by $\epsr(\rr)$ or $\epsr(\rr')$. 

Another concern is that, by construction, RPA theory accounts only 
for the lowest-order 
polymer density fluctuations beyond the mean-field homogeneous 
density. In contrast, some of the proposed IDR-concentration-dependent 
form of $\epsr=\epsr(\phi_m)$, such as the ``slab''\cite{bragg} and 
Clasusius-Mossotti\cite{Jackson} models and the  effective medium 
approximations of Maxwell Garnett and of Bruggeman\cite{MGB}
considered in Refs.~\citen{linJML,njp2017}
involve higher-order dependence on $\phi_m$, raising questions as to
whether application of these $\epsr(\phi_m)$ formula in the
context of RPA is consistent with the basic premises of RPA. We address
these issues below.


$\null$

\noindent 
{\bf D{\footnotesize{ERIVATION}}}
{\bf {\footnotesize{OF}} RPA}
{\bf {\footnotesize{WITH}}}
{\bf C{\footnotesize{ONCENTRATION}}-D{\footnotesize{EPENDENT}}}
{\bf P{\footnotesize{ERMITTIVITY}}}

Unless specified otherwise, the notation in this subsection follows that 
of Ref.~\citen{kings2020}, as the following formal development is, on
one hand, a restricted case of the theory in Ref.~\citen{kings2020} in
that here we do not consider salt, counterions or Kuhn-length 
renormalization. On the other hand, the present analysis is an extension of 
the theory in Ref.~\citen{kings2020}, which is limited to constant $\epsr$s,
to case with a position-dependent $\epsr(\rr)$. Accordingly, we note
that the number of chains in the system, which is symbolized by $n$ in 
the main text and elsewhere in this Supporting Information, is denoted 
by $n_{\rm p}$ (following Ref.~\citen{kings2020}) in the derivation below.

In general, the Boltzmann factor for the electrostatic interaction energy 
of a system with charge density $\rho(\rr)$ is given by
$\exp[-(1/2) \int d\rr d\rr' \rho (\rr) \UU(\rr,\rr')\rho (\rr')]$.
(Note that the electric charge density $\rho(\rr)$ here and in subsequent 
development in this section should not be confused with the matter density
operator $\hat{\rho}$ or its matrix elements.)
We focus first on obtaining an equivalent mathematical form of this factor
that is amenable to RPA analyses.
By standard field-theoretic Hubbard-Stratonovich transformation, this
factor may be expressed as a functional integral
over a conjugate field $\psi(\rr)$:
\begin{equation}
\frac {1}{(\det \hat{\UU})^{1/2}}
\biggl\{\prod_{\rr} \int \frac {d\psi(\rr)}{\sqrt{2\pi}}\biggr\}
\;
\exp \biggl[-\frac {1}{2}\int d\rr' d\rr'' 
\psi (\rr') \UU^{-1}(\rr',\rr'')\psi (\rr'')
-i\int d\rr' \rho(\rr')\psi(\rr')\biggr] \; ,
\label{eq:HST}
\end{equation}
where $\hat{\UU}$ denotes, in matrix notation, the operator
$\UU(\rr,\rr')$ [i.e., the matrix element $\hat{\UU}_{\rr,\rr'}=\UU(\rr,\rr')$],
$\UU^{-1}(\rr',\rr'')$ is the $\rr',\rr''$ matrix element of the inverse 
operator $\hat{\UU}^{-1}$ of $\hat{\UU}$. By definition, 
$\int d\rr'' \UU^{-1}(\rr,\rr'')\UU(\rr'',\rr')=\delta(\rr - \rr')$.
Consider now the operator
$-\nabla_{\rr''}\cdot[\epsr(\rr'') \nabla_{\rr''}\delta(\rr -\rr'')]/(4\pi\lb)$.
Since
\begin{equation}
\begin{aligned}
\int d\rr'' 
\{\nabla_{\rr''}\cdot[\epsr(\rr'') \nabla_{\rr''}\delta(\rr -\rr'')]\}
\UU(\rr'',\rr') & =
\int d\rr'' 
[\epsr(\rr'') \nabla_{\rr''}\delta(\rr -\rr'')]\cdot
\nabla_{\rr''}\UU(\rr'',\rr') \\
& =
\int d\rr'' 
\delta(\rr -\rr'')
\{\nabla_{\rr''}\cdot [\epsr(\rr'')\nabla_{\rr''}\UU(\rr'',\rr')]\} \\
\end{aligned}
\label{eq:delta}
\end{equation}
follows from repeated applications of integration by parts under the 
reasonable assumption that the values of the integrand cancel or 
vanish at the pertinent boundaries of integration, and by
Eq.~(\ref{eq:poisson})
the quantity in curly brackets in the last term in Eq.~(\ref{eq:delta})
is $-4\pi\lb \delta(\rr'' -\rr')$, Eq.~(\ref{eq:delta}) is evaluated as
$-4\pi\lb\int d\rr'' \; \delta(\rr -\rr'')\delta(\rr'' -\rr')=$
$-4\pi\lb\delta(\rr -\rr')$ and therefore 
$-\nabla_{\rr''}\cdot[\epsr(\rr'') \nabla_{\rr''}\delta(\rr -\rr'')]/(4\pi\lb)$
is the $\rr,\rr''$ matrix element of the inverse of $\hat{\UU}$, viz.,
\begin{equation}
\UU^{-1}(\rr,\rr'') = -\frac {1}{4\pi\lb}
\nabla_{\rr''}\cdot[\epsr(\rr'') \nabla_{\rr''}\delta(\rr -\rr'')] \; .
\label{eq:U-1}
\end{equation}
Equivalently, the $\rr'',\rr$ matrix element of $\hat{\UU}^{-1}$ takes the form
\begin{equation}
\UU^{-1}(\rr'',\rr) = -\frac {1}{4\pi\lb}
\nabla_{\rr}\cdot[\epsr(\rr) \nabla_{\rr}\delta(\rr'' -\rr)] \; .
\end{equation}
It follows that the 
$-(1/2)\int d\rr' d\rr'' \psi (\rr') \UU^{-1}(\rr',\rr'')\psi (\rr'')$ factor
in Eq.~(\ref{eq:HST}) is given by
\begin{equation}
\begin{aligned}
- \frac{1}{2}\int d\rr' d\rr'' \psi (\rr') \UU^{-1}(\rr',\rr'')\psi (\rr'')
& = 
\frac{1}{8\pi\lb}
\int d\rr d\rr' \psi(\rr)
\{\nabla_{\rr'}\cdot[\epsr(\rr') \nabla_{\rr'}\delta(\rr -\rr')]\}\psi(\rr')\\
& = - \frac{1}{8\pi\lb}
\int d\rr d\rr' \psi(\rr)
[\epsr(\rr') \nabla_{\rr'}\delta(\rr -\rr')]\cdot [\nabla_{\rr'}\psi(\rr')]\\
& = \frac {1}{8\pi\lb}
\int d\rr d\rr' \psi(\rr)
[\epsr(\rr') \nabla_{\rr}\delta(\rr -\rr')]\cdot [\nabla_{\rr'}\psi(\rr')]\\
& = - \frac{1}{8\pi\lb}
\int d\rr d\rr' \epsr(\rr') \delta(\rr -\rr')
[\nabla_{\rr}\psi(\rr)] \cdot [\nabla_{\rr'}\psi(\rr')]\\
& = - \frac {1}{8\pi\lb}
\int d\rr \; \epsr(\rr) [\nabla_{\rr}\psi(\rr)]
\cdot [\nabla_{\rr}\psi(\rr)]\\
& = - \frac{1}{8\pi\lb}
\int d\rr \; \epsr(\rr) \vert\nabla\psi(\rr)\vert^2 \; ,\\
\end{aligned}
\label{eq:gradpsi}
\end{equation}
where the first equality follows from a mere change in the integration
variable, the second and fourth equalities from integration 
by parts assuming that boundary contribution vanishes, the 
third equality from
{\hbox{
$\nabla_{\rr'}\delta(\rr -\rr')=-\nabla_{\rr}\delta(\rr -\rr')$,
}}
and the $\rr$ subscript of $\nabla_{\rr}$ is dropped in the final 
expression because there is little danger of notational ambiguity.
Equation~(\ref{eq:gradpsi}) is identical to the corresponding terms in
the Hamiltonians in
Eq.~(3) of Ref.~\citen{Wang2004} and Eq.~(2.7) of Ref.~\citen{Wang2010}
for systems with an inhomogeneous dielectric medium.

We turn next to the $(\det {\hat\UU})^{-1/2}$ factor in Eq.~(\ref{eq:HST}).
For any matrices $A$ and $B$, $(\det A)^{-1}=(\det A^{-1})$ and
$(\det AB)=(\det A)(\det B)$, we write
$(\det{\hat\UU})^{-1/2}=(\det{\hat\UU}^{-1})^{1/2}$
$=(\det{\hat\epsr})^{1/2}(\det{\hat\UU}_0^{-1})^{1/2}$, where
${\hat\UU}^{-1}$'s matrix elements 
$\UU^{-1}_{\rr\rr'}\equiv\UU^{-1}(\rr,\rr')$ is given by Eq.~(\ref{eq:U-1}),
the $\rr,\rr'$ matrix elements of the operators ${\hat\epsr}$ 
and ${\hat\UU}_0^{-1}$ are defined, respectively, by
\begin{equation}
({\hat\epsr})_{\rr\rr'}\equiv \epsr(\rr)\delta(\rr-\rr') \; ,
\label{eq:epsr}
\end{equation}
\begin{equation}
({\hat\UU}^{-1}_0)_{\rr\rr'}\equiv -\frac {1}{4\pi\lb} 
\nabla^2_{\rr}\delta(\rr -\rr')
\; .
\end{equation}
Then, ${\hat\epsr}\;{\hat\UU}_0^{-1}= {\hat\UU}^{-1}$
can be ready verified using integration by parts:
\begin{equation}
\begin{aligned}
({\hat\epsr}\;{\hat\UU}_0^{-1})_{\rr\rr'} & =
\int d\rr'' ({\hat\epsr})_{\rr\rr''} ({\hat\UU}_0^{-1})_{\rr''\rr'}
\\
& = -\frac {1}{4\pi\lb} \int d\rr'' \epsr(\rr)\delta(\rr-\rr'')
\nabla^2_{\rr''}\delta(\rr'' -\rr')
\\
& = \frac {1}{4\pi\lb} \int d\rr'' \epsr(\rr'')[\nabla_{\rr''}\delta(\rr-\rr'')]
\cdot
[\nabla_{\rr''}\delta(\rr'' -\rr')]
\\
& = - \frac {1}{4\pi\lb} \int d\rr'' 
\nabla_{\rr''}\cdot
\{\epsr(\rr'')[\nabla_{\rr''}\delta(\rr-\rr'')]\}
\delta(\rr'' -\rr')
\\
& = -\frac {1}{4\pi\lb}
\nabla_{\rr'}\cdot[\epsr(\rr') \nabla_{\rr'}\delta(\rr -\rr')]
\\
& = {\hat\UU}^{-1}_{\rr\rr'} \; ,
\; {\rm Q.E.D.}
\end{aligned}
\end{equation}
Because ${\hat\epsr}$ in Eq.~(\ref{eq:epsr}) is a diagonal matrix, 
\begin{equation}
(\det{\hat\epsr}) = \prod_\rr \epsr(\rr) \; .
\label{eq:det_epsilon}
\end{equation}
Using Fourier transformation from $\rr$ to $\kk$ space,\cite{kings2020}
\begin{equation}
(\det {\hat\UU}^{-1}_0) = \prod_{\kk\ne {\bf 0}}\frac {k^2}{4\pi\lb} \; ,
\label{eq:det_U0}
\end{equation}
where $k^2\equiv |\kk|^2$. Note that the $\kk ={\bf 0}$ term is excluded
in the above and subsequent considerations because it does not
contribute to the exponential factor in Eq.~(\ref{eq:HST})
for our electrically neutral system of overall neutral 
polyampholytes.

The free energy per unit volume $l^3$ in units of $\kB T$ of our system
is given by
\begin{equation}
f = \frac{\phi_m}{N}\ln\phi_m + 
(1-\phi_m)\ln(1-\phi_m) - \frac{l^3}{\Omega}\ln \Zel,
\label{eq:free-energy}
\end{equation}
where $N$ is the chain length (number of monomers) of the polyampholyte,
$\Omega$ is solution (system) volume, 
$\phi_m\equiv l^3\np N/\Omega$ is monomer volume fraction
with $\np$ being the total number of
identical polyampholyte chains in the solution
[$\np$ corresponds to the variable $n$ used above in the formulation
for explicit-chain simulations; it should also be noted here that the 
alternately defined $\phi_m=a^3\np N/\Omega$ 
in Eq.~(3) of Ref.~\citen{linJML}---which applies to 
Eqs.~(\ref{eq:fel_dimless_dic}) and (\ref{eq:G_factors}) 
in the present work---is equal to polyampholyte volume fraction 
only when the size of a monomer $\approx l^3$ is 
equal to the model volume unit $a^3$ of the model, i.e., when $r_m=1$; 
whereas polyampholyte volume fraction is given by $r_m\phi_m$ 
in general \cite{linJML}; for simplicity, 
$r_m = 1$ is assumed below unless specified otherwise],
and $\Zel$ is the electrostatic partition function, which may be viewed 
as a special case of $\Z^\prime$ in Eq.~(A9) of Ref.~\citen{kings2020}
with no salt, no counterion, and $v_2=0$, but now extended to 
$\epsr=\epsr(\rr)$. 
$\Zel$ is given by integrals over monomer coordinates,
\begin{equation}
\Zel = \int \prod_{\alpha=1}^{\np} 
\prod_{\tau=1}^N d\R_{\alpha, \tau} e^{-\Eng[\R]} \; ,
        \label{eq:Zel_intR}
\end{equation}
where $\R_{\alpha, \tau}$ denotes the coordinate of the $\tau$th monomer 
in the $\alpha$th polyampholyte [$\R_{\alpha, \tau}$ corresponds to the 
position variable ${\bf R}_{\mu i}$ defined before Eq.~(\ref{eq:U-T})
in the formulation for explicit-chain 
simulations; the monomer label $\tau$ corresponds also to the label
$i$ in Eq.~(\ref{eq:G_factors}b)], and 
\begin{equation}
\Eng[\R] = \frac{3}{2l^2}\sum_{\alpha=1}^\np \sum_{\tau=1}^{N-1}\left( \R_{\alpha,\tau+1}-\R_{\alpha,\tau} \right)^2
+ \frac {1}{2} \sum_{\alpha,\beta=1}^\np
\sum_{\tau,\mu=1}^N \VV_{\alpha\beta}^{\tau\mu}\left( \R_{\alpha,\tau}, 
\R_{\beta,\mu} \right) \; .
        \label{eq:Ham_ori}
\end{equation}
The first term of $\Eng[\R]$ is for Gaussian-chain connectivity of the
polyampholyte chains and 
$\VV_{\alpha\beta}^{\tau\mu}$ in the second term is 
the interaction potential energy between the $\tau$th monomer 
in the $\alpha$th chain and the $\mu$th monomer in the $\beta$th chain,
viz.,
\begin{equation}
\VV_{\alpha\beta}^{\tau\mu}\left( \rr, \rr' \right) = 
\lb\sigma_\tau\sigma_\mu\UU(\rr,\rr') \; ,
\end{equation}
where $\sigma_\tau$, $\sigma_\mu$ are the charges, 
respectively, of monomers $\tau$, $\mu$ along each of the $\np$ polyampholyte
chains.
We may now rewrite
Eq.~(\ref{eq:Zel_intR}) as
a functional integral over the charge density $\rho(\rr)$
by including in the integrand a $\delta$-functional for $\rho(\rr)$:
\begin{equation}
\Zel = \int \prod_\rr d\rho(\rr) 
\int\prod_{\alpha=1}^{\np} 
\prod_{\tau=1}^N d\R_{\alpha, \tau} 
e^{-\Eng[\rho,\R]} \; 
\delta[\rho(\rr)-
\sum_{\alpha=1}^\np \sum_{\tau=1}^N \sigma_\tau\delta(\rr - \R_{\alpha,\tau})]
\; ,
        \label{eq:Zel_rho}
\end{equation}
which follows from $\rho(\rr)=\sum_{\alpha=1}^\np \sum_{\tau=1}^N \sigma_\tau
\delta(\rr - \R_{\alpha,\tau})$, whereas $\Eng[\rho,\R]$ is defined as
\begin{equation}
\Eng[\rho,\R] = \frac{3}{2l^2}\sum_{\alpha=1}^\np 
\sum_{\tau=1}^{N-1}\left( \R_{\alpha,\tau+1}-\R_{\alpha,\tau} \right)^2
+ \frac {1}{2}\int d\rr d\rr' \; \rho(\rr)\UU(\rr,\rr')\rho(\rr')
\; .
        \label{eq:Ham_rho}
\end{equation}
Now, by applying
Eqs.~(\ref{eq:HST}) and (\ref{eq:gradpsi}), 
the partition function $\Zel$ in Eq.~(\ref{eq:Zel_rho})
may be expressed as
a functional integral over $\rho(\rr)$, $\R_{\alpha, \tau}$, and
the conjugate fields $\psi(\rr)$:
\begin{equation}
\begin{aligned}
\Zel = & \int \prod_\rr d\rho(\rr) 
\int\prod_{\alpha=1}^{\np} 
\prod_{\tau=1}^N d\R_{\alpha, \tau} 
\exp \biggl[
-\frac{3}{2l^2}\sum_{\alpha=1}^\np 
\sum_{\tau=1}^{N-1}\left( \R_{\alpha,\tau+1}-\R_{\alpha,\tau} \right)^2
\biggr] \\
&\times
\frac {1}{(\det \hat{\UU})^{1/2}}
\biggl\{\prod_{\rr} \int \frac {d\psi(\rr)}{\sqrt{2\pi}}\biggr\}
\;
\exp \biggl[
 - \frac{1}{8\pi\lb}
\int d\rr \; \epsr(\rr) \vert\nabla\psi(\rr)\vert^2 
-i\int d\rr' \rho(\rr')\psi(\rr')\biggr] \\
&\quad \times
\delta[\rho(\rr)-
\sum_{\alpha=1}^\np \sum_{\tau=1}^N \sigma_\tau\delta(\rr - \R_{\alpha,\tau})]
\; .
\end{aligned}
        \label{eq:pathingetal_psi}
\end{equation}
After performing the $\prod_\rr d\rho(\rr)$ functional integrals in the
above expression, $\Zel$ becomes 
\begin{equation}
\Zel = 
\int\prod_{\alpha=1}^{\np} 
\prod_{\tau=1}^N d\R_{\alpha, \tau} 
\frac {1}{(\det \hat{\UU})^{1/2}}
\biggl\{\prod_{\rr} \int \frac {d\psi(\rr)}{\sqrt{2\pi}}\biggr\}
e^{-\Eng[\psi,\R]} \; , 
\label{eq:after_drho}
\end{equation}
where
\begin{equation}
\Eng[\psi,\R]= 
\frac{3}{2l^2}\sum_{\alpha=1}^\np
\sum_{\tau=1}^{N-1}\left( \R_{\alpha,\tau+1}-\R_{\alpha,\tau} \right)^2
+ \frac{1}{8\pi\lb}
\int d\rr \; \epsr(\rr) \vert\nabla\psi(\rr)\vert^2
+i\sum_{\alpha=1}^\np\sum_{\tau=1}^N \sigma_\tau \psi(\R_{\alpha, \tau})
\label{eq:HpsiR}
\; .
\end{equation}
We now proceed to evaluate the $(\det \hat\UU)^{-1/2}$ factor in
Eq.~(\ref{eq:after_drho}) via the aforementioned relations 
$(\det \hat\UU)^{-1/2}$ $=(\det \hat\UU^{-1})^{1/2}$ and 
${\hat\UU}^{-1}={\hat\epsr}\;{\hat\UU}_0^{-1}$. Using 
Eq.~(\ref{eq:det_epsilon}) and  applying the correspondence
\begin{equation}
\sum_\rr \to \frac{{\cal N}_\rr}{\Omega} \int d\rr
\label{eq:corr1}
\end{equation}
where ${\cal N}_\rr$ is formally the number of $\rr$ positions in the 
system, we may write
\begin{equation}
\sqrt{\det\hat{\epsr}} = \prod_\rr \sqrt{\epsr(\rr)} = 
\exp\biggl\{  \frac {1}{2} \sum_\rr \ln [\epsr(\rr)] \biggr\}
= \exp\left\{  \frac{{\cal N}_\rr}{2\Omega}\int d\rr \ln [\epsr(\rr)] \right\}
\; .
\label{eq:root_det}
\end{equation}
For reasons to be enunciated below, consider the case in which
$\epsr(\rr)$ is a linear combination of polyampholyte and water 
relative permittivites, 
i.e.,
\begin{equation}
\epsr(\rr) = \epsp \phi_m(\rr) + \epsw[1-\phi_m(\rr)] = 
\epsw + \deps \phi_m(\rr) \; ,
\label{eq:def_eps}
\end{equation}
where 
$\epsp$ and $\epsw$ are, respectively, the relative permittivities of
polymer and water, and $\deps = \epsp - \epsw$. Since
the position-dependent monomer density
\begin{equation}
\phi_m(\rr) = l^3\sum_{\alpha=1}^\np\sum_{\tau=1}^N \delta 
\left(\rr-\R_{\alpha,\tau}\right) \; ,
\label{eq:phi_rr}
\end{equation}
\begin{equation}
\begin{aligned}
\ln[\epsr(\rr)] = & \ln \epsw + \ln\left[ 1 + \frac{\deps}{\epsw}\phi_m(\rr)
\right] \\
= & \ln \epsw + \ln \left[ 1+ \frac{\deps l^3}{\epsw}\sum_{\alpha=1}^\np\sum_{\tau=1}^N \delta \left(\rr-\R_{\alpha,\tau}\right)\right] \; .
\end{aligned}
\end{equation}
To be consistent with RPA which accounts only for lowest-order polymer density
fluctuations, we approximate the above expression for $\ln[\epsr(\rr)]$ by
including terms only up to the one linear in $\phi_m$, viz.,
\begin{equation}
\ln[\epsr(\rr)] \approx
\ln \epsw + \frac{\deps l^3}{\epsw}\sum_{\alpha=1}^\np\sum_{\tau=1}^N 
\delta \left(\rr-\R_{\alpha,\tau}\right) \; .
\label{eq:er_approx}
\end{equation}
Hence the argument of the exponential function in Eq.~(\ref{eq:root_det}) 
is given by
\begin{equation}
\begin{aligned}
\frac{{\cal N}_\rr}{2\Omega}\int d\rr \ln[\epsr(\rr)] \approx &
\frac{{\cal N}_\rr}{2}\ln\epsw +  \frac{{\cal N}_\rr\deps 
l^3 \np N}{2\epsw\Omega} 
= \frac{{\cal N}_\rr}{2}\ln\epsw + \frac{{\cal N}_\rr\deps}{2\epsw}\phi_m \\
\approx & \frac{{\cal N}_\rr}{2}\ln\epsw + \frac{{\cal N}_\rr}{2}\ln\left( 1 + \frac{\deps}{\epsw} \phi_m \right) \\
= &  \frac{{\cal N}_\rr}{2}\ln [\epsr(\phi_m)]
\; ,
\end{aligned}
\label{eq:epsilon_r_int}
\end{equation}
where the position-independent $\phi_m\equiv (l^3/\Omega)\int d\rr 
\sum_{\alpha=1}^\np\sum_{\tau=1}^N \delta(\rr-\R_{\alpha,\tau})
= l^3\np N/\Omega$ is the overall average monomer volume fraction, 
the second approximate relation is in line with that in 
Eq.~(\ref{eq:er_approx}), and the
last equality follows from definition Eq.~(\ref{eq:def_eps}).
In formulations involving a size-dependent mean-field lattice model 
with $\phi_m$ defined in terms of unit volume $a^3\ne l^3$
(Ref.~\citen{linJML}), the actual average monomer volume fraction $\phi$
is given by $\phi=r_m\phi_m$ where $r_m$ is the monomer size factor,
in which case $\epsr(\phi_m)$ is understood to represent 
the $\epsr$ expression
in which all $\phi_m$ is replaced by $\phi=r_m\phi_m$; i.e.,
$\epsr(\phi_m)\rightarrow \epsr(\phi_m\to\phi=r_m\phi_m)$.
With Eq.~(\ref{eq:epsilon_r_int}),
further application of Eqs.~(\ref{eq:det_U0}) and (\ref{eq:root_det}) yields 
\begin{equation}
\begin{aligned}
(\det \hat\UU)^{-1/2}  & = \sqrt{\det\hat\epsr}\sqrt{\det{\hat\UU}_0^{-1}}
\approx \left[\sqrt{\epsr(\phi_m)}\right]^{{\cal N}_\rr}
\prod_{\kk\ne {\bf 0}}\sqrt{\frac {k^2}{4\pi\lb}}\\
&
= \prod_{\kk\ne {\bf 0}}\sqrt{\frac {k^2[\epsr(\phi_m)]^{{\cal N}_\rr/({\cal N}_\rr-1)}}{4\pi\lb}}
\approx
\prod_{\kk\ne {\bf 0}}\sqrt{\frac {k^2\epsr(\phi_m)}{4\pi\lb}}
\;
\end{aligned}
\end{equation}
for the $(\det \hat\UU)^{-1/2}$ factor in Eq.~(\ref{eq:after_drho}). To
arrive at this expression,
we made use of the fact that the total number of reciprocal space 
positions $\kk$ is ${\cal N}_\rr$ (same as the total number of coordinate space
positions $\rr$ when $\kk ={\bf 0}$ is included in the count), 
and that ${\cal N}_\rr \gg 1$. It follows that $\Zel$ in 
Eq.~(\ref{eq:after_drho}) may be written as
\begin{equation}
\Zel =\left\{\prod_{\kk\ne {\bf 0}}\sqrt{\frac {k^2\epsr(\phi_m)}{4\pi\lb}}
\right\}
\int\prod_{\alpha=1}^{\np} 
\prod_{\tau=1}^N d\R_{\alpha, \tau} 
\biggl\{\prod_{\rr} \int \frac {d\psi(\rr)}{\sqrt{2\pi}}\biggr\}
e^{-\Eng[\psi,\R]} \; ,
\label{eq:after_det}
\end{equation}
where $\Eng[\psi,\R]$ is given by Eq.~(\ref{eq:HpsiR}) with
$\epsr(\rr)$ given by Eq.~(\ref{eq:def_eps}):
\begin{equation}
\begin{aligned}
\Eng[\psi,\R] = 
& \frac{\epsw}{8\pi\lb}\int d\rr \left[ \nabla \psi(\rr) \right]^2
                + \frac{\deps}{8\pi\lb}\int d\rr \;
\phi_m(\rr)\left[ \nabla \psi(\rr) \right]^2  \\
                & + \frac{3}{2l^2} \sum_{\alpha=1}^\np
                                \sum_{\tau=1}^{N-1}\left( \R_{\alpha,\tau+1}-\R_{\alpha,\tau} \right)^2
                    + i\sum_{\alpha=1}^\np\sum_{\tau=1}^N \sigma_\tau\psi(\R_{\alpha,\tau}) \\
= & \frac{\epsw}{8\pi\lb}\int d\rr \left[ \nabla \psi(\rr) \right]^2
        + \sum_{\alpha=1}^\np \Bigg\{
                \frac{3}{2l^2} \sum_{\tau=1}^{N-1}\left( \R_{\alpha,\tau+1}-\R_{\alpha,\tau} \right)^2  \\
& \qquad\qquad\qquad\qquad\qquad\qquad\quad
                     + \sum_{\tau=1}^N\left[
                        i \sigma_\tau\psi(\R_{\alpha,\tau})
                + \frac{\deps l^3}{8\pi\lb} \left[ \nabla \psi(\R_{\alpha,\tau}) \right]^2
                     \right]
        \Bigg\} \; ,
\end{aligned}
        \label{eq:Ham_nabla_psi}
\end{equation}
where Eq.~(\ref{eq:phi_rr}) for $\phi_m(\rr)$
has been applied to yield the last equality.
Utilizing the Fourier transformation 
$\psi_\kk = (\Omega/{\cal N}_\rr)\sum_\rr \psi(\rr)\exp(-i\kk\cdot\rr)$ of
the conjugate field $\psi(\rr)$ [which may then be expressed as
the inverse transformation of $\psi_\kk$, i.e., 
{\hbox{$\psi(\rr)=(1/\Omega)\sum_\kk \psi_\kk \exp(i\kk\cdot\rr)$]}} and
the $\sum_\rr \leftrightarrow ({\cal N}_\rr/\Omega) \int d\rr$
correspondence in Eq.~(\ref{eq:corr1}),
the first term in the above Eq.~(\ref{eq:Ham_nabla_psi}) can
be rewritten as
\begin{equation}
\begin{aligned}
\frac{\epsw}{8\pi\lb}\int d\rr \left[ \nabla \psi(\rr) \right]^2  
 & \rightarrow \frac{\epsw}{8\pi\lb} \left(\frac {\Omega}{{\cal N}_\rr}\right)
\sum_\rr \left[ \left(
\frac {1}{\Omega}\sum_\kk \psi_\kk \nabla e^{-i\kk\cdot\rr}\right)\cdot
\left(
\frac {1}{\Omega}\sum_{\kk'} \psi_{\kk'} \nabla e^{-i{\kk'}\cdot\rr}
\right) \right]
\\
& =
 - \frac{\epsw\Omega}{8\pi\lb} 
\frac {1}{\Omega^2}\sum_\kk \sum_{\kk'} 
\psi_\kk (\kk\cdot\kk') \psi_{\kk'} \delta_{\kk+{\kk'}} 
\\
& = \frac{1}{2\Omega} \sum_{\kk}\frac{\epsw k^2}{4\pi\lb}\psi_\kk\psi_{-\kk}
= \frac{1}{2\Omega} \sum_{\kk\ne{\bf 0}}
\frac{\epsw k^2}{4\pi\lb}\psi_\kk\psi_{-\kk}
\; , 
\end{aligned}
\label{eq:psi-k}
\end{equation}
where the last equality follows because the $\kk={\bf 0}$ term vanishes
by virtue of the $k^2$ factor.
The remaining terms of $\Eng[\psi,\R]$ in Eq.~(\ref{eq:Ham_nabla_psi})
can be rewritten as the summation of contributions from $\np$ 
independent polymers, as follows. Consider the partition function
\begin{equation}
\Qp[\psi] = \int \DD[\R] e^{-\Hp[\psi,\R]}
\label{eq:Qp}
\end{equation}
for a single polymer, where
$\DD[\R] = \prod_{\tau=1}^N d \R_\tau$, and 
\begin{equation}
\begin{aligned}
\Hp[\psi,\R] \equiv & \frac{3}{2l^2} \sum_{\tau=1}^{N-1}\left( \R_{\tau+1}-\R_{\tau} \right)^2  \\
        & + \sum_{\tau=1}^N\left[\frac{i}{\Omega} \sum_{\kk}
                         \sigma_\tau\psi_\kk e^{-i\kk\cdot\R_\tau}
                        - \frac{\deps l^3}{8\pi\lb} \frac{1}{\Omega^2}
\sum_\kk \sum_{\kk'}
                                (\kk\cdot{\kk'}) \psi_{\kk}\psi_{\kk'}
e^{-i(\kk+{\kk'})\cdot\R_\tau}
                     \right] 
\\
= &
\frac{3}{2l^2} \sum_{\tau=1}^{N-1}\left( \R_{\tau+1}-\R_{\tau} \right)^2  \\
& + \sum_{\tau=1}^N\left[\frac{i}{\Omega} \sum_{\kk\ne {\bf 0}}
                         \sigma_\tau\psi_\kk e^{-i\kk\cdot\R_\tau}
                        - \frac{\deps l^3}{8\pi\lb} \frac{1}{\Omega^2}
\sum_{\kk,{\kk'}\ne {\bf 0}}
                                (\kk\cdot{\kk'}) \psi_{\kk}\psi_{\kk'}
e^{-i(\kk+{\kk'})\cdot\R_\tau}
                     \right]
\; .
\end{aligned}
        \label{eq:Hp_psi_dpsi}
\end{equation}
Note that the label $\alpha$ in $\R_{\alpha,\tau}$ is dropped in
$\prod_{\tau=1}^N d \R_\tau$
and Eq.~(\ref{eq:Hp_psi_dpsi}) because the 
pertinent integration is only over the monomer coordinates of a single 
polymer chain. 
The $\kk, {\kk'} = {\bf 0}$ terms can be excluded in the summations
of the last line of Eq.~(\ref{eq:Hp_psi_dpsi})
because in the first summation
$\sum_{\tau=1}^N \sigma_\tau =0$ for the overall neutral polyampholytes
considered here and the $(\kk\cdot{\kk'})$ factor in the second
summation means that the  $\kk, {\kk'} = {\bf 0}$ terms are identically zero.

Utilizing the definition of $\psi(\rr)$ to $\psi_\kk$ Fourier 
transformation stated after Eq.~(\ref{eq:Ham_nabla_psi}), it can readily be 
verified that $\Hp[\psi,\R]$ is precisely the $\kk$-space version of the 
quantity enclosed in curly brackets on the right hand side of 
Eq.~(\ref{eq:Ham_nabla_psi}).
Upon changing the functional integration variables $\psi(\rr)$ in
Eq.~(\ref{eq:after_det})
to $\psi_\kk$ and including the $\kk$-independent
functional Jacobian $|\delta \psi(\rr)/\delta \psi_\kk|$ 
(which have no effect on the configurational distribution of the system),
\begin{equation}
\biggl\{\prod_{\rr} \int \frac {d\psi(\rr)}{\sqrt{2\pi}}\biggr\}
\rightarrow
\left\{\prod_{\kk} \int 
\sqrt{\frac {{\cal N}_\rr}{2\pi\Omega^2}} \; 
d\psi_\kk\right\} \; 
\label{eq:change_variables}
\end{equation}
formally, and thus
Eq.~(\ref{eq:after_det}) can now be recast in the equivalent form
\begin{equation}
\begin{aligned}
\Zel = & \left\{\prod_{\kk\ne {\bf 0}}\sqrt{\frac {k^2\epsr(\phi_m)}{4\pi\lb}}
\right\}
\biggl\{\prod_{\kk} \int 
\sqrt{\frac {{\cal N}_\rr}{2\pi\Omega^2}} \; 
d\psi_\kk\biggr\} 
\exp\left[-\frac{1}{2\Omega} \sum_{\kk\ne{\bf 0}}
\frac{\epsw k^2}{4\pi\lb}\psi_\kk\psi_{-\kk}\right]
\\
& \times
\int\prod_{\alpha=1}^{\np}
\left\{\prod_{\tau=1}^N d\R_\tau
\exp \left(-\Hp[\psi,\R] \right)
\right\}
\; ,
\label{eq:after_change_variable}
\end{aligned}
\end{equation}
where we have made use of the fact that in the above expression, the 
first exponential factor [from Eq.~(\ref{eq:psi-k})] is independent of 
$\R_{\alpha,\tau}$, and the quantity enclosed in the last
set of curly brakets [from Eq.~(\ref{eq:Hp_psi_dpsi})] 
is identical for all $\np$ values of $\alpha$, thus the
entire last line of Eq.~(\ref{eq:after_change_variable})
is equal to $\np\ln\Qp[\psi]$ in accordance with Eq.~(\ref{eq:Qp}).
Because, as argued above, there is no $\kk = {\bf 0}$ contribution to 
$\Hp[\psi,\R]$, the $\prod_\kk 
({\cal N}_\rr/2\pi\Omega^2)^{1/2}
\int d\psi_\kk$ functional integral
in Eq.~(\ref{eq:after_change_variable}) may be restricted to 
$\prod_{\kk\ne {\bf 0}} 
({\cal N}_\rr/2\pi\Omega^2)^{1/2}
\int d\psi_\kk$ with no impact on configurational
distribution. Therefore, $\Zel$ takes the simplified form:
\begin{equation}
\Zel = 
\left\{
\prod_{\kk\ne {\bf 0}}  
\int 
\sqrt{\frac {{\cal N}_\rr}{2\pi\Omega^2}} \; 
d\psi_\kk 
\sqrt{\frac{\epsr(\phi_m)k^2}{4\pi\lb}}\right\} e^{-\Eng[\psi_\kk]}
\; ,
        \label{eq:Zel_psik}
\end{equation}
where
\begin{equation}
\Eng[\psi_\kk] = \frac{1}{2\Omega} 
\sum_{\kk\ne{\bf 0}}\frac{\epsw k^2}{4\pi\lb}\psi_\kk\psi_{-\kk} 
- \np\ln\Qp[\psi]
\: .
        \label{eq:Eng_Qp}
\end{equation}
%
%

We are now in a position to apply RPA by expanding $\ln\Qp$ around 
$\psi_\kk=0$ up to second order in $\psi_\kk$,\cite{kings2020} namely
\begin{equation}
\ln\Qp[\psi] \approx \ln\Qp[\psi=0] + \sum_{\kk}\left( \frac{\delta\ln\Qp}{\delta\psi_\kk}\right)_{\psi=0} \psi_\kk
        + \frac{1}{2}
                \sum_{\kk,\kk'}\left(\frac{\delta^2\ln\Qp}{\delta\psi_\kk\delta\psi_{\kk'}}\right)_{\psi=0}\psi_\kk\psi_{\kk'}
\; ,
        \label{eq:lnQp_quad}
\end{equation}
wherein the zeroth order term (first term on the right hand side) is a 
constant that plays no role in determining configurational distribution.
The first order term
\begin{equation}
\begin{aligned}
\left( \frac{\delta\ln\Qp}{\delta\psi_\kk}\right)_{\psi=0} =
        & \frac{1}{\Qp[\psi=0]}\left.\frac{\delta \Qp}{\delta \psi_\kk}\right|_{\psi=0} \\
= & \sum_{\tau=1}^N \avg{ -\frac{i}{\Omega}
                         \sigma_\tau e^{-i\kk\cdot\R_\tau}
                        + 2 \times \frac{\deps l^3}{8\pi\lb} \frac{1}{\Omega^2} 
\kk \cdot \sum_{{\kk'}\ne {\bf 0}} \kk'
                                \psi_{\kk'}e^{-i(\kk + \kk')\cdot\R_\tau}
                               }_{\psi=0} \\
= & -\frac{i}{\Omega}\sum_{\tau=1}^N\sigma_\tau\avg{ e^{-i\kk\cdot\R_\tau} }_{\psi=0} \\
= & 0
\end{aligned}
\label{eq:first_zero}
\end{equation}
as well. Here, the average $\langle ... \rangle_{\psi=0}$ is over
monomer coordinates $[\R]$ and evaluated at $\psi_\kk = 0$, the
third equality follows because the second term in
the second line of the above equation contains a factor of $\psi$ that
is set to zero, and the last equality is a consequence of the
overall neutrality of the polyampholytes in the system we considered
($\sum_{\tau = 1}^N\sigma_\tau = 0$).
The second order term in the above Eq.~(\ref{eq:lnQp_quad}) is given by 
\begin{equation}
\begin{aligned}
\left(\frac{\delta^2\ln\Qp}{\delta\psi_\kk\delta\psi_{\kk'}}\right)_{\psi=0}
        = & 
\frac{1}{\Qp[\psi=0]}\left.\frac{\delta^2 \Qp}{\delta \psi_\kk\delta \psi_\kk'}\right|_{\psi=0}
                - \frac{1}{\Qp[\psi=0]}\left.\frac{\delta \Qp}{\delta \psi_\kk}\right|_{\psi=0} \times
                  \frac{1}{\Qp[\psi=0]}\left.\frac{\delta \Qp}{\delta \psi_\kk'}\right|_{\psi=0} \\
= &
\frac{1}{\Qp[\psi=0]}\left.\frac{\delta^2 \Qp}{\delta \psi_\kk\delta \psi_\kk'}\right|_{\psi=0}
\\
= & 
\frac{1}{\Omega^2} \frac{\deps l^3}{4\pi\lb} \kk \cdot \kk'
           \sum_{\tau=1}^N\avg{ e^{-i(\kk + \kk')\cdot\R_\tau}}_{\psi=0} 
-\frac{1}{\Omega^2}\sum_{\tau,\mu=1}^N\sigma_\tau\sigma_\mu
                        \avg{ e^{-i (\kk\cdot\R_\tau +\kk'\R_\mu)} }_{\psi=0}
\; ,
\end{aligned}
        \label{eq:ddQp}
\end{equation}
where the second equality follows from Eq.~(\ref{eq:first_zero}).
The two $\R$-averages over Gaussian chain configurations in the 
above Eq.~(\ref{eq:ddQp}) may be evaluated as follows.
For $\avg{ e^{-i(\kk + \kk')\cdot\R_\tau}}_{\psi=0}$, only a single
monomer coordinate variable $\R_\tau$ is involved and thus it is
uncontrained and $\R$-averaging entails only a single integration of 
$\R_\tau$ over the entire system volume $\Omega$.
The correspondence $\int d\R_\tau \leftrightarrow (\Omega/{\cal N}_\rr)
\sum_{\R_\tau}$ yields $\avg{ e^{-i(\kk + \kk')\cdot\R_\tau}}_{\psi=0}$
$= \delta_{\kk,-{\kk'}}$. Next, to compute
$\avg{ e^{-i (\kk\cdot\R_\tau +\kk'\R_\mu)} }_{\psi=0}$,
we rewrite it as
$\avg{ e^{-i\kk\cdot(\R_\tau-\R_\mu)} e^{-i(\kk +\kk')\cdot\R_\mu)} }_{\psi=0}$,
which indicates that the $\R$-averaging involves integrating over two
monomer coordinates, one is unconstrained and the other is constrained by
the Gaussian chain statistics for two points separated by a contour length
$l|\tau-\mu|$. Without loss of generality, we select $\R_\mu$ to be
the unconstrained coordinates. As for the first average, summing over 
$\R_\mu$ using the 
$\int d\R_\mu \leftrightarrow (\Omega/{\cal N}_\rr) \sum_{\R_\mu}$
correspondence yields the Kronecker $\delta_{\kk,-{\kk'}}$.
In accordance with the Gaussian statistics governed by the
$3/2l^2 \sum_{\tau=1}^{N-1} (\R_{\tau+1}-\R_\tau)^2$ term of
$\Hp[\psi,\R]$ in Eq.~(\ref{eq:Hp_psi_dpsi}), $\R_\tau-\R_\mu$ is
weighted by $\exp(-3|\R_\tau-\R_\mu|^2/2l^2|\tau-\mu|)$, and therefore
the $\R$-averaging 
of $e^{-i\kk\cdot(\R_\tau-\R_\mu)}$ yields $\exp(-k^2l^2|\tau-\mu|/6)$.
These considerations allow us to arrive at the expression
\begin{equation}
\left(\frac{\delta^2\ln\Qp}{\delta\psi_\kk\delta\psi_{\kk'}}\right)_{\psi=0}
=  -\frac{\delta_{\kk,-\kk'}}{\Omega^2}
                \left[
                  \frac{\deps N l^3 k^2}{4\pi\lb}
                + \bra{\sigma} \GM(kl)  \ket{\sigma}
                \right]
\; 
\end{equation}
for Eq.~(\ref{eq:ddQp}),
where $[\GM(kl)]_{\tau\mu}=\exp[-(kl)^2|\tau-\mu|/6]$ as defined above.
Therefore, according to Eqs.~(\ref{eq:lnQp_quad}) and
(\ref{eq:first_zero}), the $\np \ln\Qp[\psi]$ term in 
Eq.~(\ref{eq:Eng_Qp}) is given by
\begin{equation}
\begin{aligned}
\np
\ln\Qp[\psi] & \approx
\frac{1}{2} \np
\sum_{\kk,\kk'}\left(
\frac{\delta^2\ln\Qp}{\delta\psi_\kk\delta\psi_{\kk'}}\right)_{\psi=0}
\psi_\kk\psi_{\kk'} \\
& = - \frac {\np}{2\Omega^2} \sum_{\kk,\kk'}\delta_{\kk,-\kk'}
                \left[
                  \frac{\deps N l^3 k^2}{4\pi\lb}
                + \bra{\sigma} \GM(kl)  \ket{\sigma}
                \right] \psi_\kk\psi_{\kk'}
\\
& = 
-\frac {1}{2\Omega}
\sum_{\kk\ne 0} 
\left[ \frac{\deps \phi_m k^2}{4\pi\lb}
                + \frac {\phi_m}{Nl^3} \bra{\sigma} \GM(kl)  \ket{\sigma}
                \right] \psi_\kk\psi_{-\kk}
\; ,
\end{aligned}
\label{eq:nplnnp}
\end{equation}
where we have used the definition of polymer volume fraction
$\phi_m = l^3 \np N/\Omega$, and the fact that the $\kk = {\bf 0}$
terms vanishes: the first term because of the $k^2$ factor and the
second term because of the overall neutrality of the polyampholytes, i.e.,
$\sum_\tau \sigma_\tau = 0$, and $[\GM(0)]_{\tau\mu}=1$. 
Combining this result with Eq.~(\ref{eq:Eng_Qp}), we arrive at
\begin{equation}
\begin{aligned}
\Eng[\psi_\kk] & \approx
\frac{1}{2\Omega}
\sum_{\kk\ne{\bf 0}} \left[
\frac{(\epsw + \deps \phi_m) k^2}{4\pi\lb}
                + \frac {\phi_m}{Nl^3} \bra{\sigma} \GM(kl)  \ket{\sigma}
                \right] \psi_\kk\psi_{-\kk} \\
& = \frac{1}{2\Omega}
\sum_{\kk\ne{\bf 0}} \left[
\frac{\epsr(\phi_m) k^2}{4\pi\lb}
                + \frac {\phi_m}{Nl^3} \bra{\sigma} \GM(kl)  \ket{\sigma}
                \right] \psi_\kk\psi_{-\kk} 
\; ,
        \label{eq:Hpsi}
\end{aligned}
\end{equation}
where we have made use of the above definition of $\epsr(\phi_m)$ which is 
linear in $\phi_m$. We may now evaluate 
$\Zel$ by performing the functional integral 
$\prod_{\kk\ne {\bf 0}} \int d\psi_\kk$ in Eq.~(\ref{eq:Zel_psik}). Because the
$\psi_\kk$s are Fourier transformations of the real-valued field 
$\psi(\rr)$, $\psi_\kk^*=\psi_{-\kk}$ and 
$\prod_{\kk\ne {\bf 0}} \int d\psi_\kk$ $=$
$\prod_{\kk > {\bf 0}} \int d\psi_\kk \int d\psi_\kk^*$, 
where the $\kk > {\bf 0}$ notation 
means that the product or summation excludes
the origin and is over $\kk=(k_1,k_2,k_3)$ but not $-\kk=(-k_1,-k_2,-k_3)$.
This can be effectuated by first excluding $(k_1,k_2,k_3)=(0,0,0)$ 
and then restricting the product or sum to $k_1\ge 0$ (or to $k_2\ge 0$ or to 
$k_3 \ge 0$). 
Expressing
$\psi_\kk$ in terms of its real part $\psi_\kk^{\rm R}$ and imaginary
part $\psi_\kk^{\rm I}$, 
i.e.,
$\psi_\kk = \psi_\kk^{\rm R} + i\psi_\kk^{\rm I}$ and 
$\psi_\kk^* = \psi_\kk^{\rm R} - i\psi_\kk^{\rm I}$ 
where $\psi_\kk^{\rm R}$ and $\psi_\kk^{\rm I}$ are real numbers,
one obtains
$\prod_{\kk > {\bf 0}} \int d\psi_\kk \int d\psi_\kk^*$
$=$
$\prod_{\kk > {\bf 0}} 2
\int_{-\infty}^\infty 
d\psi_\kk^{\rm R} \int_{-\infty}^\infty d\psi_\kk^{\rm I}$.
Since $\psi_\kk\psi_{-\kk}$ $=(\psi_\kk^{\rm R})^2 + (\psi_\kk^{\rm I})^2$,
\begin{equation}
\begin{aligned}
\Zel = & \left\{
\prod_{\kk > {\bf 0}}
\left(\frac {{\cal N}_\rr}{\pi\Omega^2}\right)
\left[\frac{\epsr(\phi_m)k^2}{4\pi\lb}\right] \; 
\int_{-\infty}^\infty 
d\psi_\kk^{\rm R} \int_{-\infty}^\infty d\psi_\kk^{\rm I}
\right\}
\\
& \quad\quad \times
\exp \left\{ \frac{1}{\Omega}\sum_{\kk > {\bf 0}}
\left[
\frac{\epsr(\phi_m) k^2}{4\pi\lb}
                + \frac {\phi_m}{Nl^3} \bra{\sigma} \GM(kl) \ket{\sigma}
                \right]
\biggl[(\psi_\kk^{\rm R})^2 + (\psi_\kk^{\rm I})^2\biggr]\right\}\\
= &
\prod_{\kk > {\bf 0}}
\left(\frac {{\cal N}_\rr}{\pi\Omega^2}\right)
\left[\frac{\epsr(\phi_m)k^2}{4\pi\lb}\right] \times
\pi\Omega
\left[
\frac{\epsr(\phi_m) k^2}{4\pi\lb}
                + \frac {\phi_m}{Nl^3} \bra{\sigma} \GM(kl) \ket{\sigma}
                \right]^{-1}
\\
= &
\prod_{\kk \ne {\bf 0}}
\sqrt{\frac {{\cal N}_\rr}{\Omega}}
\left[ 1 + \frac {4\pi\lb}{\epsr(\phi_m)k^2}
\frac {\phi_m}{Nl^3} \bra{\sigma} \GM(kl) \ket{\sigma}
                \right]^{-1/2} 
\; .
\end{aligned}
\end{equation}
Hence, up to an additive constant 
$\propto{\cal N}_\rr\ln ({\cal N}_\rr/\Omega)$ that does not affect
configurational distribution, the electrostatic contribution to
the free energy in Eq.~(\ref{eq:free-energy}) is equal to
\begin{equation}
\begin{aligned}
f_{\rm el} \equiv -\frac{l^3}{\Omega}\ln \Zel
& =
-\frac {l^3}{\Omega}\sum_{\kk\ne {\bf 0}}
\ln 
\left[ 1 + \frac {4\pi\lb}{\epsr(\phi_m)k^2}
\frac {\phi_m}{Nl^3} \bra{\sigma} \GM(kl) \ket{\sigma}
                \right]^{-1/2} 
\\
& \rightarrow\frac{l^3}{2}\int \frac{d^3 k }{(2\pi)^3} \; 
\ln \left[ 1 + \frac{4\pi\lb}{\epsr(\phi_m)k^2}\frac{\phi_m}{N l^3} 
\bra{\sigma} \GM(kl) \ket{\sigma}  \right] \; ,
        \label{eq:fel_final}
\end{aligned}
\end{equation}
where we have applied the correspondence
\begin{equation}
\frac{1}{\Omega}\sum_\kk \rightarrow \int \frac{d^3 k }{(2\pi)^3}
\end{equation}
and noted that the $k\to 0$ contribution vanishes
inside the integral in Eq.~(\ref{eq:fel_final}) 
because $d^3k \propto k^2 dk$ and thus 
$\sum_{\kk\ne {\bf 0}}$ may be approximated by
$\Omega \int d^3k/(2\pi)^3$ for this quantity.
The last expression in Eq.~(\ref{eq:fel_final}) is formally identical
to the one we obtained previously by heuristically replacing the
position- and $\phi_m$-independent $\epsr$ in simple RPA theory
with $\epsr(\phi_m)$ [Eq.~(\ref{eq:fel_dimless_dic})]. This can be readily 
verified by setting
$b=a=l$, hence $\eta=1$ in Eqs.~(\ref{eq:fel_dimless_dic}) and
(\ref{eq:G_factors}), and noting that $(1/2)d^3k/(2\pi)^3=k^2dk/4\pi^2$,
in which case the last line of Eq.~(\ref{eq:fel_final})
is seen to be equal to Eq.~(\ref{eq:fel_dimless_dic}) with 
the ${\cal G}_1(\kr)$ term [Eq.~(\ref{eq:G_factors}a)] present but 
the ${\cal G}_2(\kr)$ term [Eq.~(\ref{eq:G_factors}b)] 
omitted (no subtraction of self interaction) as well as $\kr^2(1+\kr^2)$ 
$\rightarrow$ $\kr^2$ (no short-range cutoff for Coulomb interaction).

In other words, the heuristic RPA formulas for $\epsr \to \epsr(\phi_m)$
in Eqs.~(\ref{eq:fel_dimless_dic}) and (\ref{eq:G_factors}) can be
rigorously established in the context of RPA approximation provided that
$\epsr$ is a linear function of $\phi_m$. Indeed, if $\epsr$ was a more 
complicated function of $\phi_m$, the last term in Eq.~(\ref{eq:Ham_nabla_psi})
would have individual interaction terms, such as 
$\delta (\R_{\alpha,\tau}-\R_{\beta,\mu})$, etc.,
that involve different polymer chains, and that would necessitate
an additional summation $\sum_\alpha$ over polymer chains instead of
a single $\sum_\tau$ over monomers on a single chain. In that case, the 
subsequent simplification in terms of the single-chain partition function
$\Qp$ [Eq.~(\ref{eq:after_change_variable})] 
and thus the RPA expansion of $\ln\Qp$ [Eq.~(\ref{eq:lnQp_quad})]
cannot proceed in the manner described above. Therefore, it remains
unclear whether Eq.~(\ref{eq:fel_final}) 
holds in general for $\epsr(\phi_m)$ that is not linear in $\phi_m$.

In our previous applications, we considered a Coulomb potential with
a physical short-range cutoff by the modification
\begin{equation}
\UU(\rr,\rr')=\frac {\lb}{\epsr |\rr-\rr'|}
\rightarrow
\UU(\rr,\rr')=\frac {\lb}{\epsr |\rr-\rr'|}\biggl (1-e^{-|\rr-\rr'|/l}\biggr)
\label{eq:short_range}
\end{equation}
[cf. Eq.~(6) of Ref.~\citen{linPRL}; Eq.~(34) of Ref.~\citen{linJML}],
which for constant, position-independent $\epsr$ results in a 
$f_{\rm el}$ with $1/k^2$ replaced by $1/[k^2(1+k^2)]$. In the context
of a general position-dependent $\epsr$, this feature can in principle
be accounted for by introducing an $\epsr(|\rr-\rr'|)$, but the necessary
formalism has not been developed. In the present work, we incorporate
this feature by simply replacing the $1/k^2$ factor by $1/[k^2(1+k^2)]$ in 
Eq.~(\ref{eq:fel_final}) so as to capture this physical property as
much as possible and place our present results on an essentially equal
footing with our earlier results for position-independent $\epsr$. 
Mathematically, this procedure may be viewed as a regularization for 
``ultraviolet'' large-$k$ (i.e., small-$|\rr-\rr'|$) divergence. As such,
it does serve to impart a physical short-spatial-range 
cutoff, though it may not correspond exactly to any particularly modified
form of $f_{\rm el}$ in Eq.~(\ref{eq:short_range}) 
that is applicable to a general position-dependent $\epsr(\rr)$.

Taking all of the above into consideration, we use the general formula
in Eqs.~(\ref{eq:fel_dimless_dic}) and (\ref{eq:G_factors}) above
(which allows for $a\ne b=l$ and thus $\eta=(b/a)^3\ne 1$ and $r_m\ne 1$)
for comparing RPA theory against explicit-chain simulation,
with the understanding that $\epsr$ must be a linear function of 
polymer volume fraction
$\phi=r_m\phi_m$. Following previous practice,\cite{linJML,linPRL} the
electrostatic self-interaction term ${\cal G}_2(\kr)$ 
$=$ $4\pi\lb\phi_m/[k^2(1+k^2b^2)\epsr(\phi_m)Nb^3]\sum_{\tau=1}^N 
|\sigma_\tau|$
is subtracted in Eq.~(\ref{eq:fel_dimless_dic}).
In the context of a position-dependent $\epsr(\rr)$, however, we recognize
that this term can be physically significant for capturing
the polyampholyte chains' varying preference for different dielectric 
environments.\cite{warshel2001}
Hence we consider also an electrostatic free energy 
\begin{equation}
f_{\rm el}^{\rm [self]} \equiv \int \frac{d\kr \kr^2}{4\pi^2\eta} \ln\Bigl[
1 + \eta{\cal G}_1(\kr)\Bigr] 
= 
\; a^3 
\int \frac{dk k^2}{4\pi^2} \ln\biggl[
1 + \frac{b^3}{a^3}{\cal G}_1(kb)\biggr]
        \label{eq:fel_dimless_dic_self}
\end{equation}
that includes (does not substract) electrostatic self-interaction, and use 
both Eq.~(\ref{eq:fel_dimless_dic}) and Eq.~(\ref{eq:fel_dimless_dic_self})
in our comparison of analytical theory with chain simulation.


$\null$

\noindent 
{\bf U{\footnotesize{NIT}}}
{\bf C{\footnotesize{ONVERSION}}}
{\bf {\footnotesize{FOR}}}
{\bf C{\footnotesize{OMPARISON}}}
{\bf {\footnotesize{WITH}}}
{\bf E{\footnotesize{XPLICIT}}-C{\footnotesize{HAIN}}}
{\bf S{\footnotesize{IMULATIONS}}}

The theory-predicted phase diagrams (coexistence curves) 
in Fig.~S7 of the Supporting Information
for position- and IDR concentration-independent $\epsr$
are computed numerically using the RPA+FH model described in 
Ref.~\citen{linJML}. Specifically, translational and mixing entropy
is given by Eqs.~(13) and (14), the RPA formula for $f_{\rm el}$ is 
provided by Eqs.~(39) and (40), and the augmented FH term is the
one in Eq.~(61) of this reference. Values of the parameters in these formulas
are adapted to the present application, as follows: 

\begin{itemize}
\item
$a$: Unit length of the model. We set the unit volume, $a^3$, to be that
of the volume occupied by a water molecule in pure water, i.e., 
$\phi_{\rm w}^{\rm pure} 
= \rho_{\rm w}^{\rm pure} \times a^3 =1$, where the number density of
pure water $\rho_{\rm w}^{\rm pure} =10^6$ g m$^{-3}N_{\rm A}/18.01528$ g 
where $10^6$ g m$^{-3}$ is density of water, 
$N_{\rm A}=6.02214086\times10^{23}$ is Avogadro's constant
and $18.01528$ g is molar mass of water. Thus,
$a=(1/\rho_{\rm w}^{\rm pure})^{1/3} = 3.104$ \AA $=3.104\times 10^{-10}$m.
\item
$b$: The C$_\alpha$--C$_\alpha$ virtual bond length of polypeptides 
$b=l=3.8$ \AA $=3.8\times10^{-10}$m.
\item
$\eta$ [in Eq.~(39) of Ref.~\citen{linJML}]: From the above values for $a$ and
$b$, $\eta =(b/a)^3 = (3.8/3.104)^3 = 1.835$.
\item
$r_m$ (monomer size factor in Eq.~(14) of Ref.~\citen{linJML}): The $r_m$ 
ratio between the size of one amino acid residue in Ddx4 IDR and the 
unit volume $a^3$ is obtained as follows. Because the density of pure protein
$=1,587$ mg ml$^{-1}$, number of amino acid residues (monomers) in Ddx4 
IDR is $N=$ 241, and the molar mass of Ddx4 IDR is $25,833$ 
(Ref.~\citen{jacob2017}),
the monomer (amino acid residue) number density of pure protein is given by
\begin{equation}
\rho_m^{\rm pure} = (1.587\times10^6) {\rm \ g\ m}^{-3} 
\times 241 \times N_{\rm A}/25,833 {\rm \ g} \; .
\end{equation}
Since the volume fraction $\phi$ of pure protein is unity by definition, i.e.,
$\phi=\rho_m \times r_m \times a^3$, it follows that
\begin{equation}
r_m = (a^3\rho_m^{\rm pure})^{-1} = \rho_{\rm w}^{\rm pure}/\rho_m^{\rm pure}
= \frac {25833}{18.0} \cdot \frac {1}{1.587} \cdot \frac {1}{241} 
=3.752
\; .
\end{equation}
\item $r_s$ and $r_c$ [size factors for salt and counterions, respectively,
in Eq.~(14) of Ref.~\citen{linJML}]: Both $r_s$ and $r_c$ are set to 1.
\end{itemize}

The conversion between the $\phi_m=\np N a^3/\Omega$ in analytical theory to 
Ddx4 concentration, [Ddx4], in units of mg/ml (mg ml$^{-1}$), is given by
\begin{equation}
\phi_m = \Bigl\{\text{[Ddx4(mg/ml)]} \times 1000 {\rm \ g/mg} \times 
236/\text{(Ddx4 molar mass in g)}\Bigr\} \times N_{\rm A} \times a^3
\; ,
\label{eq:phi_m_match}
\end{equation}
where $N=236$ is the chain length of the Ddx4 IDRs, 
(Ddx4 molar mass in g) of the four Ddx4 IDR sequences are 
25412.48, 25412.48, 24346.80, and 24740.48, respectively, for WT, CS, FtoA, 
and RtoK.\cite{jacob2017} It should be noted that
there is a slight mismatch in the lengths of Ddx4 IDRs (236 vs 241) because
a Ddx4$^{\rm N1}$ sequence with six amino acids added to its C-terminus
as a tag was used in experiments.\cite{Nott15,jacob2017}
Nonetheless, $N=236$ is adopted in Eq.~(\ref{eq:phi_m_match}) because 
the $N=236$ sequence published in Ref.~\citen{Nott15}
is used in our simulations. In the context of our approximate analytical theory
and coarse-grained chain model, the numerical difference 
between using $N=236$ and $N=241$ is not expected to be significant.

The mean-field Flory-Huggins (FH) $\chi$ parameters of non-electrostatic 
interactions for the four Ddx4 IDR sequences are obtained from averaging 
the KH potential energies $\epsilon_{ij}(r_0)$ 
($=E_{ij}(r_0)$ [KH] in Fig.~1a of main text)
for a given sequence (seq) over all $i,j$ pairs of sequence positions
except those entailing a charge-charge interaction [i.e., 
RR (Arg-Arg), RK (Arg-Lys), RD (Arg-Asp), RE (Arg-Glu), KK (Lys-Lys), 
KD (Lys-Asp), KE (Lys-Glu), DD (Asp-Asp), DE (Asp-Glu), EE (Glu-Glu); 
see main-text], yielding
$\langle E\rangle_{\rm KH,seq}$ $=-0.1047$, $-0.1047$, $-0.0689$, and 
$-0.0924$ kcal mol$^{-1}$,
respectively, for seq $=$ WT, CS, FtoA, and RtoK. These average 
sequence-dependent mean-field non-electroatic interaction energies 
$\langle E\rangle_{\rm KH,seq}$s are converted to the FH 
$\chi = \varepsilon_h/T^*$ in Eq.~(61) of Ref.~\citen{linJML} as follows:
\begin{enumerate}
\item Convert per-mole units to per-interaction units:
\begin{equation}
\begin{aligned}
& \langle E\rangle_{\rm KH,seq}[\text{(J/amino acid pair)}] \\
& \quad =
\Bigl\{\langle E\rangle_{\rm KH,seq}[\text{(kcal/mole of amino acid pairs)}]/N_{\rm A}\Bigr\} 
\times 1000 {\rm \ cal/kcal}  \times 4.18 {\rm \ J/cal}\; .
\end{aligned}
\end{equation}
\item Convert to the reduced variables used in analytical theory:
\begin{equation}
(z/2) \times \langle E\rangle_{\rm KH,seq}[\text{(J/amino acid pair)}]/(\kB T) 
= -\varepsilon_h/T^* \; ,
\end{equation}
where $T^*$ is the reduced temperature given by Eq.~(38) in Ref.~\citen{linJML}
(see below) and $z$ is an FH geometric 
factor representing the maximal number of 
monomers (amino acid residues) that are spatial nearest neighbors to a given 
monomer; e.g., $z=6$ for three-dimensional simple cubic lattices. We obtain 
$z/2=4.3$ by fitting our RPA+FH predictions to our explicit-chain simulation 
results.
\\
\item Convert absolute temperature $T$ in K to the reduced temperature $T^*$:
\begin{equation}
\frac{1}{T^*} = \frac{e^2}{4\pi\epsilon_0 \epsr \kB b}\frac{1}{T}
\; ,
\label{eq:reduced_temp}
\end{equation}
where the electronic charge $e = 1.6\times10^{-19}$ C, 
$\epsilon_0 = 8.854\times 10^{-12}$ C V$^{-1}$m$^{-1}$, 
$b = 3.8\times 10^{-10}$ m, and $\epsr=80$, $40$, or $20$ in accordance
with the corresponding simulations with position- and 
IDR concentration-independent relative permittivities.
Note that $T^*=\epsr T_0^*$ where $T_0^*$ is defined after 
Eq.~(\ref{eq:G_factors}) above and in Eq.~(67) of Ref.~\citen{linJML}.
\\
\item Convert $\langle E\rangle_{\rm KH,seq}$ to FH $\varepsilon_h$:
\\
Based on the above consideration,
\begin{equation}
\begin{aligned}
\varepsilon_h  
& = 
-T^*\left(\frac{z}{2}\right) 
\frac {\langle E\rangle_{\rm KH,seq}[\text{(J/amino acid pair)}]}{\kB T}\\
& = - \left(\frac {4\pi\epsilon_0\epsr b}{e^2}\right)
\biggl(\frac{z}{2}\biggr)
\times
\Bigl\{\langle E\rangle_{\rm KH,seq}[\text{(kcal/mole of amino acid pairs)}]/N_{\rm A}\Bigr\}\\
& \quad\quad 
\times 1000 {\rm \ cal/kcal}  \times 4.18 {\rm \ J/cal}\\
& =
-\frac {4\pi\times(8.854\times 10^{-12})\times( 3.8\times 10^{-10})}{(1.6\times10^{-19})^2}
\frac{4.3\times1000\times4.18}{(6.02214086\times10^{23})}\\
& \quad\quad \times
\epsr \langle E\rangle_{\rm KH,seq}[\text{(kcal/mole of amino acid pairs)}]
\\
& = - 0.04929 \times \epsr \times
\langle E\rangle_{\rm KH,seq}[\text{(kcal/mole of amino acid pairs)}]
\; .
\end{aligned}
\end{equation}
\end{enumerate}
Accordingly, the $\varepsilon_h$ values for
WT, CS, FtoA, and RtoK Ddx4 IDRs are, respectively, 
$\varepsilon_h=$ 
$0.413$, $0.413$, $0.272$, and $0.364$ when $\epsr = 80$;
$\varepsilon_h=$ 
$0.206$, $0.206$, $0.136$, and $0.182$ when $\epsr = 40$; and
$\varepsilon_h=$ 
$0.103$, $0.103$, $0.068$, and $0.091$ when $\epsr = 20$.

Note that $\varepsilon_h$ decreases with decreasing $\epsr$ because the
reduced temperature $T^*$ in Eq.~(\ref{eq:reduced_temp}) 
is proportional to $\epsr$. In this formulation using $T^*$, the result of 
decreasing $\epsr$ is a reduction in the strength of favorable FH 
interactions relative to that of the electrostatic interactions, which 
is equivalent to the physical situation (with temperature measured
in K) of enhanced electrostatic interactions under a reduced $\epsr$ while 
keeping the non-electrostatic interactions unchanged.

\vfill\eject

\noindent
{\Huge\bf SI Figures}\\

$\null$\\

\begin{center}
   \includegraphics[width=0.9\columnwidth]{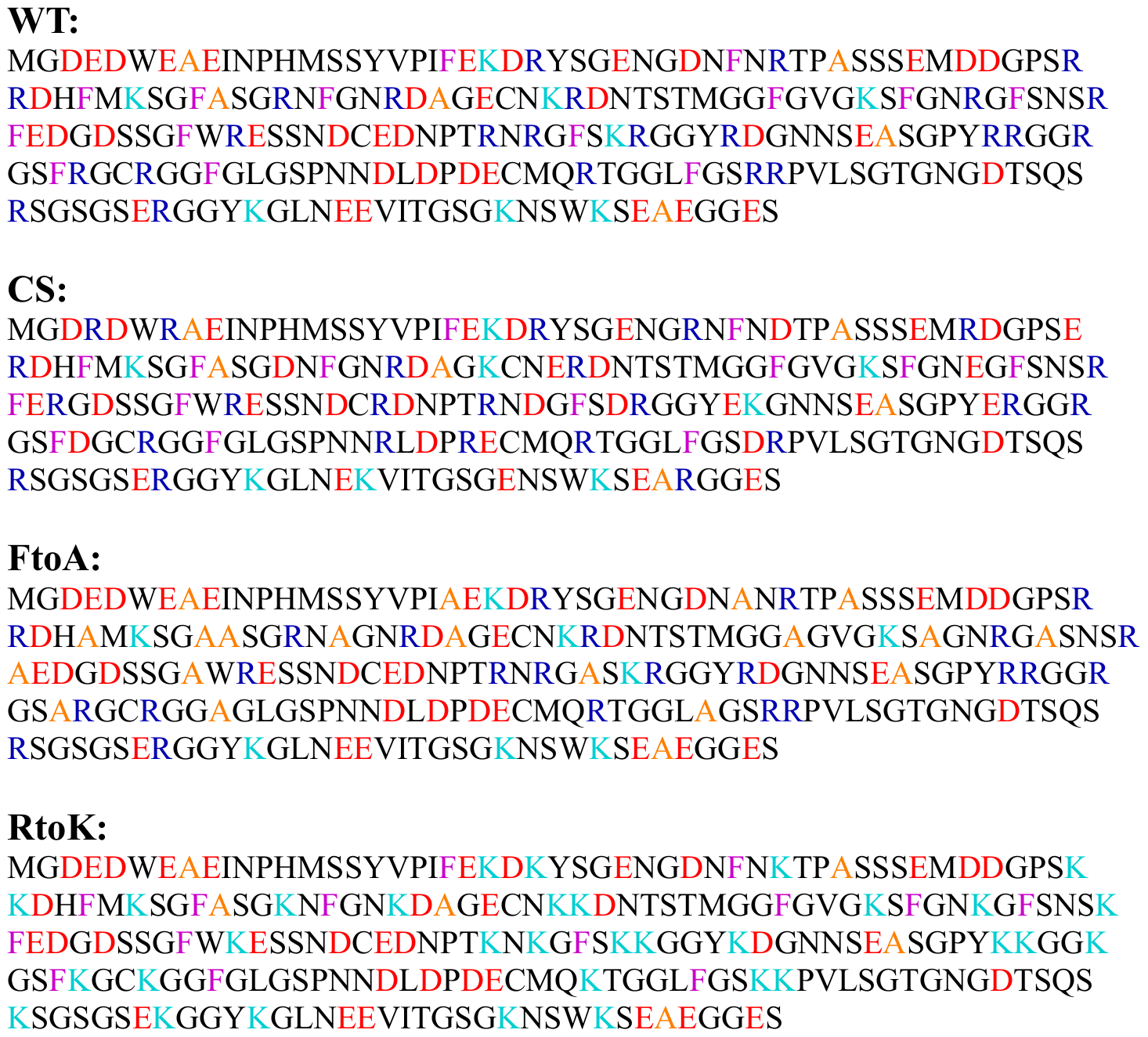}
\end{center}
\noindent
{\bf Fig.~S1:} 
The amino acid sequences (residues given by one-letter code) of
the 236-residue Ddx4 IDR (wildtype, WT) and its charge scrambled
(CS) variant (introduced by Nott et al.\cite{Nott15}), 
phenylalanine-to-alanine variant (FtoA) (corresponds to
the 14FtoA in Brady et al.\cite{jacob2017} and Vernon et al.\cite{robert})
and arginine-to-lysine (RtoK) variant\cite{robert} 
considered in the present study.
Residues of amino acids A, D, E, F, K, and R are shown in the same colors
as those in Fig.~3a of the main text.
Other residues, including Y which is considered in subsequent simulations,
are shown in black.
\\

\vfill\eject

\begin{center}
   \includegraphics[width=0.9\columnwidth]{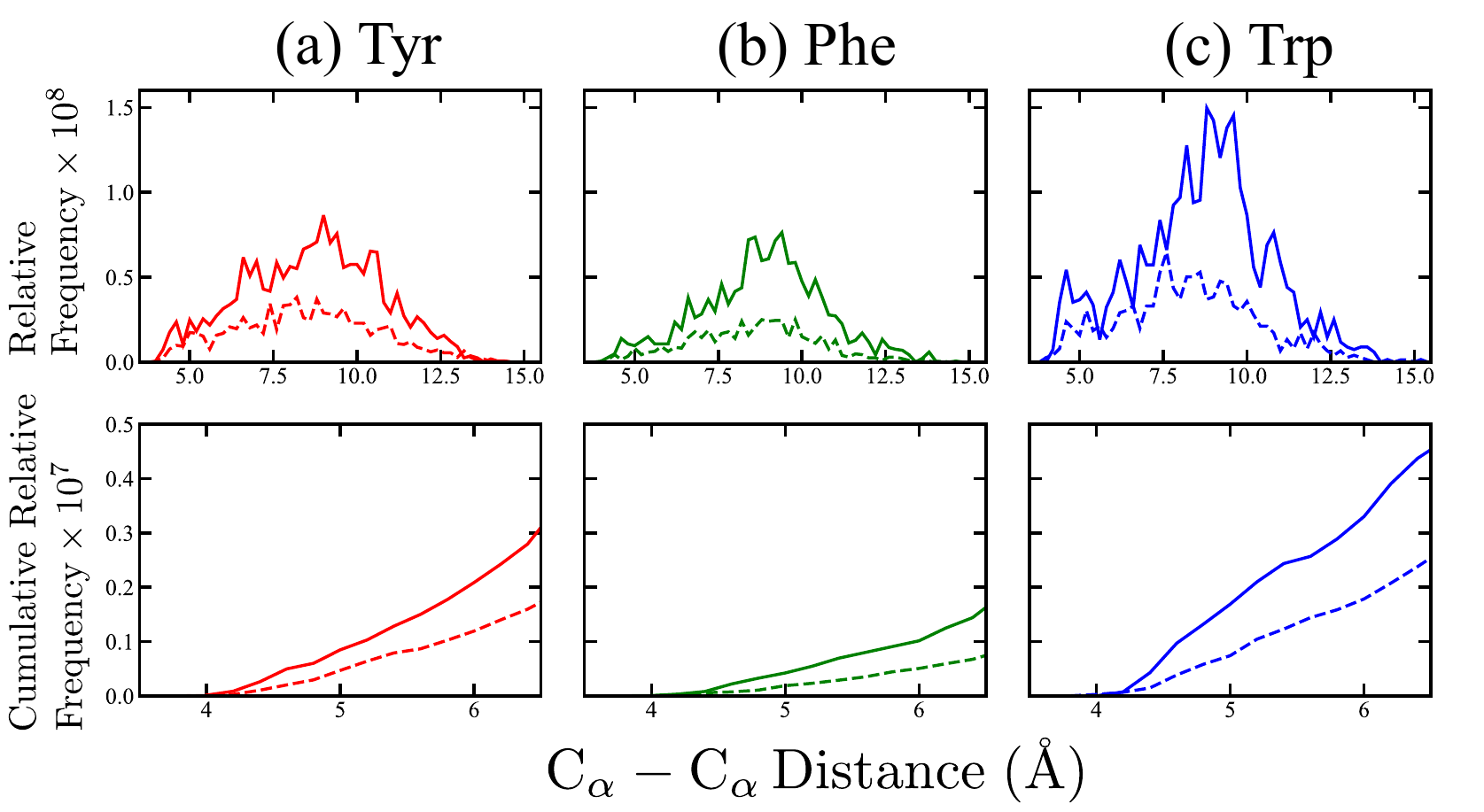}
\end{center}
\noindent
{\bf Fig.~S2:} Statistics of cation-$\pi$-like contacts.
Distributions of C$_\alpha$--C$_\alpha$ distance between
a positively charged residue [arginine (solid curve) or lysine (dashed
curve)] and an aromatic residue [tyrosine (a), phenylalanine (b), or
tryptophan (c)] are obtained from the same dataset of 6,943 high-resolution 
X-ray structures (from a non-redundant set\cite{robert}) 
used in Fig.~2 of the main text. The bin size
for C$_\alpha$--C$_\alpha$ distance and the color code for 
different residue pairs are also identical to those in Fig.~2 of the main text.
For a given residue pair [Arg-Tyr, Lys-Tyr (a); Arg-Phe, Lys-Phe (b);
or Arg-Trp, Lys-Trp (c)], the relative frequency of a given 
C$_\alpha$--C$_\alpha$ distance bin is the total number of instances
in the dataset in which the C$_\alpha$--C$_\alpha$ distance between 
the given pair of residues falls within the bin, normalized (divided) 
by the product of the two total numbers of residues in the dataset 
for the two residues making up the pair.
Cumulative relative frequency at a given distance is the sum of
relative frequencies for distances lower or equal to the given distance.
Here, cumulative relative frequencies are reported up to 
C$_\alpha$--C$_\alpha$ distance of 6.5 \AA, which is illustrative of
common criteria for a residue-residue contact.
The plotted distributions show clearly that arginine-aromatic contacts are
consistently and significantly more numerous
than lysine-aromatic contacts when compared on the same footing,
suggesting strongly that the overall arginine-aromatic 
interactions are energetically more favorable than the overall
lysine-aromatic interactions.\cite{MJ85}
\\

\vfill\eject

\begin{center}
   \includegraphics[width=0.72\columnwidth]{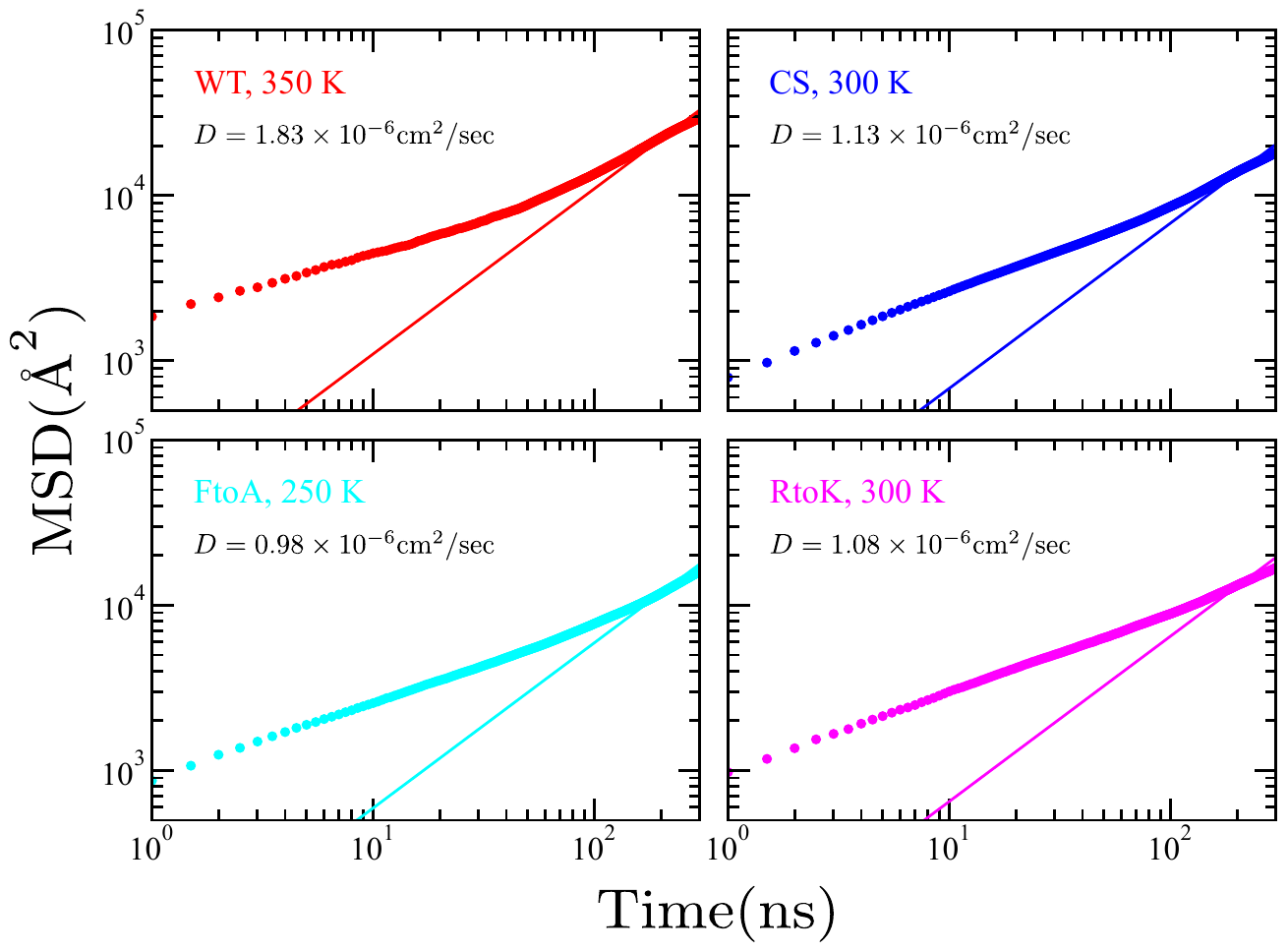}
\end{center}
\noindent
{\bf Fig.~S3:}  Verification of liquid-like dynamics of
simulated condensed phases. As in Dignon et al.,\cite{dignon18}
a relevant time-dependent mean-square deviation ${\rm MSD}(t)$ of
molecular coordinates was simulated
to provide evidence for liquid-like behavior in our model systems, 
viz.,\cite{allen1987}
\begin{equation*}
{\rm MSD}(t)= 
\frac{1}{n} \biggl \langle \sum_{\mu=1}^n 
\Bigl |\bigl[{\bf r}_{\mu,{\rm CM}}(t+t_0)-{\bf r}_{\rm CM}(t+t_0)\bigr]
-\bigl[{\bf r}_{\mu,{\rm CM}}(t_0)-{\bf r}_{\rm CM}(t_0)\bigr]\Bigr |^2 
\biggr \rangle_{t_0} \; ,
\end{equation*}
where $\mu = 1, 2,\dots,n$ labels the model IDR chains, $n$ is the total
number of IDR chains in the simulation system, ${\bf r}_{\mu,{\rm CM}}$
$=\sum_{i=1}^N m_i{\bf r}_{\mu i}/\sum_{i=1}^N m_i$ 
is the center-of-mass position of the $\mu$th chain, with $m_i$ being the
mass of the $i$th bead (residue) along an IDR chain, ${\bf r}_{\rm CM}$
$=\sum_{\mu=1}^n {\bf r}_{\mu,{\rm CM}}/n$ 
is the center-of-mass of the entire collection of $n$ chains, and the
average is over the initial time point $t_0$. 
By subtracting drifts in molecular coordinates 
arising solely from the diffusion of the entire system's center of mass
(see Fig.~S4), the above-defined
${\rm MSD}(t)$ values, which are provided by the circles in the plots,
are a useful measure of the liquidity of our simulated system.
Diffusion coefficients,
$D = \{\lim_{t\to\infty}d [{\rm MSD}(t)]/dt\}/6$,
were then estimated, as indicated by the fitted straight line in each plot.
Shown examples for the four Ddx4 IDR variants were simulated using 
the KH model with relative permittivity $\epsilon_{\rm r}=40$ at 
the indicated temperatures, each of which is lower than the 
respective system's critical temperature.
The magnitudes of our simulated $D$s
are similar to those simulated by Dignon et al.  for their model FUS systems 
(Fig.~S12 of Ref.~\citen{dignon18}). Note that our simulated $D$s
for the model Ddx4 IDR systems are, not unexpectedly, approximately three 
orders of magnitude higher than the corresponding experimental 
values\cite{jacob2017} because a unphysically low friction coefficient 
was necessitated in our Langevin dynamics simulations in order to 
accelerate sampling and also because a coarse-grained representation 
of the IDRs was used.

\vfill\eject

\begin{center}
   \includegraphics[width=0.86\columnwidth]{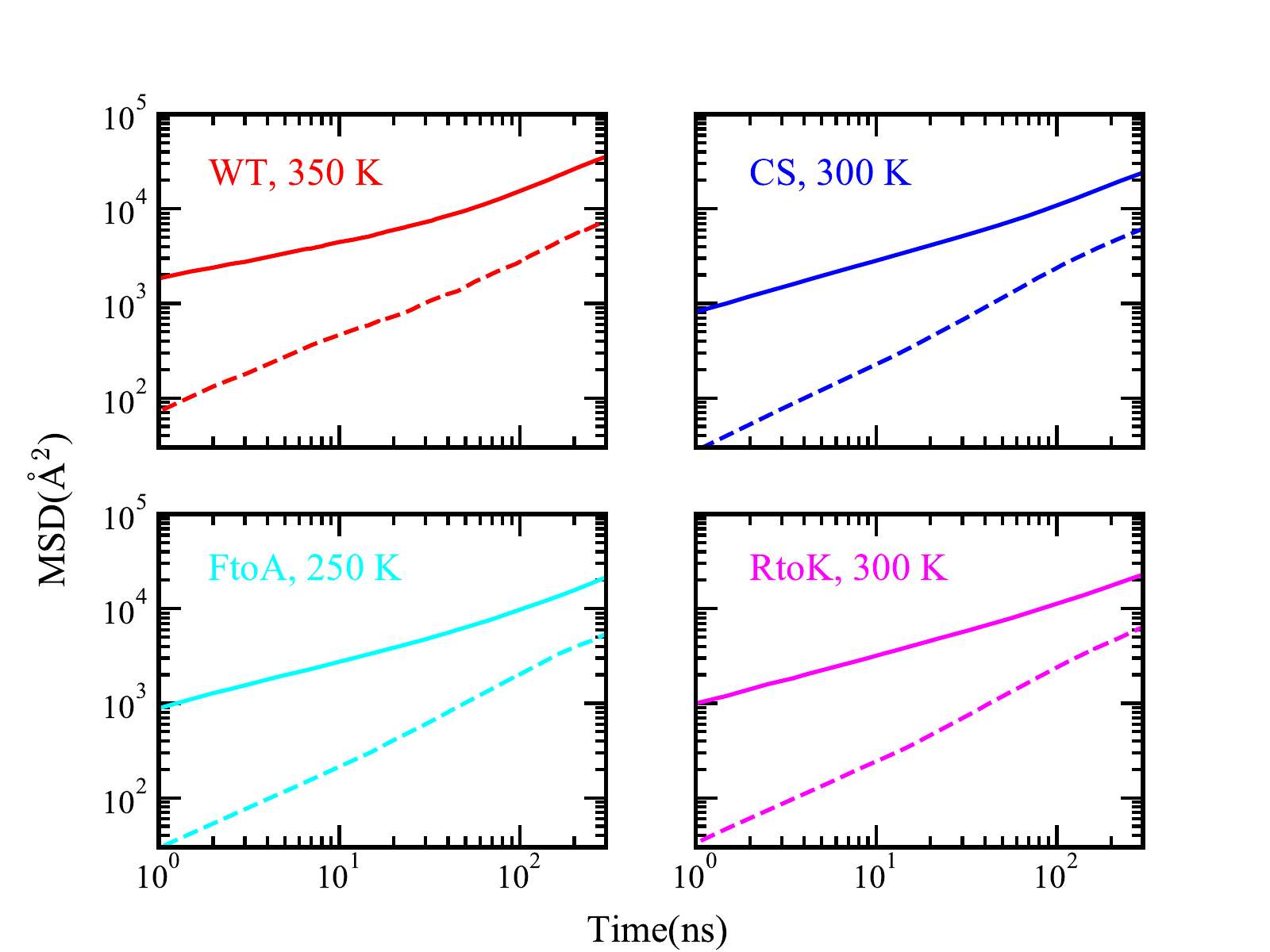}
\end{center}
{\bf Fig.~S4:} Center-of-mass diffusion of the simulated Ddx4 IDR
systems. Data are from the same systems as those in Fig.~S3.
The solid curves provide the mean-square deviation
of the center-of-mass positions of the IDRs {\it without} subtracting
the the center-of-mass position of the entire system, in which case
\begin{equation*}
{\rm MSD}(t)
=\frac{1}{n} \biggl \langle \sum_{\mu=1}^n 
\Bigl |{\bf r}_{\mu,{\rm CM}}(t+t_0)
-{\bf r}_{\mu,{\rm CM}}(t_0)\Bigr |^2 \biggr \rangle_{t_0} \; ,
\end{equation*}
whereas the dashed curves represent the diffusion of the center of mass
of the entire system of $n$ IDRs, given by
${\rm MSD}(t)= \langle \vert {\bf r}_{\rm CM}(t+t_0)-{\bf r}_{\rm CM}(t_0)
\rangle_{t_0}$. 
Echoing the findings in Fig.~S3,
a comparison of the solid and dashed curves in the present figure 
indicates that there is significant diffusion of individual IDRs relative 
to the center of mass of the entire collection of IDR chains.

\vfill\eject


\begin{center}
   \includegraphics[width=0.9\columnwidth]{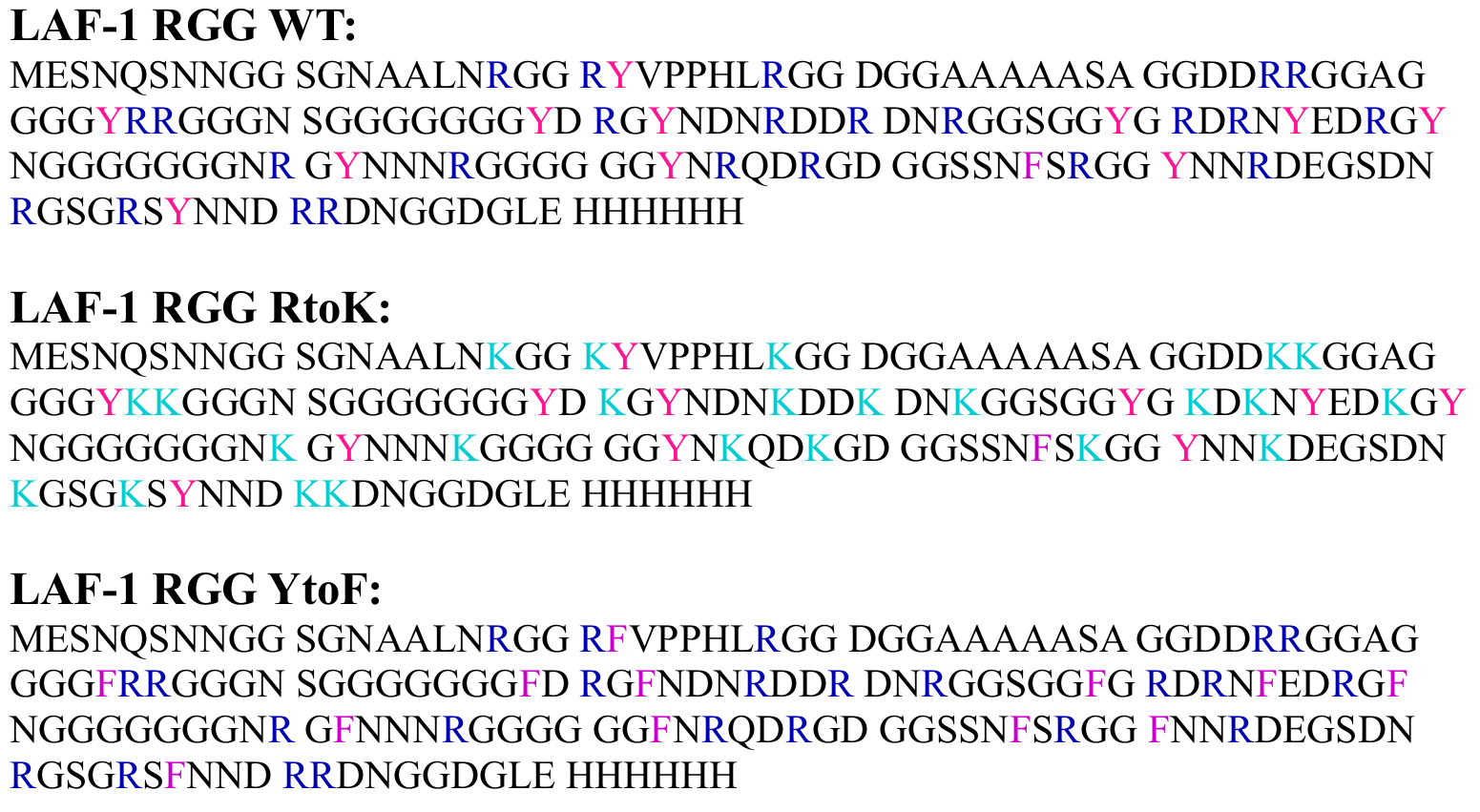}
\end{center}
{\bf Fig.~S5:} LAF-1 IDR sequences simulated in Fig.~6 of the main text.
Amino acid residues are given by one-letter code.
The 176-residue LAF-1 RGG WT sequence and its RtoK and YtoF mutants 
(all with a polyhistidine tag) are the same as those used for in-vitro
studies and given in the Supporting Information of Ref.~\citen{Schuster2020}.
Underscoring the substitution positions considered in our simulations, 
K, F, R, and Y residues are shown, respectively, in cyan, magenta, blue, 
and pink, as in Fig.~6a of the main text. Other residues, including A, D, 
and E which are highlighted in Fig.~3a and Fig.~S1 for Ddx4 IDRs, are 
shown here in black.

\vfill\eject


\begin{center}
   \includegraphics[width=0.96\columnwidth]{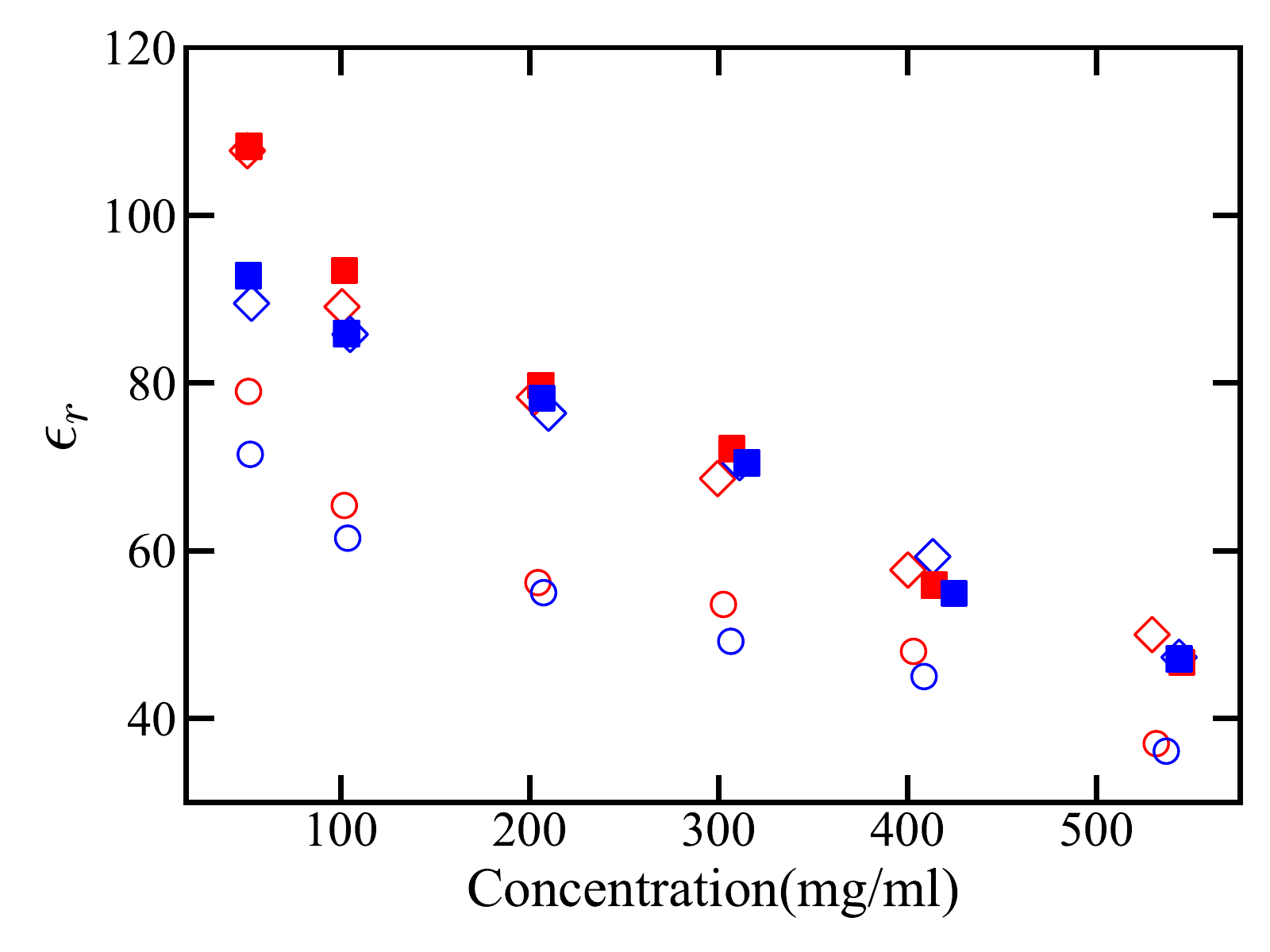}
\end{center}
{\bf Fig.~S6:} Simulated IDR-concentration-dependent relative permittivity.
Shown results---part of which are also provided in Fig.~7a of the main 
text---are for the WT Ddx4 IDR. 
Simulations were conducted using the SPC/E water model with 100 mM NaCl
(circles), the TIP3P water model without salt (squares), and the TIP3P model
with 100 mM NaCl (diamonds).
Red symbols represent $\epsilon_{\rm r}$ values simulated using the full 
force field, whereas blue symbols denote $\epsilon_{\rm r}$ values simulated 
while the electric charges on the sidechains of arginine, lysine, 
glutamic acid, and aspartic acid are artifically turned off.
The $\epsilon_{\rm r}$ values plotted here are tabulated in Table~S1.

\vfill\eject

\begin{center}
   \includegraphics[width=0.96\columnwidth]{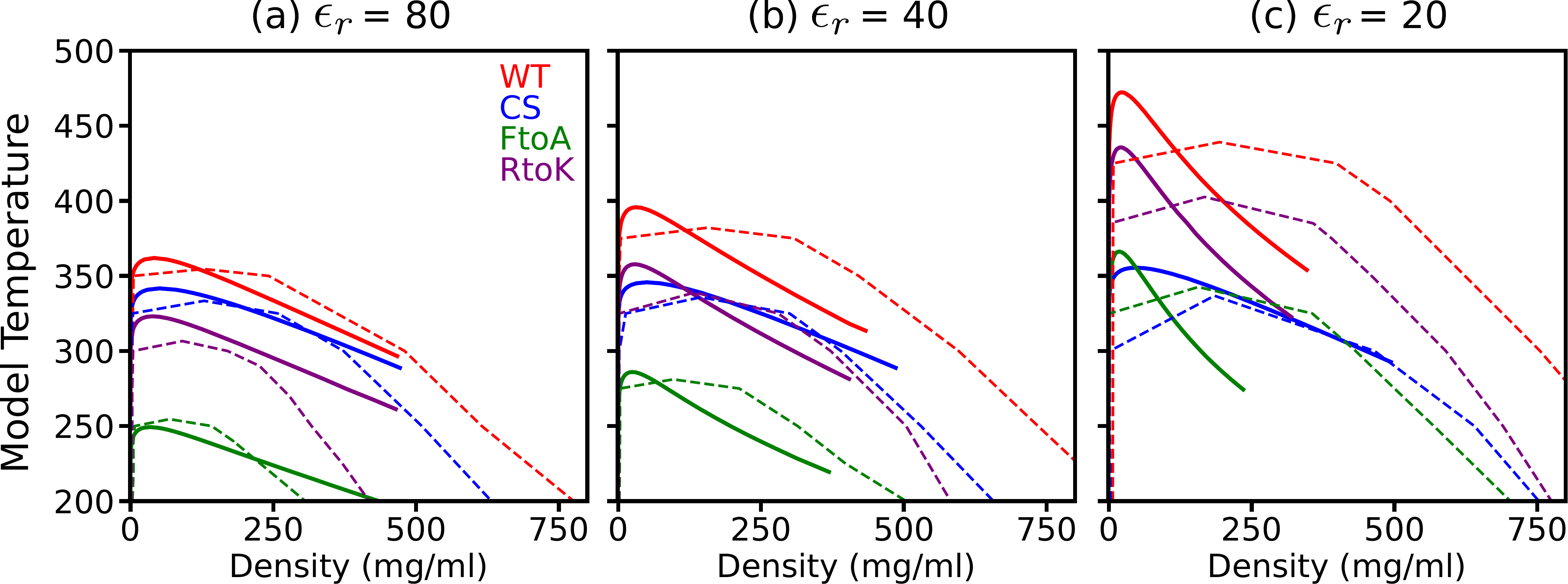}
\end{center}
{\bf Fig.~S7:} Comparing analytical theory with
simulation for sequence-dependent liquid-liquid phase separation
of model Ddx4 systems. Phase diagrams simulated using the explicit-chain 
KH model under different permittivities ($\epsilon_{\rm r}$)
for the four Ddx4 IDRs from Fig.~4 of the main text are replotted here
as dashed curves. Predicted phase diagrams by the RPA+FH theory that 
afford the best overall fit, at $z/2=4.3$, are shown as solid curves.

\vfill\eject
\noindent
{\Huge\bf SI Table}\\

\noindent
{\bf Table S1:} IDR-concentration-dependent relative permittivity,
$\epsilon_{\rm r}$, simulated for WT Ddx4 IDR using the SPC/E and
TIP3P atomic models of water at $T=300$ K.

\begin{center}
\begin{tabular}{|c|c||c|c||c|c|}
\hline
\multicolumn{2}{|c||}{$\quad$ {\bf SPC/E + salt$^{\rm b}$} $\quad$} &
\multicolumn{2}{|c||}{$\quad$ {\bf TIP3P, no salt} $\quad$} &
\multicolumn{2}{|c|}{$\quad$ {\bf TIP3P + salt$^{\rm b}$} $\quad$}\\
\hline
$\quad$ [Ddx4]$^{\rm a}$ $\quad$ & $\epsilon_{\rm r}$ &
$\quad$ [Ddx4]$^{\rm a}$ $\quad$ & $\epsilon_{\rm r}$ &
$\quad$ [Ddx4]$^{\rm a}$ $\quad$ & $\epsilon_{\rm r}$ \\
\hline
\hline
{\bf 51.1} & {\bf 79.0} & {\bf 51.3} & {\bf 108.2} & {\bf 50.5} & {\bf 107.7}\\
(52.04) & (71.5) & (51.1) & (92.8) & (52.7) & (89.5)\\ 
\hline
{\bf 101.8} & {\bf 65.4} & {\bf 101.9} & {\bf 93.4} & {\bf 100.6} & {\bf 89.1}\\
(103.6) & (61.5) & (103.0) & (85.9) & (104.9) & (85.8)\\ 
\hline
{\bf 204.3} & {\bf 56.2} & {\bf 205.8} & {\bf 79.7} & {\bf 202.4} & {\bf 78.3}\\
(207.3) & (55.0) & (206.5) & (78.2) & (209.9) & (76.4)\\ 
\hline
{\bf 302.5} & {\bf 53.6} & {\bf 307.0} & {\bf 72.2} & {\bf 299.4} & {\bf 68.6}\\
(306.4) & (49.2) & (315.0) & (70.5) & (311.0) & (70.5)\\ 
\hline
{\bf 403.1} & {\bf 48.0} & {\bf 414.1} & {\bf 55.9} & {\bf 400.1} & {\bf 57.7}\\
(408.7) & (45.0) & (424.6) & (54.9) & (413.3) & (59.3)\\ 
\hline
{\bf 531.6} & {\bf 37.0} & {\bf 545.2} & {\bf 46.7} & {\bf 529.4} & {\bf 50.0}\\
(536.9) & (36.1) & (543.8) & (47.1) & (543.7) & (47.3)\\ 
\hline
\end{tabular}
\end{center}

$^{\rm a}$
Concentrations (in mg ml$^{-1}$) and simulated $\epsilon_{\rm r}$ values
given in bold font are for systems that apply the full force field;
those given in ordinary roman (non-bold) font and in parentheses are for
systems in which the electric charges on the
sidechains of arginine, lysine, glutamic acid, and aspartic acid
of WT Ddx4 IDR are artifically turned off.

$^{\rm b}$ [NaCl] = 100 mM.

\vfill\eject

\vfill\eject \noindent
{\Large\bf References}\\

\vfill\eject

\vfill\eject

\end{document}